\documentclass[fleqn,usenatbib]{mnras}

\usepackage{newtxtext,newtxmath}

\usepackage[T1]{fontenc}

\DeclareRobustCommand{\VAN}[3]{#2}
\let\VANthebibliography\thebibliography
\def\thebibliography{\DeclareRobustCommand{\VAN}[3]{##3}\VANthebibliography}

\usepackage{graphicx}	
\usepackage{amsmath}	

\newcommand{\modelA}{{\tt constant$\ast$tbabs$\ast$(diskbb+cutoffpl$_{\rm 1}$+cutoffpl$_{\rm 2}$+gaussian)}}

\newcommand{\modelC}{{\tt\string constant$\ast$tbabs$\ast$(thcomp*diskbb+cutoffpl$_{\rm 2}$+gaussian)}}
\newcommand{\modelD}{{\tt\string constant$\ast$tbabs$\ast$(diskbb+relxill+cutoffpl$_{\rm 2}$)}}

\newcommand{\SwiftJ}{Swift~J1727.8--1613 }

\title[Spectral Properties during Rising Phases of Swift J1727.8–1613]{The Broadband X-ray Spectral Properties during the Rising Phases of the Outburst of the New Black Hole X-ray Binary Candidate Swift J1727.8--1613}

\author[He-Xin Liu et al.]{
He-Xin Liu,$^{1,2,3,4}$
Yan-Jun Xu,$^{3,5}$\thanks{E-mail: xuyj@ihep.ac.cn}
Wei Yu$^{6}$
Shuang-Nan Zhang$^{3,5}$
and Jin-Lu Qu$^{3,5}$
\\
$^{1}$Research Center of Astronomy, QingHai University, Xining, 810016, PR China\\
$^{2}$Department of Physics and Astronomy, QingHai University, Xining, 810016, PR China\\
$^{3}$Key Laboratory for Particle Astrophysics, Institute of High Energy Physics, Chinese Academy of Sciences, 19B Yuquan Road, Beijing 100049, People's Republic of China\\
$^{4}$Dongguan Neutron Science Center, Zhongziyuan Road, Dongguan 523808, People's Republic of China\\
$^{5}$University of Chinese Academy of Sciences, Chinese Academy of Sciences, Beijing 100049, People's Republic of China\\
$^{6}$Institut f\"ur Astronomie und Astrophysik, Kepler Center for Astro and Particle Physics, Eberhard Karls Universit\"at, T\"ubingen, 72076, Germany\\
}

\date{Accepted XXX. Received YYY; in original form ZZZ}

\pubyear{\the\year{}}

\begin{document}
\label{firstpage}
\pagerange{\pageref{firstpage}--\pageref{lastpage}}
\maketitle

\begin{abstract}
We report data analysis results about the outburst evolution and spectral properties during the hard state of the recently discovered X-ray transient Swift J1727.8--163 as observed by \emph{Insight}-HXMT.
The broadband X-ray spectrum of Swift J1727.8--163 is more complex than the most typical spectral patterns of black hole X-ray binary systems, with not only a comparatively weaker reflection component but also an additional high-energy component, manifesting itself as a hard X-ray tail beyond the thermal Comptonization description detectable below 100 keV. Using reflection models combining thermal ({\tt reflkerr}) and hybrid ({\tt reflkerr\_bb}) Comptonization, we find that the emission arises from a plasma containing both thermal and non-thermal electrons, with the non-thermal tail forming a smooth extension of the thermal continuum. This supports a compact hybrid corona. We find that the inner disk radius remains at the ISCO and the reflection fraction stays low, with no abrupt spectral evolution across epochs. These results highlight the role of hybrid corona in shaping hard-state emission in black hole X-ray binaries.
\end{abstract}




\section{Introduction}
Low-mass Black Hole X-ray Binaries (BHXRBs) are usually transient sources that undergo active periods accompanied by the increase of X-ray flux, commonly referred to as ``outbursts". The radiation from a black hole X-ray binary during an outburst is usually thought to consist of several components. One is thermal radiation from the accretion disk, others are non-thermal radiation produced by Compton scattering of seed photons from the disk by hot electrons in the corona and a reflection component which come from disk material reprocessing the illuminating coronal photons \citep{1988MNRASGuilbert,1988ApJLightman,1976ApJShapiro}

The radiative properties of the different components observed during the outbursts of BHXRBs evolve over time. Outburst evolution can be distinguished into different spectral states based on the variations in the spectral properties and the accompanying timing nature, generally: low hard state (LHS); hard intermediate state (HIMS); soft intermediate state (SIMS); high soft state (HSS) (see the \citealt{2010LNPBelloni,2006ARA&ARemillard,2009MNRASMott}).

The relation of the corona and jet in accreting black hole systems (either stellar-mass or supermassive ones) has long been an interesting research topic. On one hand, it is widely known that X-rays dominated by emission from the corona and radio emission from the large scale jet is well correlated for a large sample of black hole transients (BHTs) and supermassive black holes, indicating that the corona and jet are ubiquitously connected \citep{2003A&ACorbel,2003MNRASMerloni,2012MNRASGallo}. On the other hand, the possible miscrophysics involved tying the two parts together are far from clear and it is most likely over-simplistic to assume that the corona and the jet base are synonymous, although it is probably reasonable to assume that the corona could be outflowing at mildly relativistic velocities and some observational evidence have been found (e.g., \citealt{1999ApJBeloborodov,2021NatCoYouBei}). In addition, the origin of the X-ray emission in black hole X-ray binaries remains debated. While coronal emission is generally attributed to inverse Compton scattering of disk photons, there is also evidence that optically thin synchrotron emission from jets can provide seed photons and contribute significantly to the hard X-ray emission at very low accretion rates \citep{2004A&AFalcke,2006MNRASKording}.
In particular, the timing studies as well as multi-band studies, provide further evidence that jet and corona may coexist and evolve in conjunction with one another \citep{2022NatAsMendez,2023MNRASMaRuican,2022ApJLiuHX,2023MNRASYangZX,2023SciYou}.

The evolution of BHTs during outbursts has been extensively
studied since early days of X-ray astronomy (e.g., see the references in the \citealt{2006ARA&ARemillard}). Their full outburst light curves in general follow a
fast rise-slow decay profile, with the X-ray flux reaching the peak within a few days. The corresponding properties during the rising phases are comparatively less well investigated, and are somehow of special interest since they are believed to encode important information about the process of instability propagation in the accretion flow that leads to the full outburst. Significant changes have been reported in the physical properties of the corona based on multi-epoch X-ray observations during the early phases of the outbursts of several BHTs. The coronal spectral continuum in the X-ray band can be well approximated by a power-law model with an exponential cutoff at high energies, a phenomenological representation for Comptonization by thermally-distributed electrons. The characteristic high energy cutoff is commonly observed to undergo a gradual decrease during the rising phases of the hard state (e.g., \citealt{2009MNRASMott,2010ApJTitarchuk,2017ApJXu,2020ApJYan,2023ApJYou}). At energies above several hundred keV, an additional “hard tail” is often observed in the soft $\gamma$-ray band. This component may originate either from synchrotron or synchrotron self-Compton emission in the jet, or from Comptonization of seed photons by non-thermal electrons in the corona, and its exact origin remains debated (e.g., \citealt{2015A&APepe} and see the references \citealt{2021NewARMotta}).

The new X-ray source \SwiftJ discovered by Swift/BAT was initially identified as GRB 230824A. However, the subsequent MAXI/GSC observations revealed that the source exhibited a rapid flux increase and is identified as a new galactic X-ray transient \citep{2023ATel16205Negoro}. In addition, the optical counterpart of \SwiftJ has been found and indicates that it is a black hole X-ray binary candidate with a distance estimate of $2.7\pm0.3$ kpc based on empirical correlations \citep{2024A&AMata}.
 
The source is of special interest because of its outstanding brightness in X-rays (reaching about 7 Crab near the peak; \citealt{2023ATel16215Palmer}, and also because it exhibited rare large amplitude flaring behaviors that was only previously known in a small number of X-ray binaries, i.e., V404 Cygni, V4641 Sgr, and Swift~J1858.6--0814, see the references in \citealt{2020ApJHare}). The source was thought to be a V404 Cygni like object due to repeating bursts triggering INTEGRAL and Swift/BAT \citep{2023GCNKennea}, as well as rapid sub-second variations detected by NICER \citep{2023GCNO'Connor} during the onset of the outburst on Aug 24 and 25, 2023. These earliest flaring were missed by \emph{Insight}-HXMT. Prominent flaring states were subsequently found and extensively monitored by \emph{Insight}-HXMT later on during the outburst. But those were observed during the development of the outburst, and different from the active phases of V404 Cygni occurred in 2015 when the source switched between giant flaring states and quasi-quiescent states \citep{2017ApJWalton}. 

\citet{2024ApJPeng} found that there are two hard components in the spectrum and measured a high spin of $0.98_{-0.07}^{+0.02}$.
\citet{2023arXivMereminskiy} reported on the detection of type-C QPO during the initial stages of this outburst and also found that an additional power-law tail extending at least to 400~keV based on INTERGRAL observations. Using IXPE data, \citet{2024ApJZhao} firstly present the polarimetric analysis QPOs in a black hole binary of \SwiftJ, and they found that the polarization degree (PD) and polarization angle (PA) exhibit no modulations in relation to the QPO phase, which is inconsistent with the expectation of the Lense–Thirring precession of the inner flow. 
Type-C LFQPOs were detected throughout most of the \emph{Insight}-HXMT observation period\citep{2024MNRASYW}, covering nearly the entire majority outburst, while no Type-B or Type-A QPOs were found. \citet{2024ApJYang} reported a high-energy rms excess in the QPO spectra, linked to a possible jet-related hard component and jet base precession, though its origin and connection to the QPO require further investigation.
The X-ray polarization measurements for \SwiftJ with the Imaging X-ray Polarimetry Explorer (IXPE) in the bright hard state indicate that the corona geometry is elongated orthogonal to the jet \citep{2023ApJVeledina}, more in favor of sandwich-like corona covering the accretion disk rather than a lamppost jet-like corona, similar to the case of Cygnus X-1 with recent reports based on IXPE observations \citep{2022SciKrawczynski}. Furthermore, \citet{2024ApJIngram} traced the evolution of polarisation across a hard to soft state transition of \SwiftJ and suggested that the X-ray corona is extended in the disk plane for the entire hard intermediate state. \citet{2024A&ABouchet} first detected a soft $\gamma$-ray polarized signal with INTEGRAL/IBIS, showing a polarization angle aligned to the jet axis, coinciding with a radio flare during the SIMS transition. 

These studies highlight the intricate nature of this source's spectrum and timing, emphasizing the challenges in unraveling the relationships between the disk, corona, and jet. This calls for a meticulous approach to investigating the source, especially when modeling its spectrum.

This paper focuses on the spectral evolution of \SwiftJ during the rising phases of its outburst, while the results concerning the flaring period have been reported by \citet{2025ApJCao}. 
This paper is structured as follows: in Section~\ref{sec:data}, we describe the observations used in this paper and details about the data reduction process; in Section~\ref{sec:spectra}, we present our spectral modeling process; in Section~\ref{sec:dis} we discuss about the results with their physical implications; we summarize and conclude our paper in Section~\ref{sec:summary}.


\section{Observations and Data Reduction} \label{sec:data}
\subsection{Insight-HXMT} 
\emph{Insight}-HXMT, China's first X-ray astronomical satellite launched on June 15, 2017, carries three slat-collimated instruments: the Low\citep{2020SCPMA..6349505C}/Medium\citep{2020SCPMA..6349504C}/High\citep{2020SCPMA..6349503L} Energy X-ray Telescope respectively, abbreviated as LE/ME/HE. More details about \emph{Insight}-HXMT can be found in \citet{2020SCPMA..6349502Z}. The \emph{Insight}-HXMT Data Analysis software (HXMTDAS, v2.06) is used to process and filter the data following the official recommendations: Earth elevation angle $>10^\circ$; the geometric cutoff rigidity (COR) is larger than 6 GV; the offset for the point position is smaller than $0.04^\circ$; data are used at least 300 s before and after the South Atlantic Anomaly (SAA) passage. The backgrounds are estimated with the official tools: LEBKGMAP, MEBKGMAP and HEBKGMAP in version 2.0.6. Detailed discussions of its calibrations and background are given in \citet{2020JHEAp..27...64L} and \citet{2020JHEAp..27...24L} (LE), \citet{2020JHEAp..27...44G} (ME), and \citet{2020JHEAp..27...14L} (HE), respectively. Notably, we used updated Insight-HXMT CALDB V2.07 to resolve the issues regarding the LE calibration mentioned in \citet{2024ApJPeng}.

\begin{table*}
\caption{\emph{Insight}-HXMT Observations of \SwiftJ used in this Work}
\label{tab:hxmt}
\centering
{\scriptsize
\begin{tabular}{cccccccccc}
\hline\hline
Epoch & ObsID &  Start Time & End Time &
Exposure(s) & HE rate (cts/s) & 
ME rate(cts/s) & LE rate(cts/s) & Hardness
\\
&  &  &
 &  &27--150~keV & 
10--35~keV& 2--10~keV & 2--10~keV/2--4~keV\\
\hline
Epoch 1 & P061433800104 & 2023-08-25T18:17:36 & 2023-08-25T21:27:53 & 2652 & 3171 &1518 & 1911 & 0.88 \\
Epoch 2 &P061433800108 & 2023-08-26T06:49:19 & 2023-08-26T10:09:01 & 3777 & 3132 & 1658 &  2201 & 0.87\\
Epoch 3 & P061433800203 & 2023-08-27T08:16:49 & 2023-08-27T11:31:15 & 1696 & 3059 & 1895 & 2699& 0.84\\
Epoch 4 &P061433800210 & 2023-08-28T06:23:23 & 2023-08-28T09:43:13& 3574 & 2884 & 1850 & 3061 & 0.81\\
Epoch 5 &P061433800301 & 2023-08-29T07:22:51 & 2023-08-29T11:05:25 & 2145 & 2609 & 1902 & 3158 & 0.77\\
\hline
\end{tabular}
}
\end{table*}

\subsection{NuSTAR}
NuSTAR conducted its first observation of the new black hole X-ray binary candidate \SwiftJ on 2023 August 26 starting from UT 07:16:09 (ObsID: 90501337002) for an exposure of $\sim1$ ks. This observation is contemporary with the second \emph{Insight}-HXMT observation in Table~\ref{tab:hxmt}.  We include this NuSTAR observation in this paper to check the consistency between \emph{Insight}-HXMT and NuSTAR spectral measurements.  We reduced the data using NuSTARDAS pipeline v$2.2.1$ and CALDBv20230918. This source is exceptionally bright, with a count rate significantly exceeding 100 cts/s. Consequently, the {\tt statusexpr} parameter was set to \textsl{`STATUS==b0000xxx00xxxxxxxx000\&\&(SHIELD==0)'}, when processing the data using {\tt nupipeline}. The source spectra were extracted from a circular region with the radius of $90 ^{\prime \prime}$ from the two focal plane modules (FPMA and FPMB). Corresponding background spectra were extracted from a circular region of radius $120 ^{\prime \prime}$ away from the source. 

The XSPEC software package v12.13.1 \citep{1996Arnaud} is used to fit the spectra. Uncertainty estimated for each spectral parameter is quoted at a 90\% confidence level. We use the cross-sections from \citet{1996ApJVerner} and abundances from \citet{2000ApJWilms} during the spectral fitting procedures in this paper.
The energy bands, adopted for spectral analysis, are 2--10~keV (LE), 10--20, 22--35~keV (ME), and 35--120~keV (HE) in this work. Data between 20 and 22~keV are ignored because of a calibration related silver line structure. The HE data start to become background dominated above 120~keV, therefore we only include hard X-ray coverage up to 120~keV. And we added 0.5\% systematic error to each channel. Parameter uncertainties were estimated using Markov Chain Monte Carlo (MCMC) simulations within XSPEC. The chains were initialized around the best-fitting solution and run until convergence. After removing the burn-in phase, the posterior distributions were used to derive the reported parameter uncertainties.
The error bars are given and plotted at the 90 percent level for all parameters of interest. The acquisition of the fluxes for each component was calculated in XSPEC using the convolutional model {\tt cflux}. 

\section{Spectral Analysis} \label{sec:spectra}
\subsection{Evolutionary}
\begin{figure*}
\centering
 \includegraphics[width=0.45\textwidth]{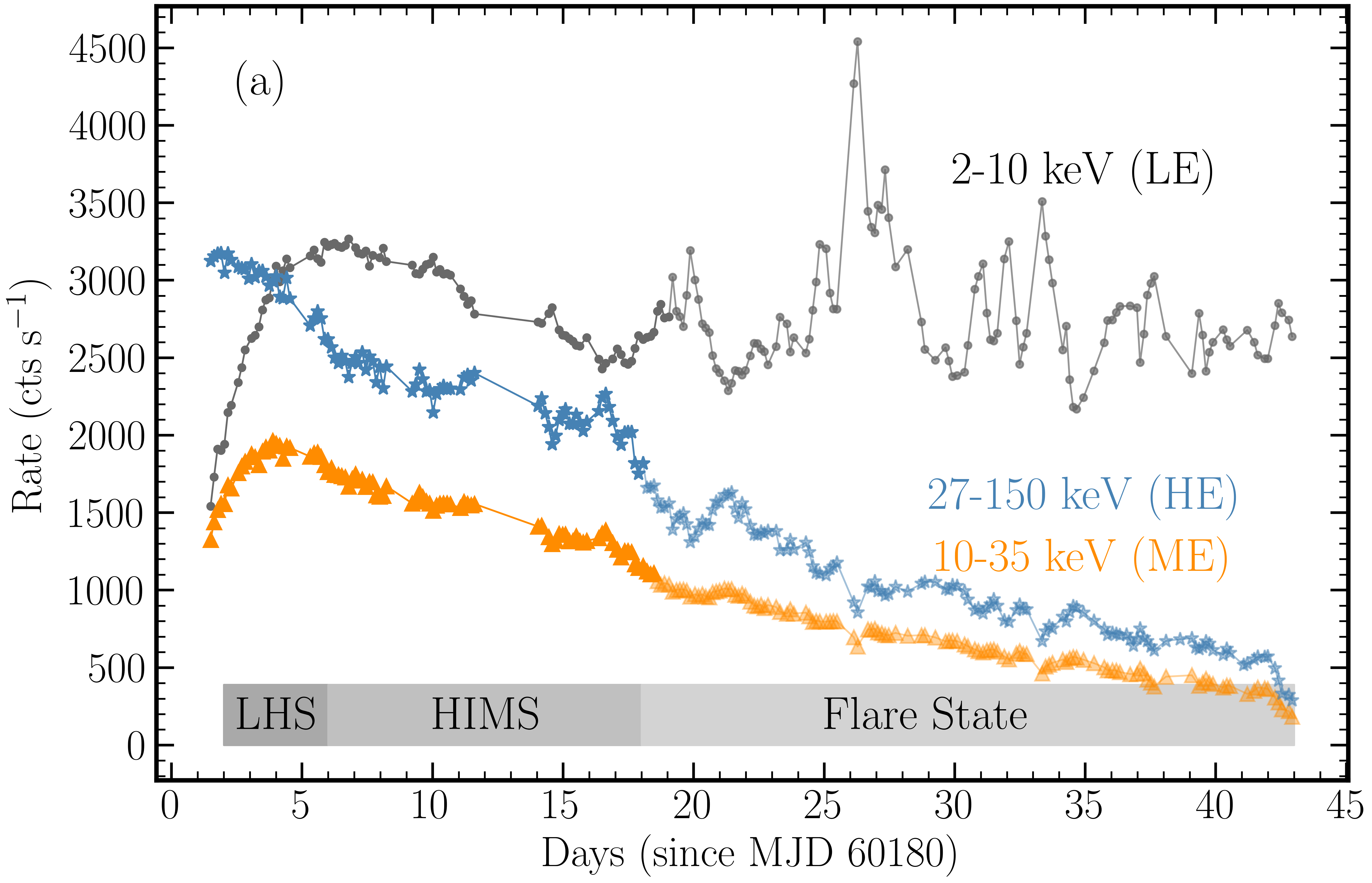}
 \includegraphics[width=0.45\textwidth]{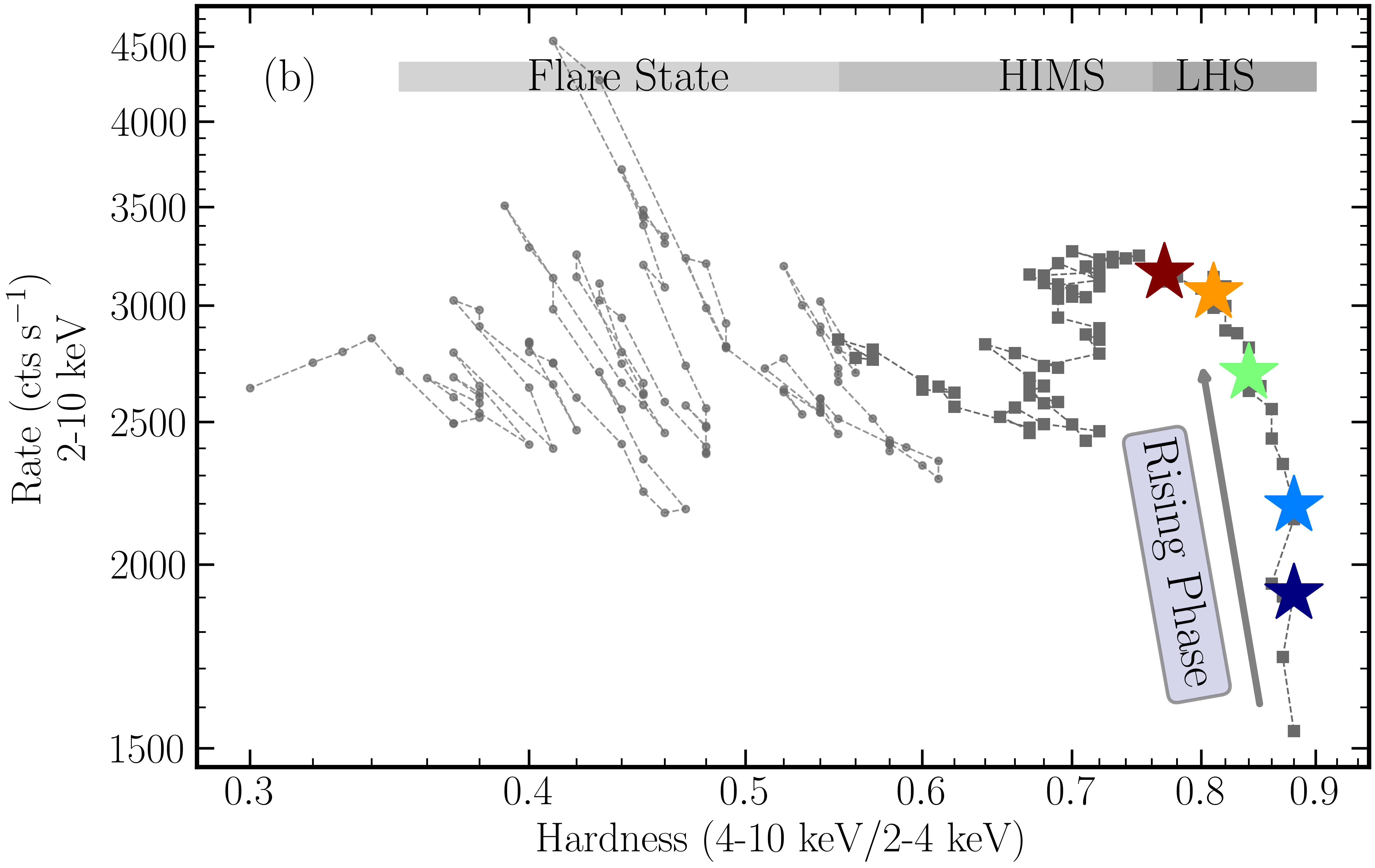}
\caption{(a) Light curves of \SwiftJ during its 2023 outburst from three instruments of \emph{Insight}-HXMT: LE (black), ME (orange) and HE (blue). Each point corresponds to one exposure. (b) Hardness--Intensity Diagram of (HID) \SwiftJ based on LE. Hardness is defined as the ratio of count rates between 4--10 and 2--4~keV. Colored stars denotes the observations analyzed in this work.}
\label{fig:hxmt-lc}
\end{figure*}

\SwiftJ was observed regularly with \emph{Insight}-HXMT from 25th August 2023 to 4th October 2023, covering almost the entire outburst. We show the count rate evolution for this outburst using the three detectors of the \emph{Insight}-HXMT in the left panel of Figure~\ref{fig:hxmt-lc}. We see that there are roughly two phases, a normal outburst state before $\sim$MJD 60199 and a following state with multiple flares. Also, the Hardness Intensity Diagram (HID) using the LE detector data is shown in the right panel of Figure~\ref{fig:hxmt-lc}. The hardness ratio is defined as the ratio of LE count rates between 2--10~keV and 2--4~keV bands. Its flux rose quickly during the rising phase, the LE count rate increased from $\sim$1500 cts/s from the start of our observations to the maximum of $\sim 3200$ cts/s, while the hardness ratio displayed a small decrease($\sim$0.88-0.75). A photon index of less than 2 (see Section~\ref{sec:spectra}) also indicates that the source was in its LHS before $\sim$MJD~60186. From about MJD 60186 to MJD 60198, the hardness ratio decreased and occasionally rebounced, but there were always type-C QPOs, so we classify this period as the HIMS \citep{2024MNRASYW}. 
The type-B QPO is an indication that the source enters the SIMS, but based on our timing study, the type-C QPO was found to be present until MJD~60222 \citep{2024MNRASYW}. However, after MJD~60198, the LE count rate as well as the hardness ratio indicate that the source underwent multiple flares before transitioning into the proper soft state, and therefore we classify MJD~60198 to MJD~60222 as the flare state.

For this extremely bright outburst of \SwiftJ, \emph{Insight}-HXMT captured the progress of its rapid rise in flux, providing us with an invaluable example to study the properties of the rising phase X-ray spectrum in BHTs. 
We choose five observations from the initial rise in the hard state, as shown by the colored stars in the HID in the right panel of Figure~\ref{fig:hxmt-lc}. The basic properties of selected spectral data used in this paper are detailed in Table~\ref{tab:hxmt}. Since the first two observations of \emph{Insight}-HXMT lack sufficiently long good time for LE, we select the third observation as the first spectrum to be fitted. For the remaining observations, our preferences for selection take into account the exposure time and the degree of flux variation, etc., forming a homogeneous \emph{Insight}-HXMT monitoring dataset covering the rising phase of \SwiftJ at roughly daily cadence.

\subsection{Spectral Comparison} \label{subsec:compar}
We first analyse our \emph{Insight}-HXMT observations at the five representative epochs in the rising phase to check the very basic spectral properties. 
In the top panel of Figure~\ref{fig:ratio} the shapes of the all spectra are modeled with an absorbed cutoff power-law model ({\tt tbabs*cutoffpl} in XSPEC notation), and we highlight any secondary features on top of the spectral continuum in the ratio plot in the bottom panel. As can be seen in Figure~\ref{fig:ratio}, the high energy part of the broadband X-ray spectra (20--120~keV) are much less variable than the low energy energy part (2--20~keV). Based on this simple estimate, the spectral indexes are about $1.52$, $1.51$, $1.57$, $1.64$, $1.71$, and the cutoff energies are about $35.5$, $30.2$, $28.5$, $28.6$, $29.7$ keV,  for Epoch 1, 2, 3, 4, 5, respectively. 
The simple cutoff power-law model fails to provide a satisfactory fit with all $\chi^{2}/d.o.f>2$ for the our datasets, and several prominent features can be seen in the ratio plots.
Similar to other BHTs, a broad Fe K emission line is found, believed to arise from reflection of the corona emission by the accretion disk.  
A clear exceeding tendency at the high energy end can be seen above 30 keV in the fitting residuals in every epoch and is observed to become more prominent from Epoch 1 to Epoch 5, which is very similar to the "hard X-ray tail" found in a number of BHTs in several hundred keV to MeV, such as Cyg~X-1\citep{2002ApJMcConnell}, GX~339--4\citep{2002MNRASWardziski}, and MAXI~J1820+070\citep{2021ApJZdziarski}, but is rare in the sense that this feature have not been reported to be detectable at such low energies before.  
Furthermore, we compare the spectral patterns in the LHS of several black hole systems that have recently undergone active periods based on \emph{Insight}-HXMT and NuSTAR observations (see Figure~\ref{fig:nustar}). The spectra are presented as ratio plots with respect to a simple cutoff power-law model. For \SwiftJ, given that the cutoff energy significantly affects the shape of the ratio spectra, the photon index and cutoff energy of the \emph{Insight}-HXMT spectra were fixed to the values (1.56 and 21.6 keV) obtained from the free fit to the NuSTAR data. For the other sources included  for comparison here, we use data from the following observation IDs:\\
{\footnotesize MAXI~J1348--630: NuSTAR:80402315002; HXMT:P021400200601}, \\
{\footnotesize MAXI~J1820+070: NuSTAR:90401309006; HXMT:P011466100301}, \\
{\footnotesize MAXI~J1535--571: NuSTAR:90301013002; HXMT:P011453500144},\\
{\footnotesize Swift~J1727.8--1613: NuSTAR:90902330002; HXMT:{P061433800301}.

Similar to other bright sources, the spectrum of \SwiftJ presents broad iron lines, but differs in a weaker reflection component. The Compton reflection hump (typically peaking around 30~keV) is not evident in the NuSTAR ratio plot of \SwiftJ, further confirming that the the disk reflection component is indeed weaker in \SwiftJ when compared with other black hole systems. We note although a hump-like feature is shown in the \emph{Insight}-HXMT ratio plot of Swift J1727.8--1613,the centroid energy around 20~keV is lower than that of other BHTs, therefore the feature might at least be partially artificial due to the shape of the high energy roll-over not well modeled with a simple cutoff power-law model. The Compton reflection hump of \SwiftJ (if exist) is deeply coupled with the shape of the spectral roll-over in the high energy end of the X-ray spectrum, and thus the strength of its true contribution is uncertain but should be considered to be weaker than typical cases. In addition, one of the most notable features that distinguishes \SwiftJ\ from the other sources is its excess emission at high energies. Whereas the ratios of the other sources generally begin to decrease at around 20–30 keV, the ratio of \SwiftJ\ tends to increase instead. We note that a ``hard X-ray tail" can be seen in the \emph{Insight}-HXMT data of MAXI~J1535--571 above 80~keV as well,  meanwhile being still elusive to the NuSTAR band. The hard X-ray excess starts to emerge from 30~keV in \SwiftJ, lower than that in MAXI~J1535--571, and much lower than those observed in MAXI~J1820+070 and MAXI~J1348--630 by INTEGRAL \citep{2023A&ACangemi}.  The weak reflection feature and the hard X-ray excess are consistently detected by \emph{Insight}-HXMT and NuSTAR in Swift~J1727.8--1613. Therefore, during the following more detailed spectral modeling, we only focus on the \emph{Insight}-HXMT data to study its evolution from a multi-epoch perspective.

\begin{figure}
    \centering
     \includegraphics[width=0.45\textwidth]{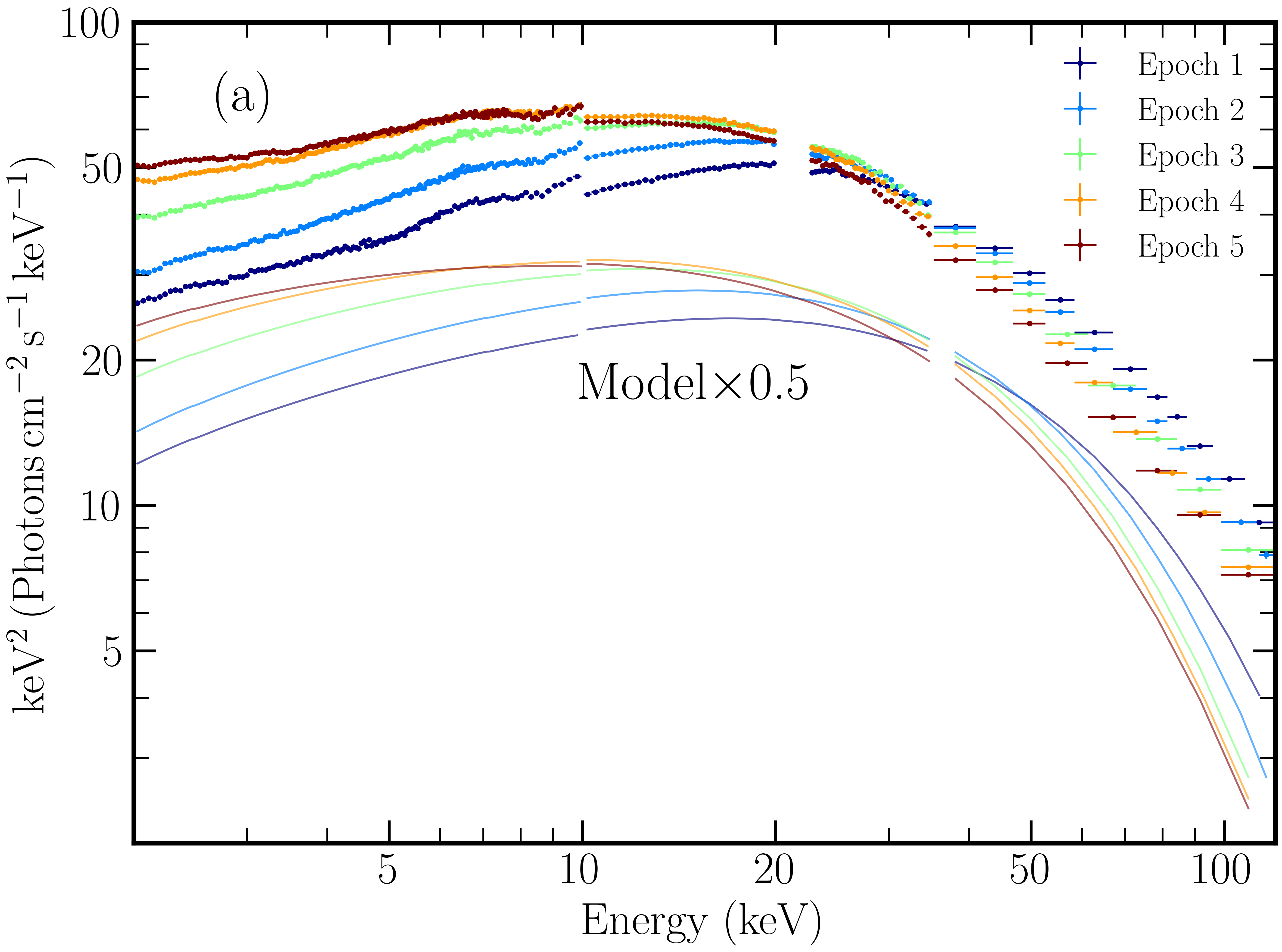}
     \includegraphics[width=0.45\textwidth]{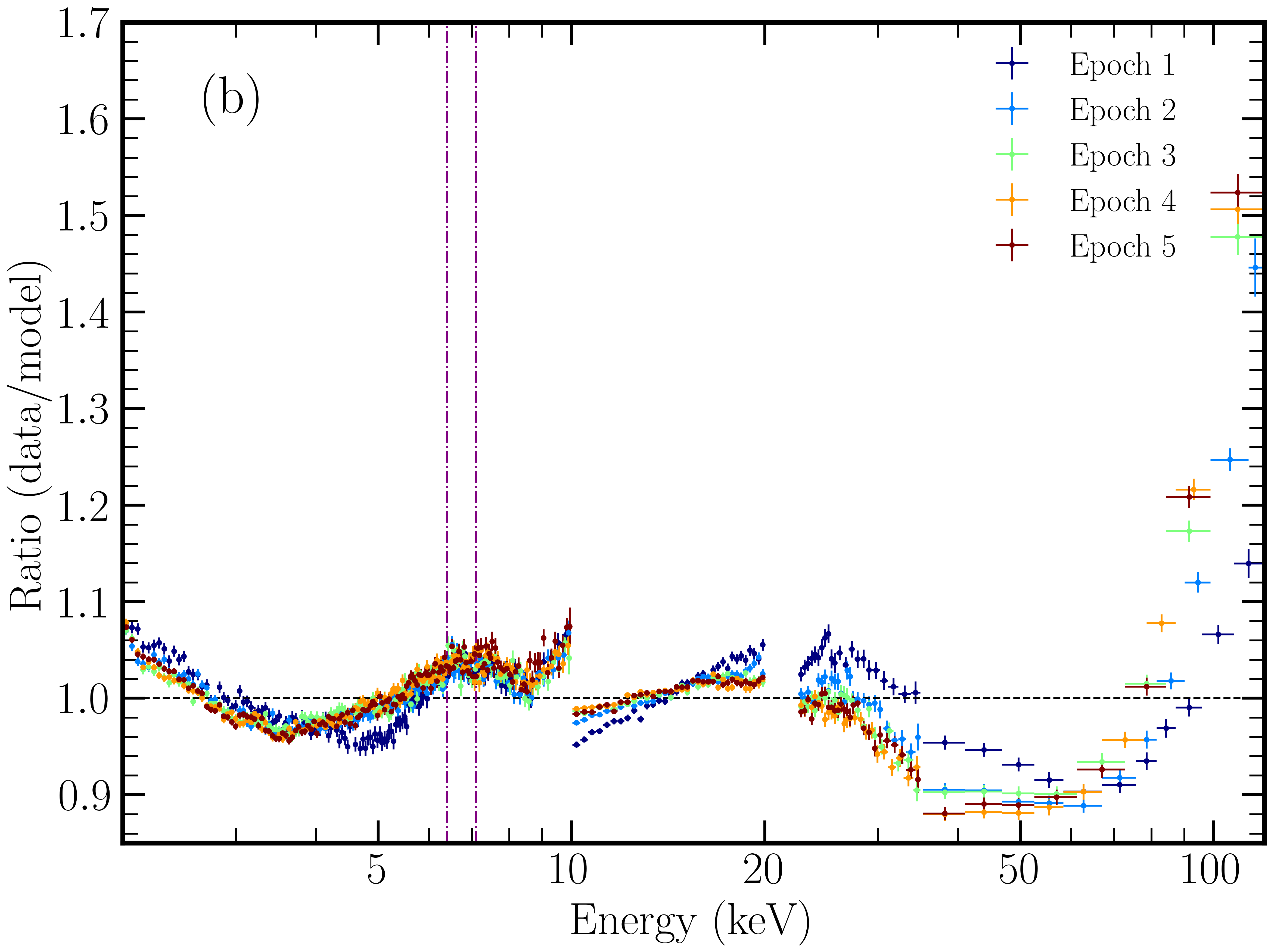}
\caption{(a) Unfolded \emph{Insight}-HXMT spectra for five epochs of the hard state during the 2023 outburst of Swift~J1727.8--1613. The spectra are fitted with the {\tt tbabs*cutoffpl} model. Solid lines of different colors represent models (For clarity, the normalization is multiplied by $\times 0.5$). (b) Data-to-model ratios after fit. The purple dotted lines represent 6.4~keV and 7.1~keV respectively.}
\label{fig:ratio}
\end{figure}

\begin{figure*}
 \centering
     \includegraphics[width=0.45\textwidth]{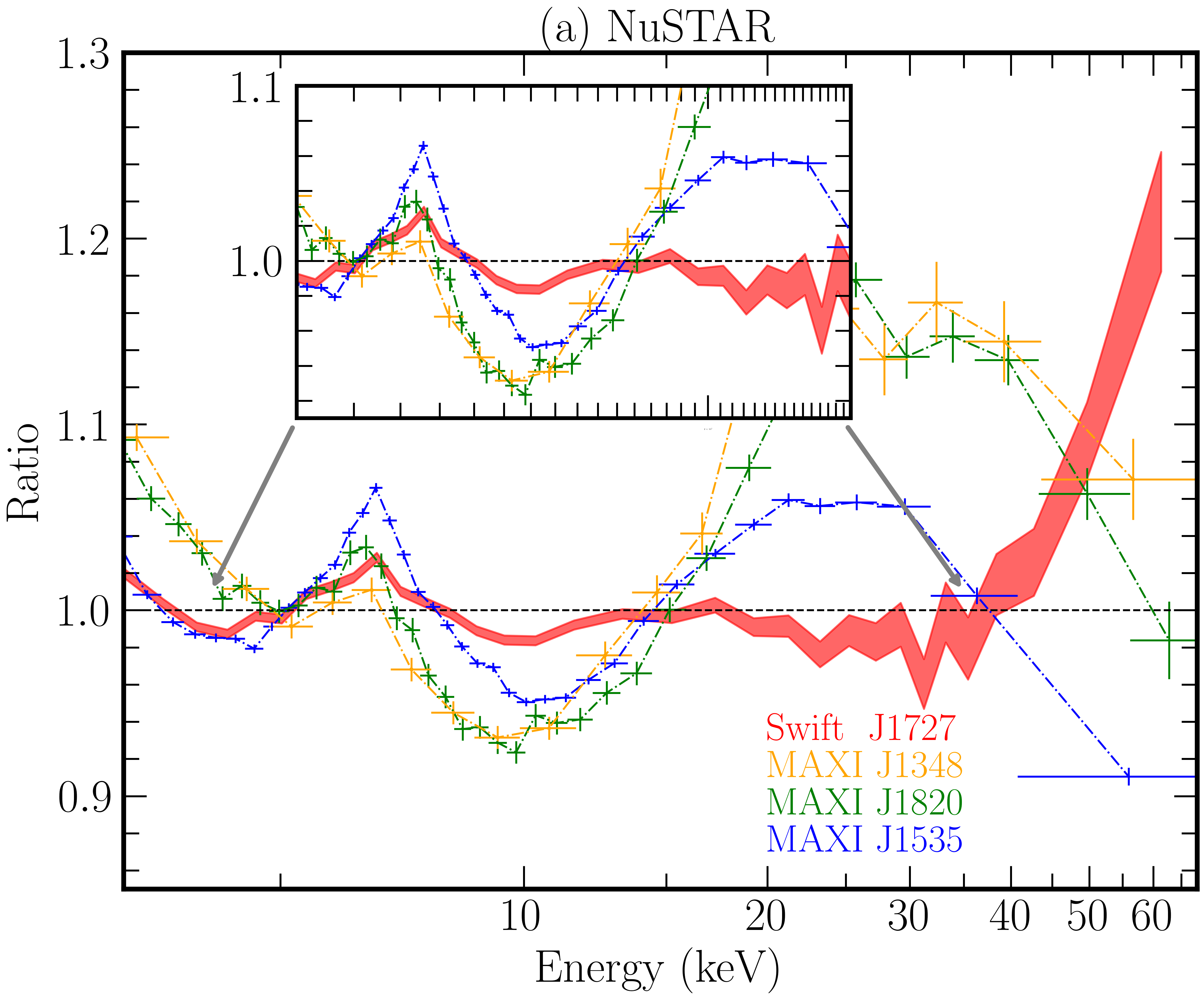}
     \includegraphics[width=0.45\textwidth]{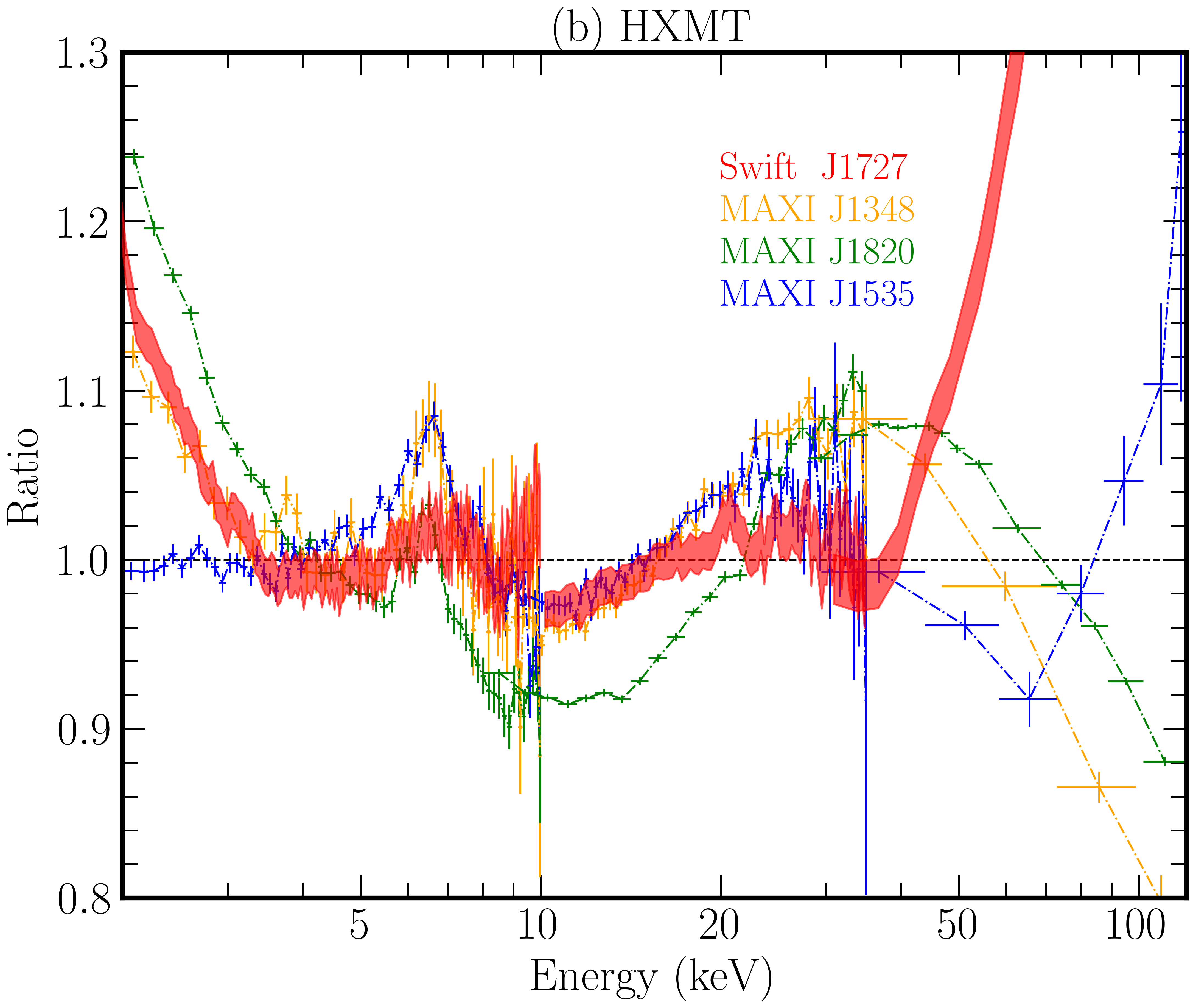}
\caption{Data/model ratios to an absorbed cutoff power-law model from of four bright BHTs in the LHS observed by NuSTAR and \emph{Insight}-HXMT: Swift~J1727.8--1613, MAXI~J1348--630, MAXI~J1820+070, and MAXI~J1535--571, respectively. }
    \label{fig:nustar}
\end{figure*}

\subsection{Spectral Continuum Fitting and Results} \label{subsec:cont_model}

To quantitatively study the evolution of spectral parameters, we apply different models to the spectra.
We choose Epoch 2 as an example, and the plot the corresponding spectral residuals to the different models in Figure~\ref{fig:model}. We fix the neutral absorption column parameter $N_{\rm H}$ at $0.3\times10^{22}$ cm$^{-2}$, which is slightly higher than that reported by NICER \citep{2024ApJPeng}. This is based on the average results obtained after fitting using the \emph{Insight}-HXMT broad-band data, and is consistent with the results reported by \citet{2024A&AMata}. The low absorption column density probably arises from absorption from the ISM rather than being intrinsic to the source, thus it is unlikely to be variable. We fixed it in order to show changes in the other components. Like some transient sources, a single cutoff power-law model cannot describe the LHS spectral continuum of the \SwiftJ well, and there is a clear structure in the fitted residuals (see the left panels of Figure~\ref{fig:model}). Then we tried to add another cutoff power-law model and found that the fit can be greatly improved, leaving a broad iron line and a slight excess in the low energy end as the most prominent features in the residuals (see the panel (b), the full model is {\tt\small tbabs*(cutoffpl$_{\rm 1}$+cutoffpl$_{\rm 2}$)} in XSPEC). The panel (c) and (d) of Figure~\ref{fig:model} show the components represented by each of the two cutoff power-law components, by subtracting the corresponding {\tt cutoffpl} component from the total model. The {\tt cutoffpl$_{\rm 2}$} with a photon index at $\sim0.3$ and a cutoff at $\sim10$~keV mainly help account for the spectrum of Epoch 2 at $<100$~keV. And {\tt cutoffpl$_{\rm 1}$} with a photon index around 1.8 smooths out the hard tail greater than 100~keV. Moreover, given the persistent nature of both components during the LHS, the utilization of two cutoff power-law models is imperative for achieving an accurate fit. An additional cutoff power-law component has been widely used to model the hard X-ray tails observed in black hole X-ray binary systems, either explained as an additional Comptonization component or contribution from the jet \citep{2006A&AMalzac,2011SciLaurent}. The including of this extra component is well motivated in our spectral modeling, but we note that the small index of {\tt cutoffpl$_{\rm 2}$} is untypical for the Comptonization medium in BHTs.

We consider this as the first phenomenological model that provides a good-fit for the data (Model 1: {\tt\small constant$\ast$tbabs$\ast$(diskbb+cutoffpl$_{\rm 1}$+cutoffpl$_{\rm 2}$+gaussian)}), and the best-fit parameters for Model 1 (and Model 2--4) are listed in Table~\ref{tab:model-par-elv}. The spectral residuals after fitting the Epoch 2 data are shown in the right panel of Figure~\ref{fig:model}. The {\tt constant} parameter is to reconcile the slight inconsistency in the normalisation of the different detectors. For the five epochs, we obtain statistically acceptable fits with the reduced chi-squared $\chi/d.o.f$ $\sim 0.82-1.05$ on 414 degrees of freedom. The two {\tt cutoffpl} models represent two non-thermal components, one softer, with a photon index of $\gtrsim1.7$ and $E_{\rm cut}$ above 50~keV, and the other harder, with a photon index $< 1$ and $E_{\rm cut}$ about 10~keV.  This model, while well constraining the two non-thermal parameters, does not constrain the normalization of the accretion disk well, which may be due to the fact that the energy used is larger than 2~keV and that the disk component cutoff is around 1 keV for the temperature of $kT_{\rm in} \sim 0.3-0.6$~keV. Also, in this model, the disk flux is less than $\sim 1\%$ of the total flux. The photon index of {\tt cutoffpl$_{\rm 1}$} increases from $\sim1.7$ to $\sim2.0$ with a concomitant increase in $E_{\rm cut}$ ($\sim52.4$~keV to $\sim66.9$~keV). The increase of the photo index is to be expected during the rising phase of the LHS as the spectrum is softening.  The fact that the photon index is close to 2 also indicates that the source is about to enter its HIMS.  The photon index of {\tt cutoffpl$_{\rm 2}$} is also found to increase, together with a rise in $E_{\rm cut}$. However, the parameters of {\tt cutoffpl$_{\rm 2}$} are not well constrained, as they are partially degenerate with those of {\tt cutoffpl$_{\rm 1}$}. In addition, the reflection hump may further complicate the decomposition of the two {\tt cutoffpl} components.

Soft photons may be hardened to higher energies, so we use a convolution model {\tt thcomp} \citep{2020ZdziarskiMN} to replace {\tt cutoffpl$_{\rm 1}$}, which agrees much better utilizing {\tt nthcomp} with actual Monte Carlo spectra from Comptonization. During the fitting, we link the seed photon temperature ($T_{\rm bb}$) of the {\tt thcomp} model with the inner disk temperature ($kT_{\rm in}$) of the {\tt diskbb} component. Despite replacing the continuum spectrum, we find that an additional {\tt cutoffpl} component is still needed (Model 2: {\tt\small constant$\ast$tbabs$\ast$(thcomp$\ast$diskbb+cutoffpl$_{\rm 2}$+gaussian)}), and it has the same spectral index as Model 1. It is worth mentioning that the {\tt cutoffpl$_{\rm 2}$} cannot be replaced with a simple thermal Comptonization model due to $\Gamma <1$ and the small truncation energy. The model gives the covering fraction close to 1 at the earliest part of the outburst, indicating that all soft photons are Comptonized, and as the transition state approaches, the fraction decreases slightly, but is still greater than 0.8.

\subsection{Reflection Spectra and the Additional Component} \label{subsec:ref_spec}

To simulate the relativistic reflection characteristics, we adopt the widely used reflection model {\tt relxill} v2.3 \citep{2014GarciaApJ}. We first fit the disk reflection features with the standard version of the relativistic reflection model {\tt relxill}, which use the phenomenological {\tt cutoffpl} model as the input for the coronal continuum emission that illuminates the accretion disk and cause the disk reflection features around black holes, parameterized by the inner radius ($R_{\rm in}$) and the inclination of the accretion disk ($i$), the black hole spin ($a$), etc.

During our spectral modeling in the Section~\ref{subsec:cont_model}, the {\tt cutoffpl$_{\rm 2}$} truncated at around 10~keV also seems to contribute to the energy range of the reflection hump. And thus it is possible that the inclusion of the {\tt relxill} model could take over the role of {\tt cutoffpl$_{\rm 2}$} in describing the excess caused by the reflection hump, if that is the major function of {\tt cutoffpl$_{\rm 2}$} in the modeling of the broadband X-ray spectrum. However, after fitting the spectrum with a thermal disk plus disk reflection model (intrinsically including a {\tt cutoffpl} input spectrum) , there are still significant structures in the residuals and therefore the another {\tt cutoffpl} model component is still required.
 
The full model that includes a physical description of the disk reflection features and achieves a statistically good fit is {\tt\small constant$\ast$tbabs$\ast$(diskbb+relxill+cutoffpl$_{\rm 2}$)} (Model 3). Alternatively, an accretion disk with a lamppost geometry corona (where the corona is a point source located on the spin axis of the black hole at a height above the accretion disk.), {\tt\small constant$\ast$tbabs$\ast$(diskbb+relxilllp+cutoffpl$_{\rm 2}$)} is also introduced to fit the spectra. And the differences in terms of the key physical parameters yielded by the {\tt relxill} and {\tt relxilllp} model are very small. And also because X-ray polarimetry measures of this source favors a disk-like corona over a lamppost shaped corona, we only report spectral modeling results obtained with the {\tt relxill} model rather than the lamppost version of it.

First, we fix the outer edge of the accretion disk at 400 $R_{\rm g}$ ($R_{\rm g}=GM/c^{2}$ is the gravitational radius, and $M$ is the black hole mass). We assume a standard accretion disk for the general case, fixing $q_{\rm out}$ at 3 and freeing $q_{\rm in}$. The emissivity for the coronal flavor models is given as $r^{-q_{\rm in}}$ between $R_{\rm in}$ and $R_{\rm br}$, and $r^{-q_{\rm out}}$ between $R_{\rm br}$ and $R_{\rm out}$. We assume a black hole spin of $a=0.98$ in this work, keeping the inner disk radius $R_{\rm in}$ and the disk inclination inclination $i$ free. 
The best-fit parameters of Model 3 are shown in Table~\ref{tab:model-par-elv}.

As can be seen in Figure~\ref{fig:ratio}, Epoch 1 is slightly different from the other epochs in terms of its spectral performance in the low energy end, with a large thermal disk temperature $kT_{\rm in} \sim 0.7$~keV yielded by  Model 3. Although the disk inner radius $R_{\rm in}$ of Epoch 1 is roughly 1.4 times of the radius of ISCO (Inner most Stable Circular Orbit) inferred based on the disk reflection spectrum, which indicates that the optically-thick accretion disk is slightly truncated, the evolution of the disk radius with increasing flux is in general small and all are very close to the location of ISCO of a rapidly spinning black hole. The $R_{\rm in}$ parameter is most sensitive to the red wing of the broad iron line, and is well-constrained in our case and the value remain close to $R_{\rm ISCO}$ at the 3 sigma confidence level for every epoch. Although the normalization parameter of {\tt diskbb} is not well-constrained due to in the limitations of the energy band, we use the $R_{\rm in}^{2} = f^{4}D_{10}^{2}(N_{\rm diskbb}/{\rm cos}\theta)$ relation (where $f$ is the color correction factor \citep{1998PASJKubota}, $D_{10}$ is the distance to the source in units of 10 kpc, and $\theta$ is the disk inclination) to characterize the change of the inner radius, based on an independent method from the disk reflection modeling approach.

Assuming a canonical value for color correction of $f=1.3$ and a fixed distance of 2.7 kpc \citep{2024A&AMata}, $R_{\rm in}$ is calculated to be $\sim 4-10$~$R_{\rm ISCO}$. If the first three epochs are considered to have a larger colour correction factor, such as 1.7, then the radii of the first three are roughly 7-12~$R_{\rm ISCO}$. Thus we obtain an increasing inner disk radius if assuming a constant $f$, but alternatively we can get a constant disk radius by allowing $f$ to be variant. In the previous fit with Model 2 , we find that in Epoch 1, 2 and 3, the disk photons are essentially all Comptonised (see the $f_{\rm cov}$ parameter in Table~\ref{tab:model-par-elv}), which suggests that there should be a larger $f$ than the rest of the epochs (\citealt{2022RenXQ}), and so under this scenario the disk radius calculated accordingly for the five epochs should also be essentially constant.

During the five epochs of the rising phase, the disk inclination is consistently measured to be around 50 degree, indicating that based on the reflection spectrum, this source is a system with a moderate inclination. The ionization of the disk is an important parameter reflecting the nature of the disk in the evolution. We find that the ionization of the disk log($\xi$) is slightly lower in Epoch 1 ($\sim 3.0$) than those in the other epochs ($>3.3$). The log($\xi$) may have risen slightly as the outburst progressed. In any case, the log($\xi$) is always greater than 3 during the hard state, and is accompanied by a relatively high iron abundance measured for the disk material with $A_{\rm Fe} >4$.

The reflection fraction $R_{\rm f}$ is defined in the frame of the primary source as the ratio of the intensity of the illumination emitted towards the disk and that escaping to infinity. $R_{\rm f}$ obtained in the five epochs range from 0.36 to 0.47. The relatively small reflection fraction imply that the reprocessing of photons is not affected by the strong light-bending effect around a black hole, which would be a natural result of the a truncated accretion disk (so that the disk covering factor is small), but that is not the case here as we do not find the disk to be truncated. Unlike previously observed black hole transients, the X-ray continuum spectrum of \SwiftJ is special in the sense that it requires an additional additional cutoff power-law component in the spectrum, which may bring in more complex physical effects.

To separate the direct and reflected emission, we set the reflection fraction parameter to $-1$, so that the model returns only the reflected component. This allows us to estimate the irradiated and reflected fluxes independently. The resulting fluxes and their fractional contributions are listed in Table~\ref{tab:model-par-elv}. The total flux increases from $\sim2.4\times10^{-7}$ to $\sim3.1\times10^{-7}\,\mathrm{erg\,cm^{-2}\,s^{-1}}$ within less than four days, corresponding to an Eddington ratio of $\sim0.16$--$0.21\,L_{\rm Edd}$ (assuming a $10\,M_\odot$ black hole).  The flux of the {\tt cutoffpl$_{\rm 1}$} component does not exhibit a clear monotonic trend, while the reflection flux increases steadily over time. In contrast, the flux of {\tt cutoffpl$_{\rm 2}$} increases from Epoch~1 to Epoch~3 and then remains approximately constant. This apparent difference in behaviour may arise from the fact that the two continuum components are not fully independent. Their parameters are partially degenerate, and the presence of the reflection hump further complicates the decomposition of the spectral components. As a result, the exact flux partition between {\tt cutoffpl$_{\rm 1}$} and {\tt cutoffpl$_{\rm 2}$} is somewhat model-dependent.

\subsection{Hybrid Corona Modeling} \label{subsec:eqpair_spec}

In Section~\ref{subsec:cont_model} and~\ref{subsec:ref_spec}, we model the coronal continuum with a two-component model. The basic motivation for introducing a second power-law component is that, distinct from previous X-ray observations of black hole X-ray binaries in the bright hard states, one single thermal Comptonization model is insufficient to describe the high energy part of the X-ray spectrum above 30~keV. Two cutoff power-law components offer more flexibilities in modeling the complicated shape of the high energy roll-over and provide a statistically good description of the spectral continuum, and allows for a detailed modelling the disk reflection spectrum. As the key physical parameters in the disk reflection model are most sensitive to the iron line profile, we consider they are less affected by the specific choice of the spectral continuum model. However, we have discovered the flux contribution of the different spectral continuum components cannot be well distinguished in the sense that it is highly dependent on spectral continuum model used. And then the cutoff power-law model despite being widely used for various physical processes, is only a phenomenological approximation, thus does not bare clear physical implications and may not represent the most physically reasonable scenario. Therefore,  we consider more sophisticated models containing a hybrid distribution of thermal/non-thermal particles, aiming to extract more information from the broadband X-ray spectral continuum. In such a model, the high-energy tail is explained by the emission of a non-thermal population of accelerated particles, which is widely used to model the hard X-ray tails in black hole X-ray binaries. These particles cool down and eventually thermalize because of several processes such as Coulomb collisions.

As a first step, we applied the hybrid Comptonization model {\tt EQPAIR} \citep{2000HEADCoppi}, which is frequently used to describe spectra extending to several hundred keV (e.g. INTEGRAL observations; \citealt{2021A&ACangemi,2013MNRASSanto,2009ApJCaballero}). However, this approach results in an extremely high compactness parameter ($l_{\rm s}\sim2000$) and a very steep non-thermal electron distribution ($\Gamma_{\rm inj}>4$), indicating that a more self-consistent treatment of the continuum and reflection is required. Then, we therefore adopt the {\tt reflkerr} and {\tt reflkerr\_bb} models \citep{2021ApJZdziarski}, which self-consistently compute Comptonized emission, disk reflection, and relativistic effects in the Kerr metric. 

The {\tt reflkerr} assumes thermal Comptonization on Maxwellian electrons and calculates the continuum using {\tt compps} \citep{1996ApJPoutanen}, ensuring physical consistency between the incident and reflected spectra. In the {\tt reflkerr} framework, the reflection spectra are computed in the local rest frame of the accretion disk and subsequently convolved with relativistic effects in the Kerr metric. The rest-frame reflection is calculated using {\tt xillverCp} \citep{2010ApJGarca,2018ApJGarca} at energies below $\sim$10 keV, where detailed atomic features dominate, and {\tt ireflect} \citep{1995MNRASMagdziarz} at higher energies to accurately describe the Compton reflection continuum. Both the direct Comptonized emission and the reflected component are integrated over the surface of a Keplerian accretion disk, fully accounting for relativistic Doppler shifts, gravitational redshift, and light bending. 

The {\tt reflkerr\_bb} model is used to fit the pronounced high-energy non-thermal tail by adopting a modified version of {\tt compps} that allows for a hybrid electron distribution consisting of a thermal component and a high-energy power-law tail. In this framework, the non-thermal component is parameterized by a minimum Lorentz factor, $\gamma_{\rm min}$, above which the electron distribution transitions from a thermal Maxwellian to a power law in momentum space,
\begin{equation}
\frac{{\rm d}N_{\rm e}}{{\rm d}(\beta\gamma)} \propto (\beta\gamma)^{-p},
\end{equation}
where $\beta = v/c$ and $p$ is the power-law index. The distribution extends up to a maximum Lorentz factor $\gamma_{\rm max}$, which is fixed at $10^3$ throughout this work.

Thus, the Model 4 is: {\tt\small constant$\ast$tbabs$\ast$(diskbb+reflkerr+reflkerr\_bb)}. The hybrid Comptonization model is fitted over 2--200 keV in order to better constrain the high-energy tail, whereas the other models are restricted to 2--120 keV. Unlike MAXI~J1820+070 \citep{2021ApJLZdziarskiA}, \SwiftJ shows weak reflection and only a single broad Fe~K line; thus {\tt reflkerr\_bb} is adopted for the irradiating continuum, while {\tt reflkerr} is used only for the continuum modelling.  For the Comptonization geometry, we adopt ${\tt geom}=-5$, corresponding to a sinusoidal distribution of seed photons within a spherical corona. The temperature of the seed photons is fixed to the value obtained from the {\tt diskbb} fit. The spectral slope of each Comptonization component is characterized by the Compton parameter,$y=4\tau_{\rm T}kT_{\rm e}/m_{\rm e}c^{2}$, where $\tau_{\rm T}$ denotes the Thomson optical depth of the plasma. Table \ref{tab:model-reflkerr} shows the best-fit parameters of Model 4. In Figure~\ref{fig:eemodel}, we present the unfolded spectra and the data-to-model residuals for Model~3 and Model~4, respectively. For the thermal Comptonization component ({\tt reflkerr}), the Compton parameter decreases from $y_{\rm th}\sim0.58$ to $0.51$, while the electron temperature $kT_{\rm e} $ declines from $\sim6.60$ keV to $\sim5.65$ keV. At the same time, the normalization increases from $\sim 10$ to $\sim 15$. For the hybrid Comptonization component ({\tt reflkerr\_bb}), the Compton parameter shows a similar decreasing trend, from $y_{h}\sim0.70$ to $\sim0.61$. The electron temperature varies within the range $kT_{\rm e}\simeq17$--20 keV without a clear monotonic trend. The minimum Lorentz factor remains nearly constant at $\gamma_{\rm min}\simeq1.43$--1.49. In contrast, the power-law index of the non-thermal electron distribution varies between $p\sim2.16$ and $1.73$, showing a noticeable hardening in the intermediate observations, followed by a return to $p\sim2$, implying a relatively steep non-thermal electron distribution.  The outer radius of the accretion disc is fixed at $R_{\rm out}=1000 R_{\rm g}$. From the spectral fits, we obtain a disk inclination in the range of $\sim40$--$50^{\circ}$, while the inner disk radius is constrained to be close to the ISCO. The reflection fraction remains relatively low ($R_{\rm ref}\sim0.3$) at lower fluxes, but increases significantly to $R_{\rm ref}\sim0.7$--0.8 at higher fluxes.

\begin{figure*}
    \centering
     \includegraphics[width=0.4\textwidth]{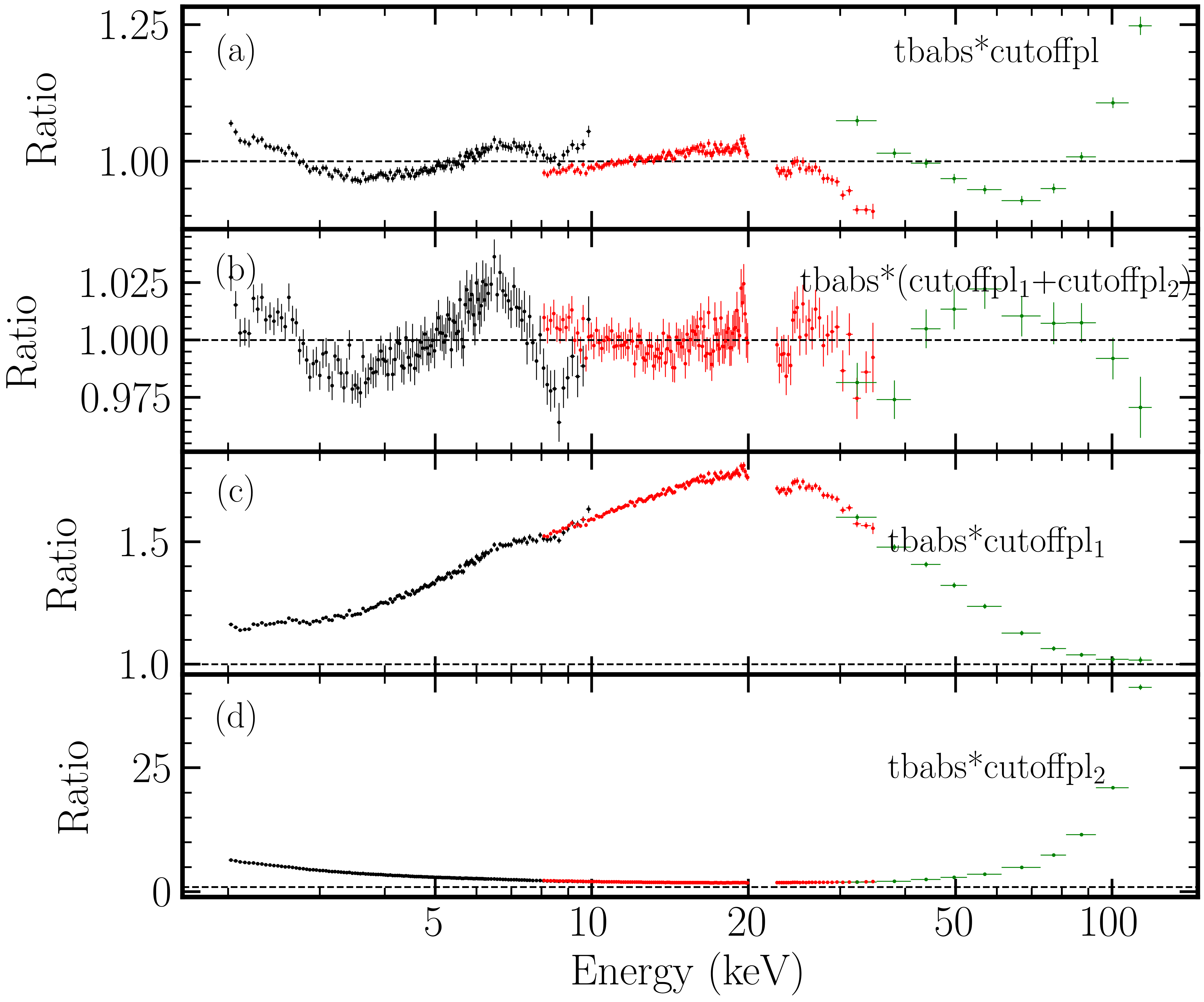}
     \includegraphics[width=0.4\textwidth]{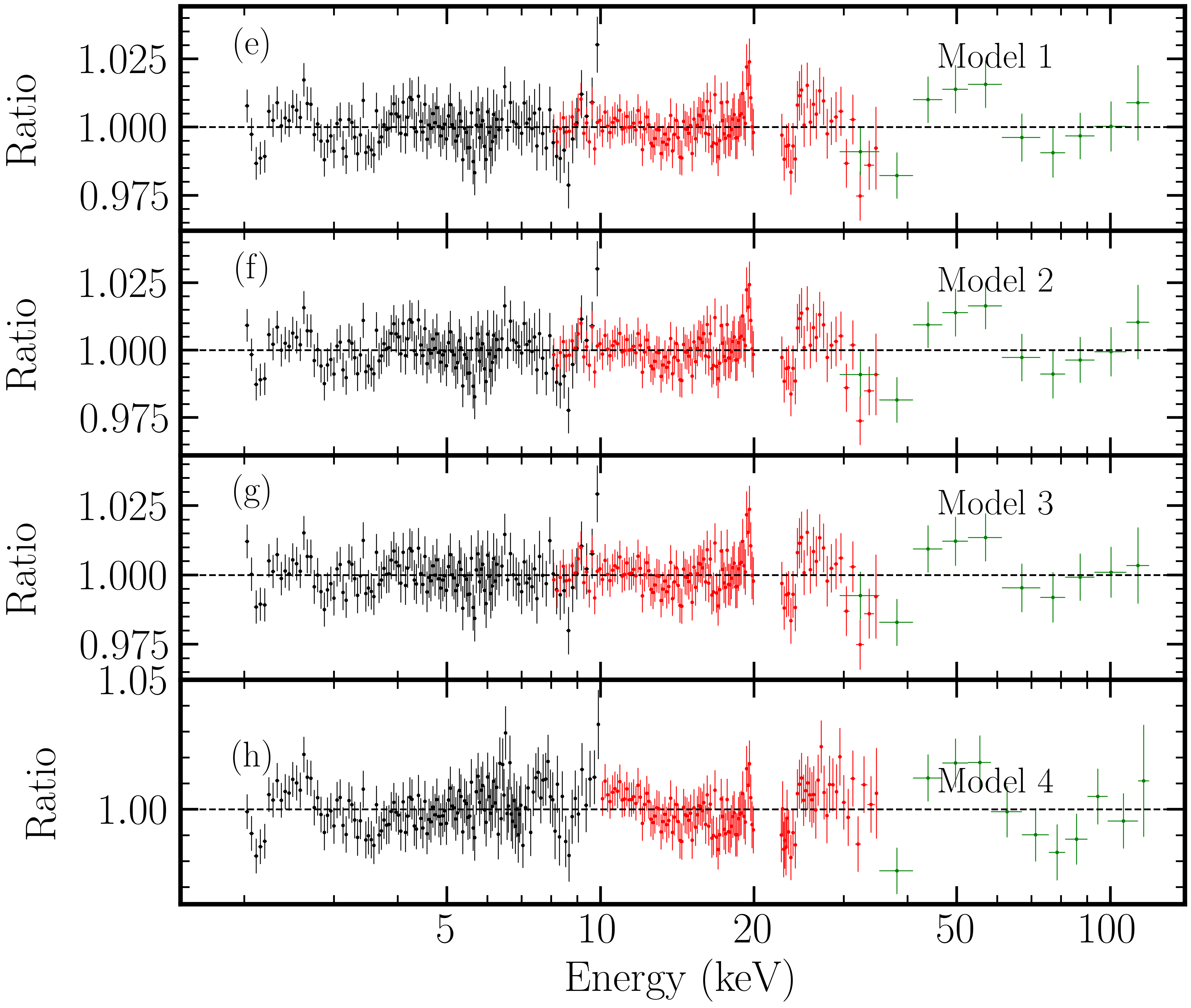}
    \caption{Left: ratio diagrams for the phenomenological models. Right: ratio plots for the four models that we considered as good fits. }
    \label{fig:model}
\end{figure*}

\begin{figure*}
    \centering
    \includegraphics[width=0.8\linewidth]{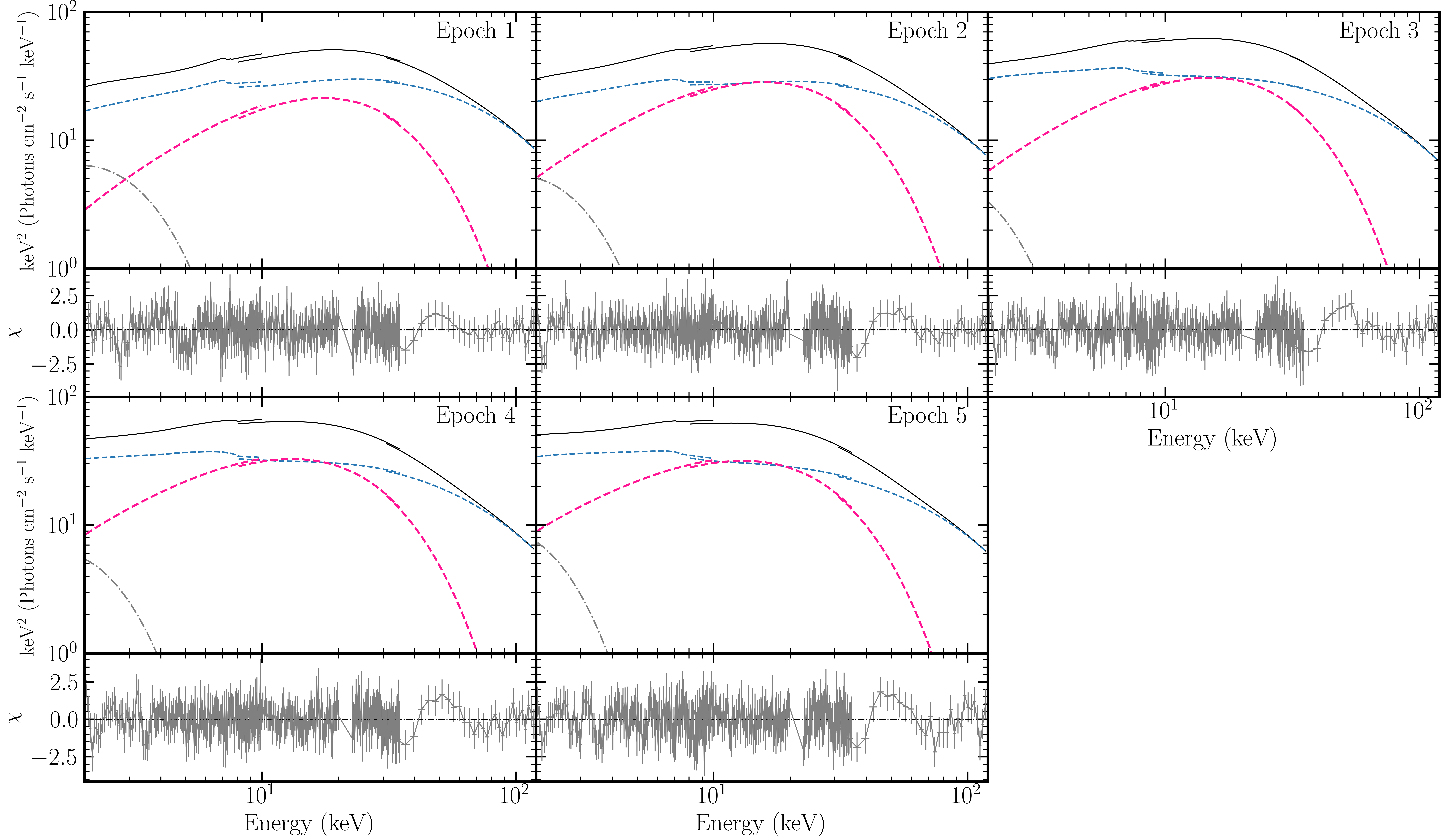}
    \includegraphics[width=0.8\linewidth]{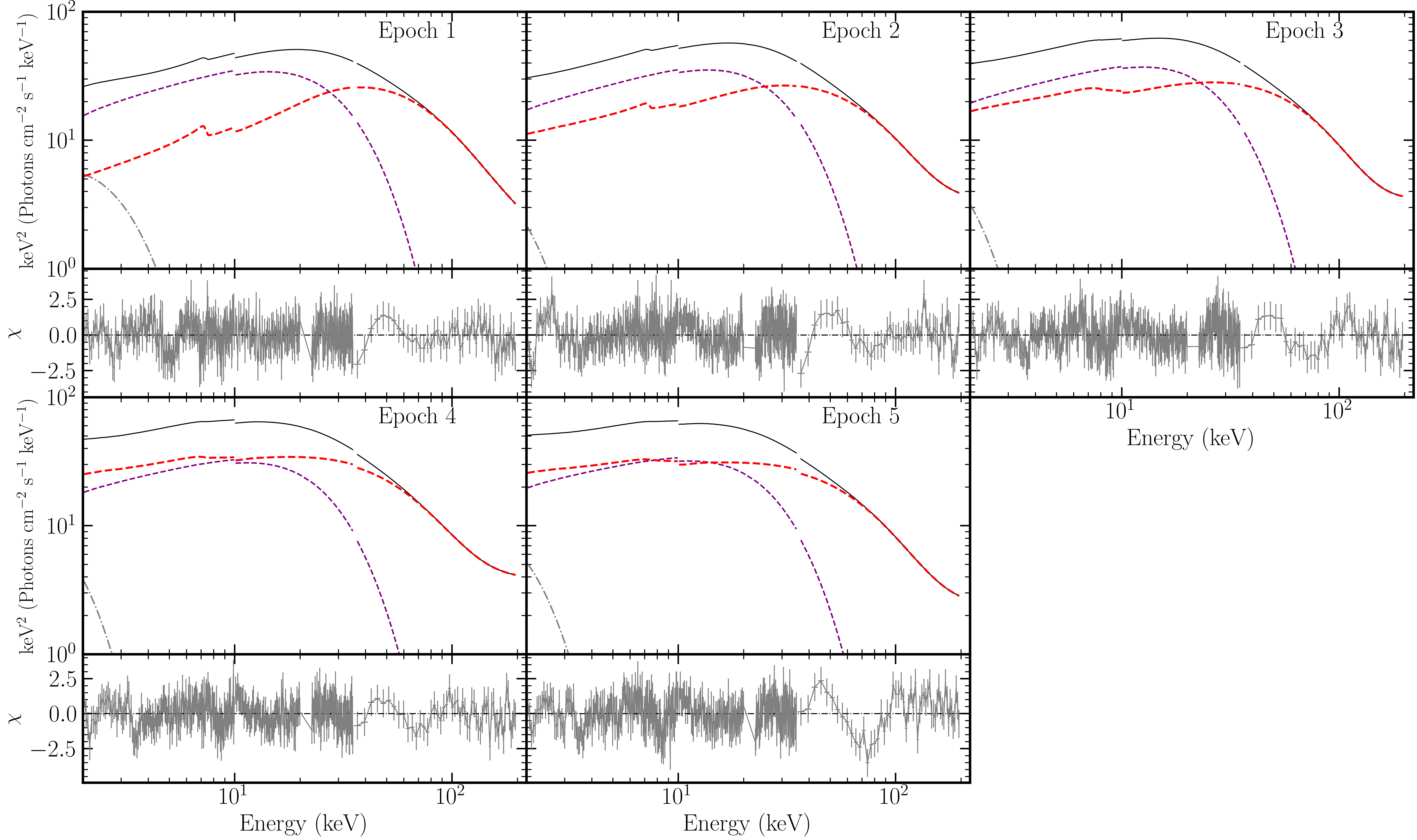}
    \caption{Plots of best-fit models with residuals of \emph{Insight}-HXMT observations of \SwiftJ from epochs 1 to 5 with Model 3 (top) and Model 4 (bottom): \modelD and {\tt\small constant$\ast$tbabs$\ast$(diskbb+reflkerr+reflkerr\_bb)} in XSPEC. The grey dash-dot lines represent the disk component. Top Panel: The light blue dashed lines are for the reflection component. The pink lines are for the additional ({\tt cutoffpl$_{\rm 2}$}) component. Bottom Panel: The purple dashed lines mark the {\tt reflkerr} component and red lines represent the {\tt reflkerr\_bb} component.}
    \label{fig:eemodel}
\end{figure*}

\section{Discussion} \label{sec:dis}
During the rising phases of this outburst, \SwiftJ rapidly brightens up to $\sim 0.21 L_{\rm Edd}$ (assuming a black hole mass of 10 $M_{\rm sun}$ and a distance of 2.7 kpc) in the X-rays with a very small decrease in the hardness ratio (0.88-0.77, defined as the ratio of the count rates between 4--10~keV and 2--4~keV). However, the flux in the higher part of the energy spectrum ($>30$~keV) decreases very slowly, when compared to the increase of flux in the lower energy part ($<30$~keV) of the spectrum (see Table~\ref{tab:hxmt} and Figure~\ref{fig:ratio}). We focus this work on presenting the evolution of the broadband X-ray spectrum of \SwiftJ during the rising hard state, by fitting spectra at five selected epochs.

\subsection{Two Hard Components in the Low Hard state}

We find that both the broadband \emph{Insight}-HXMT (2--120~keV) and the NuSTAR (3--70~keV) spectra require two cutoff power-law components for a phenomenological description of the spectral continuum, which is rare in BHTs. Using INTEGRAL observations of Swift~J1727.8--1613, \citet{2023arXivMereminskiy} also found an additional hard power-law spectral component extending at least up to 400~keV. Comparing the recently bright sources during their hard states, MAXI~J1820+070, MAXI~J1535--571, and MAXIJ~1348--630, \citet{2023A&ACangemi} found that additional power-law components are required when extending the energy to greater than 100~keV, whereas the spectra of \SwiftJ requires an additional component only at energies greater than 40~keV. During the hard state, as the luminosity increased, the properties of the two components evolved and but no abrupt changes were found, suggesting that there was no significant change in the geometry or the nature of the region producing the X-rays.

Also, distinct from the broadband spectral modeling results based on INTEGRAL and SRG/ART-XC observations of \SwiftJ during the plateau phase of the outburst, decent modeling of the spectrum taken by \emph{Insight}-HXMT requires the additional component {\tt cutoffpl$_{\rm 1}$} truncated around 40~keV, rather than an extra power-law component with no cutoff for the highest energies of the X-ray spectrum. These suggest that it may not be the ``hard tail" in the conventional sense. In addition, the decreasing trend in the photon index of {\tt cutoffpl$_{\rm 1}$} yielded by Models 1-3 suggests that the hard tail was softening, meanwhile becoming hotter (the electron temperature increased with luminosity). This tendency of the coronal temperature evolution of \SwiftJ is in agreement with the results measured by \emph{Insight}-HXMT in GX~339--4 (e.g., \citealt{2023ApJLiuhh}) during the hard-to-soft transition process. However, this is different from the decreasing trend observed in GX~339-4 during the rising hard states (e.g., \citealt{2009MNRASMotta,2015ApJGarc}), and a number of other BHTs during similar phases of their outbursts \citep{2010ApJTitarchuk,2017ApJXu}, which is explained by the cooling of the corona by soft photons. On the other hand, the decrease in optical depth leads to less scattering of soft photons by the corona, which could increase the temperature of the electrons. And  this is believed to be limited by the creation of $e^\pm$ pairs, which may be a runaway process that prevents further rise in temperature \citep{1971SvABisnovatyi-Kogan}. The reason that we did not find obvious change in the corona temperature in \SwiftJ during the rising hard state, as seen in other BHTs, could be related to the fact that the source did not immediately underwent a hard-to-soft state transition following the bright hard state (instead, the source experienced prolonged intermediate states with flares). 

The slight increase in the photon-index and the slight decrease in $E_{\rm cut}$ of {\tt cuotffpl$_{\rm 1}$}, indicate that there is no significant rapid shift in the spectral shape of the this component. These suggest that the region producing this emission underwent no dramatic change during our monitoring observations, which is comparatively more stable than the emission region contributing {\tt cuotffpl$_{\rm 2}$}. The index of {\tt cuotffpl$_{\rm 2}$} is found to be around 1, which is unusually small for the BHTs, but such a small photon index has been observed in Ultraluminous X-ray sources (ULXs) during hard states \citep{2013MNRASSutton}. And the small value has been explained to arise from the case when observing these systems close to face-on, we see the innermost part of the funnel wind that is expected to occur at super-Eddington accretion rates \citep{2007MNRASPoutanen,2012ApJKawashima}. Analogously, the high X-ray flux of \SwiftJ with a small distance estimate indicate that the source probably has a high accretion rate, and may lead us to see the more inner regions (harder and colder) of the corona that had not been previously observed in the BHTs. There are recent works claiming that the corona is extended and thus has a stratified structure and in that case X-ray spectral continuum of BHTs could be decomposed into multiple components (e.g., \citealt{2017MNRASBasak,2021MNRASDzielak}). However, we caution here that the best-fit values and the evolution of the photon-index and cutoff energy of the two cutoff power-law components are most likely dependent on the specific choice of the spectral continuum models. On one hand, the most salient feature in the spectral continuum here is the shape of the high energy roller-over, on the other hand the cutoff power-law model only provides the simplest approximation and not a very physical description of the plasma emission contributing the high energy cutoff. Therefore, we have also made an effort to fit the broadband X-ray spectral with more sophisticated versions of the Comptonization models.

In all epochs, we observe a high-energy tail, which may be physically interpreted as indicative of a non-thermal distribution of particles within the plasma. Whether the hard X-ray tail (or soft gamma tail in the literature) observed in BHTs originates in the corona or the jet is long-standing debate, with evidence supporting and against both sides (see the discussion in the review, \citealt{2021NewARMotta}). By first invoking an additional cutoff power-law spectral component for a phenomenological modeling, we did not specify on the physical origin of the additional component. And we note that, since corona has been proposed to be the base of jet during the LHS of BHTs \citep{2001A&AMarkoff,2003A&AMarkoff}, the distinctions between the two may not be always clearly defined or even always have to be. The relation of the two could be evolutionary depending on the status. Jet contribution (self-absorbed synchrotron radiation) has been proposed to dominate the X-ray emission at very low accretion states \citep{2001A&AMarkoff,2003A&AMarkoff}, explaining the radio/X-ray correlation, instead of the widely used disk Comptonization component. One the other hand, jet contributions in the high energies are typically known to peak in the MeV band in X-ray binaries and AGNs at relatively high accretion rates (e.g., \citealt{2018MNRASZdziarski,2019ARA&ABlandford}). Considering that accretion rate of \SwiftJ is high during the time of our observations based on a relatively conservative estimate (0.21$L_{\rm Edd}$), the total contribution of jet emission (in the canonical sense) in the \emph{Insight}-HXMT band is probably small. We consider that the hard X-ray excess we observed in \SwiftJ in the \emph{Insight}-HXMT band (beyond the thermal Comptonization description) is more closely related to the traditional concept of the corona, meanwhile retains intimate relations with the jet. 

We modeled the data using the more physically motivated {\tt reflkerr\_bb} model, and find that the hard X-ray emission can be well described by a hybrid thermal/non-thermal Comptonizing plasma. The inferred non-thermal electron distribution is broadly consistent with $p \sim 2$ across all epochs. Although a formally harder value of $p \sim 1.7$ is obtained at an intermediate stage, the overall results remain consistent with a canonical $p \sim 2$ distribution within uncertainties. From a physical perspective, such values are naturally expected from efficient particle acceleration processes. In the strong-shock limit, diffusive shock acceleration predicts an injection spectrum with $p_{\rm inj} \approx 2$ \citep{1978MNRASBell,1978ApJBlandford}, while particle-in-cell simulations show that magnetic reconnection can produce comparable slopes (e.g. \citealt{2014ApJSironi,2026MNRASSuriano}). In realistic astrophysical environments, however, radiative cooling (e.g. synchrotron and inverse Compton losses) and particle escape can modify the steady-state distribution, typically steepening it relative to the injection spectrum\citep{1970RvMPBlumenthal,1962SvAKardashev}. The values inferred here therefore suggest that the electron population remains close to the intrinsic acceleration spectrum, with limited spectral evolution. The fitted value of $\gamma_{\rm min} \simeq 1.45$ further indicates that the non-thermal electron distribution emerges already in the mildly relativistic regime. This implies a smooth transition between the thermal and non-thermal populations, rather than a distinct high-energy component. As a result, the observed hard X-ray tail is naturally produced by a population of mildly relativistic electrons, efficiently generating photons at $\gtrsim 30$--100~keV without requiring a substantial population of ultra-relativistic particles.

Compared to other systems, this behaviour points to a relatively weak degree of spectral evolution. In MAXI J1820+070, broadband spectra extending to MeV energies require significantly steeper electron distributions ($p \sim 3$--3.7; \citealt{2021ApJZdziarski}), reflecting a different balance between particle acceleration, energy redistribution, and high-energy emission. In contrast, jet-based constraints in GX 339--4 typically yield $p \sim 2.2$--2.5 \citep{2015MNRASDrappeau}, which are likely shaped by radiative cooling and transport along the outflow. These comparisons suggest that the electron population in this source is observed at an earlier stage of its evolution, where the distribution remains close to the intrinsic acceleration spectrum. The relatively low electron temperature ($kT_e \sim 20$~keV) does not require a pair-dominated plasma, but can instead be regulated by internal energy balance in the accretion flow, for example via synchrotron boiler processes. In this framework, even a small fraction of non-thermal electrons can enhance synchrotron emission and provide an efficient source of seed photons, leading to effective cooling of the thermal population. Radio observations indicate the presence of a compact jet during the hard state of \SwiftJ \citep{2025ApJHughes}. However, the jet properties evolve significantly, and the source rapidly transitions to a more variable regime with flares and ejecta during the hard-intermediate state. The overall evolution of the radio and X-ray emission suggests a coupling between the accretion flow and the jet, although the detailed correspondence between the two remains uncertain. However, the presence of a weak but significant reflection component together with a single broad Fe~K line indicates that the hard X-ray emission is still predominantly produced in the corona.

A consistent physical picture is therefore that magnetic dissipation in the corona, such as magnetic reconnection or weak shocks, simultaneously drives non-thermal particle acceleration and supplies energy to the jet base. In this scenario, the corona acts as the primary site of particle energization, while the jet represents its large-scale outflow, naturally producing the observed multiwavelength correlations.

\subsection{Accretion Geometry in the Hard State}
While the general picture for the accretion geometry in BHTs in the soft state is well recognized as a non-truncated optically thick disk extending to the ISCO, there is yet no consensus on the accretion geometry in the hard state. In the context of the advection-dominated accretion flow (ADAF) model (see the review \citealt{2014ARA&AYuan}), a paradigm emerges where the disk begins to truncate in the intermediate state, and $R_{\rm in}$ increases further in the hard state.  However, for this source, our spectral modeling results are inconsistent with the disk truncation scenario during the hard state of BHTs, and no significant temporal variation is found in the inner radius measured. As discussed in Section~\ref{subsec:ref_spec}, we argue that the inner edge of the accretion disk of \SwiftJ has been very close to ISCO, with negligible evolution from Epochs 1 to 5, which is consistently found via the disk reflection and the disk blackbody spectral continuum modeling method. We note that accurate measurements regarding the disk blackbody component is usually challenging during the hard states of BHTs, considering the weakness of the thermal disk emission, coupling between the disk and corona components, and our limited bandpass in the soft X-rays, etc.  The same results have been found in the study of several other BHTs during recent years. For MAXI~J1535--571, \citet{2018ApJXu} reported no significant disk truncation and a rapidly rotating black hole in the bright hard state of its outburst; in MAXI~J1820+070, \citet{2019Buisson} found that the disk was always at ISCO through reflection spectral modeling results; and in EXO~1846-031 \citep{2022RenXQ} and MAXI~J1348--630 \citep{2022ZhangW}, they also suggested that the inner disk was very close to ISCO during the hard states through measuring the continuum spectra. In addition, we find no significant change in the profile of the iron lines as well as a reduction in the reflection component for the five epochs, which is the same as the case observed by \emph{Insight}-HXMT in MAXI~J1820+070 \citep{2021NatCoYouBei}. By assuming an untruncated disk, \citet{2021NatCoYouBei} suggested that as the corona was observed to move closer to the black hole, the coronal material might be out-flowing faster. Using reflection models, we find a slight decreasing trend in $R_{\rm f}$. During our spectral modeling process, we first notice that two cutoff power-law components are required for a phenomenological good fit of the broadband X-ray spectral continuum, then we only link the disk reflection component with one of the cutoff power-law component (as the illuminating spectral continuum) for simplicity. In fact, the additional cutoff power-law component may or may not illuminate the accretion disk and participate in the reflection process. But either way, it is safe to consider that the reflection fraction we measured in \SwiftJ is relatively low for bright BHTs, and the numbers obtained this way are upper limits. During all five epochs, the reflection fraction is measured to be below unity, whereas the parameter has been found to be quite high ($\sim1-3$) for most other bright BHTs with reported results using the {\tt relxill} model  (e..g., \citealt{2016ApJWalton,2018ApJXu}). Our disk reflection modeling results rule out the possibility that the low reflection fraction is due to disk truncation (e.g., \citealt{2020ApJXu}), and instead it is likely to be caused by an outflowing plasma arising from magnetic flares etc. (aberration effects from mildly relativistic motion reducing X-ray emission toward the disk, \citealt{1999ApJBeloborodov}). 

For the coronal geometry or the shape of Comptonization zone in BHTs, it is debatable whether it may be in a flattened shape extended along the disk plane (disk-like) or a vertically elongated shape along with the jet (jet-like). The polarisation results \citep{2023ApJVeledina,2024ApJIngram} show that at HIMS the corona is more likely in the shape of a radial extension, perpendicular to the jet, while the jet interior is parallel to the black hole spin axis. 

Our spectral results provide further constraints on this geometric picture. The presence of a weak but significant reflection component, together with a single broad Fe~K line, indicates that the dominant hard X-ray emission originates from a region close to the inner accretion disk, rather than from a distant jet. At the same time, fits with the {\tt relxilllp} model suggest a coronal height of $h \sim 4R_{\rm g}$ with no significant evolution during the rising hard state, pointing to a relatively compact and stable emitting region.

In the context of hybrid Comptonization modeling, the corona is more likely dominated by a unified structure consisting of both thermal and non-thermal electrons, rather than multiple physically distinct regions. Although the broadband spectra can be phenomenologically described by two continuum components, this is more naturally interpreted as an inhomogeneous or radially extended hybrid corona with internal structure, where different spectral components arise from regions with different physical conditions within the same coronal framework. This interpretation is consistent with the polarization measurements, which suggest a corona extended in the disk plane.

Meanwhile, the observed correlation between X-ray spectral evolution and radio behavior suggests a dynamical coupling between the corona and the jet base, rather than a geometric identity with the large-scale jet. Taken together, these results favour a scenario in which the hard X-ray emission is dominated by a compact, radially extended hybrid corona located close to the black hole, while the jet contributes only indirectly or at a subdominant level. Further constraints on the coronal geometry will require joint spectral, timing, and polarimetric studies.

\section{Summary and Conclusions} \label{sec:summary}

In this paper, we report broadband X-ray spectral analyses results from the \emph{Insight}-HXMT monitoring observations of \SwiftJ in the rising bright hard state, during which period the X-ray flux of \SwiftJ increased by $\sim30\%$. The source continuously softened and there were more prominent changes in the soft than the hard X-ray band. We find that an additional cutoff power-law component is required to fit the non-thermal X-ray spectral continuum in both \emph{Insight}-HXMT and NuSTAR band. Interpreting the spectra using physically motivated reflection models, we model the emission with a combination of thermal Comptonization ({\tt reflkerr}) and hybrid Comptonization ({\tt reflkerr\_bb}). The results indicate that the hard X-ray emission arises from a plasma consisting of both thermal and non-thermal electrons, where the non-thermal component contributes to the high-energy spectral tail. Within this framework, the emission is most naturally explained by a compact hybrid corona in which thermal and non-thermal electrons coexist. The non-thermal tail appears as a smooth extension of the thermal Comptonization component, rather than originating from a separate high-energy emission region. In addition, we investigate the evolution in the accretion geometry based on the spectral modeling parameters during the five epochs, and find that the inner radius of the optically-thick accretion disk remains stable at the location of ISCO based on the disk reflection modeling and the reflection fraction remains low. But in general, we observed no abrupt changes in its characteristic physical parameters during our monitoring observations represented here. It would be important to catch BHTs earlier in the future at the onset of the outburst, in order to better characterize the early development and propagation of the disk instabilities that lead to the outburst. Such efforts would greatly benefit from the Einstein Probe mission \citep{2022hxga.bookYuan}, which achieves more than one-order-of-magnitude higher sensitivities than the previous generations of all-sky X-ray monitors and can make autonomous follow-up observations with its two onboard X-ray telescopes. 

\begin{table*}
\caption{Best-fit Parameters from Model~1, Model~2,and Model~3.}
\label{tab:model-par-elv}
\centering
{\small
\begin{tabular}{ccccccc}
\hline\hline
Component &  Parameter & Epoch 1 & Epoch 2 &  Epoch 3 & Epoch 4 & Epoch 5\\
\hline
TBabs&$N_{\rm H}(\times10^{22}\rm cm^{-2})$ & &&0.3(fixed)&&\\
\hline
\multicolumn{2}{c}{} & \multicolumn{5}{l}{Model 1: \modelA} \\
\hline
Diskbb&$T_{\rm in}$&$0.60_{-0.06}^{+0.04}$ &$0.47_{-0.07}^{+0.10}$ &$0.27_{-0.04}^{+0.06}$ &$0.39\pm0.05$ &$0.42\pm0.05$ \\
&$N_{\rm disk}$($10^4$)&$0.37_{-0.1}^{+0.2}$ &$0.8_{-0.4}^{+0.9}$ &$3.7_{-3.1}^{+1.2}$ &$4.7_{-2.1}^{+5.3}$ &$4.1_{-1.7}^{+3.9}$ \\
\cline{2-7}
&$\Gamma$&$1.69_{-0.02}^{+0.03}$ &$1.80_{-0.03}^{+0.02}$ &$1.91_{-0.02}^{+0.03}$ &$1.97_{-0.02}^{+0.03}$ &$2.03_{-0.03}^{+0.03}$ \\
cutoffpl$_{\rm 1}$&$E_{\rm cut}$ (keV)&$52.41_{-2.19}^{+2.86}$ &$55.98_{-3.05}^{+2.81}$ &$58.30_{-3.39}^{+3.88}$ &$61.10_{-2.57}^{+4.06}$ &$66.94_{-4.65}^{+5.33}$ \\
&$N_{\rm c1}$&$20.70_{-0.76}^{+0.80}$ &$25.85_{-1.25}^{+0.77}$ &$36.57_{-0.64}^{+0.92}$ &$42.17_{-1.40}^{+1.27}$ &$45.83_{-1.88}^{+1.63}$ \\
\cline{2-7}
& $\Gamma$ &$-0.08_{-0.09}^{+0.11}$ &$0.26_{-0.09}^{+0.05}$ &$0.35_{-0.09}^{+0.06}$ &$0.53_{-0.07}^{+0.06}$ &$0.64_{-0.06}^{+0.06}$ \\
cutoffpl$_{\rm 2}$ &$E_{\rm cut}$ (keV)&$9.72_{-0.40}^{+0.53}$ &$10.28_{-0.48}^{+0.28}$ &$9.81_{-0.46}^{+0.37}$ &$10.06_{-0.34}^{+0.37}$ &$10.46_{-0.38}^{+0.37}$ \\
&$N_{\rm c2}$&$0.29_{-0.06}^{+0.09}$ &$0.98_{-0.19}^{+0.13}$ &$1.55_{-0.28}^{+0.28}$ &$2.62_{-0.35}^{+0.43}$ &$3.27_{-0.45}^{+0.52}$ \\
\cline{2-7}
& LineE (keV)&$6.53_{-0.08}^{+0.07}$ &$6.06_{-0.31}^{+0.14}$ &$5.94_{-0.20}^{+0.14}$ &$5.70_{-0.26}^{+0.18}$ &$5.90_{-0.24}^{+0.15}$ \\
Gaussian &Sigma (keV)&$0.66_{-0.08}^{+0.11}$ &$1.17_{-0.14}^{+0.31}$ &$1.27_{-0.16}^{+0.21}$ &$1.38_{-0.15}^{+0.21}$ &$1.21_{-0.15}^{+0.21}$ \\
&$N_{\rm Fe}$&$0.09_{-0.01}^{+0.02}$ &$0.16_{-0.03}^{+0.08}$ &$0.22_{-0.04}^{+0.07}$ &$0.29_{-0.05}^{+0.09}$ &$0.21_{-0.04}^{+0.07}$ \\
\cline{2-7}
Factor & ME &$0.93\pm{0.01}$ &$0.91\pm{0.01}$ &$0.95\pm{0.01}$ &$0.93\pm{0.01}$ &$0.96\pm{0.01}$ \\
&HE&$0.95\pm{0.01}$ &$0.95\pm{0.01}$ &$0.93\pm{0.01}$ &$0.95\pm{0.01}$ &$0.92\pm{0.01}$ \\
\hline
&$\chi^2/d.o.f$&395/414&427/414&436/414&340/414&419/414\\
\hline
\multicolumn{2}{c}{} & \multicolumn{5}{l}{Model 2: \modelC} \\
\hline
&$\Gamma$ & $1.97_{-0.01}^{+0.01}$ &$2.02_{-0.01}^{+0.01}$ &$2.09_{-0.01}^{+0.01}$ &$2.13_{-0.01}^{+0.01}$ &$2.17_{-0.01}^{+0.02}$ \\
Thcomp&tau&$2.14^{+0.29}_{-0.01}$ &$2.23^{+0.12}_{-0.08}$ &$1.99^{+0.13}_{-0.08}$ &$1.79^{+0.18}_{-0.02}$ &$1.50^{+0.20}_{-0.01}$\\
&$kT_{\rm e}$&$29.85_{-1.27}^{+1.72}$ &$32.43_{-2.24}^{+1.82}$ &$33.89_{-2.27}^{+2.84}$ &$35.46_{-2.24}^{+3.26}$ &$39.87_{-2.73}^{+6.94}$ \\
&$f_{\rm cov}$&$0.99_{-0.04}^{+0.01}$ &$1.00_{-0.05}^{+0.00}$ &$1.00_{-0.06}^{+0.00}$ &$0.88_{-0.04}^{+0.04}$ &$0.87_{-0.03}^{+0.05}$ \\
\cline{2-7}
Diskbb&$kT_{\rm in}$&$0.41_{-0.01}^{+0.01}$ &$0.40_{-0.01}^{+0.02}$ &$0.35_{-0.01}^{+0.01}$ &$0.38_{-0.01}^{+0.01}$ &$0.40_{-0.01}^{+0.01}$ \\
&$N_{\rm disk}$ ($10^{4}$)&$9.33_{-0.84}^{+1.11}$ &$11.4_{-1.81}^{+1.31}$ &$26.7_{-3.6}^{+2.7}$ &$24.3_{-2.07}^{+3.51}$ &$23.4_{-2.08}^{+3.12}$ \\
\cline{2-7}
& $\Gamma$&$0.29_{-0.03}^{+0.07}$ &$0.55_{-0.05}^{+0.04}$ &$0.61_{-0.03}^{+0.06}$ &$0.74\pm0.05$ &$0.81_{-0.04}^{+0.06}$ \\
cutoffpl$_{\rm 2}$&$E_{\rm cut}$(keV)&$11.42_{-0.26}^{+0.50}$ &$11.90_{-0.39}^{+0.36}$ &$11.14_{-0.24}^{+0.44}$ &$11.21_{-0.35}^{+0.42}$ &$11.48_{-0.31}^{+0.53}$ \\
&$N_{\rm c2}$&$0.94_{-0.06}^{+0.16}$ &$2.38_{-0.25}^{+0.26}$ &$3.31_{-0.23}^{+0.46}$ &$4.81_{-0.49}^{+0.59}$ &$5.45_{-0.49}^{+0.86}$ \\
\cline{2-7}
Factor& ME &$0.93\pm{0.01}$ &$0.91\pm{0.01}$ &$0.95\pm{0.01}$ &$0.93\pm{0.01}$ &$0.96\pm{0.01}$ \\
&HE&$0.95\pm{0.01}$ &$0.95\pm{0.01}$ &$0.93\pm{0.01}$ &$0.95\pm{0.01}$ &$0.92\pm{0.01}$\\
\hline
&$\chi^2/d.o.f$&398/414&430/414&444/414&344/414&421/414\\
\hline
\multicolumn{2}{c}{} & \multicolumn{5}{l}{Model 3: \modelD} \\
\hline
Diskbb & $kT_{\rm in} (\rm keV)$&$0.74_{-0.03}^{+0.02}$ &$0.63_{-0.04}^{+0.02}$ &$0.54_{-0.04}^{+0.03}$ &$0.49\pm0.01$ &$0.53_{-0.03}^{+0.02}$ \\
&$N_{\rm disk}$ ($10^{4}$)&$0.39_{-0.07}^{+0.06}$ &$0.84_{-0.12}^{+0.17}$ &$1.83_{-0.49}^{+0.29}$ &$3.55_{-0.62}^{+0.66}$ &$3.29_{-0.53}^{+0.56}$ \\
\cline{2-7}
& $a^*$ ($c{\rm J}/GM^2$) & &&0.98 (fixed)\\
&$q_{\rm in}$&$5.72_{-1.26}^{+2.55}$ &$5.00_{-0.97}^{+0.35}$ &$4.33_{-1.39}^{+0.74}$ &$10_{-5.19}^{*}$ &$4.50_{-1.38}^{+0.65}$\\
&$q_{\rm out}$ &&&3 (fixed)\\
&$R_{\rm br}$ ($R_{\rm g}$) & &&15(fixed)\\
&$R_{\rm in}(\rm ISCO)$&$1.41^{+0.33}_{-0.19}$ &$1.00^{+0.12}_{*}$ &$1.00^{+0.42}_{*}$ &$1.26^{+0.37}_{-0.16}$ &$1.00^{+0.27}_{*}$\\
& $\theta$ (deg)&$54.11_{-6.55}^{+6.42}$ &$52.97_{-7.01}^{+2.30}$ &$46.97_{-9.71}^{+6.17}$ &$57.28_{-10.17}^{+1.87}$ &$45.56_{-9.32}^{+5.33}$\\
Relxill&$\Gamma$&$1.59_{-0.03}^{+0.07}$ &$1.58_{-0.08}^{+0.07}$ &$1.72_{-0.05}^{+0.09}$ &$1.78_{-0.05}^{+0.04}$ &$1.82_{-0.02}^{+0.08}$ \\
&$E_{\rm cut} (\rm keV)$&$54.90_{-2.93}^{+6.03}$ &$57.50_{-4.70}^{+6.36}$ &$63.11_{-5.62}^{+9.21}$ &$63.22_{-3.79}^{+5.88}$ &$62.74_{-3.84}^{+8.44}$ \\
&log($\xi$)&$2.96_{-0.15}^{+0.13}$ &$3.01_{-0.16}^{+0.24}$ &$3.53_{-0.27}^{+0.30}$ &$4.25_{-0.33}^{+0.05}$ &$3.71_{-0.37}^{+0.28}$ \\
&$A_{\rm Fe}$&$5.04_{-1.38}^{+4.01}$ &$4.40_{-2.21}^{+1.31}$ &$4.03_{-2.80}^{+2.04}$ &$9.76_{-4.86}^{+0.06}$ &$9.34_{-5.33}^{+0.32}$ \\
&$R_{\rm ref}$&$0.33_{-0.08}^{+0.11}$ &$0.69_{-0.19}^{+0.21}$ &$0.58_{-0.16}^{+0.18}$ &$0.58_{-0.12}^{+0.17}$ &$0.38_{-0.11}^{+0.06}$ \\
&$N_{\rm ref}$&$0.23_{-0.01}^{+0.02}$ &$0.17\pm0.01$ &$0.21_{-0.02}^{+0.03}$ &$0.23_{-0.04}^{+0.02}$ &$0.26_{-0.02}^{+0.04}$ \\
\cline{2-7}
& $\Gamma$&$0.66_{-0.11}^{+0.07}$ &$0.97_{-0.06}^{+0.09}$ &$0.98_{-0.06}^{+0.06}$ &$1.02_{-0.06}^{+0.06}$ &$1.02_{-0.04}^{+0.07}$ \\
 cutoffpl$_{\rm 2}$&$E_{\rm cut}$ (keV)&$12.44_{-0.73}^{+0.79}$ &$13.38_{-0.63}^{+1.07}$ &$12.77_{-0.61}^{+0.74}$ &$12.57_{-0.57}^{+0.44}$ &$11.84_{-0.39}^{+0.75}$ \\
&$N_{\rm c2}$&$2.23_{-0.53}^{+0.39}$ &$6.64_{-0.86}^{+1.22}$ &$7.78_{-1.32}^{+0.98}$ &$8.93_{-1.30}^{+1.38}$ &$9.47_{-1.17}^{+1.48}$ \\
\cline{2-7}
Factor & ME &$0.93\pm{0.01}$ &$0.91\pm{0.01}$ &$0.95\pm{-0.01}$ &$0.93\pm{0.01}$ &$0.96\pm{0.01}$ \\
&HE&$0.95\pm{0.01}$ &$0.95\pm{0.01}$ &$0.92\pm{-0.01}$ &$0.95_{-0.00}^{+0.00}$ &$0.92\pm{0.01}$ \\
\hline
&$\chi^2/d.o.f$&406/411&404/411&433/411&342/411&409/411\\
\cline{2-7}
\multicolumn{3}{c}{} & \multicolumn{4}{l}{Flux from Model 3} \\
Flux ($10^{-7} \rm ergs/\rm cm^{2}/\rm s$) & Total&$2.47\pm0.01$ &$2.68\pm0.01$ &$2.94\pm0.01$ &$3.10\pm0.01$ &$3.07\pm0.01$\\
$10^{-8} \rm ergs/\rm cm^{2}/\rm s$& Diskbb&$0.80\pm0.02$ &$0.61\pm0.02$ &$0.61\pm0.02$ &$0.69\pm0.01$ &$0.83\pm0.03$ \\
$10^{-7} \rm ergs/\rm cm^{2}/\rm s$&cutoffpl$_{\rm 1}$&$0.88_{-0.04}^{+0.03}$ &$0.79\pm0.02$ &$0.82\pm0.02$ &$0.94\pm0.02$ &$0.89\pm0.02$ \\
$10^{-7} \rm ergs/\rm cm^{2}/\rm s$&Relxill&$0.27_{-0.03}^{+0.03}$ &$0.33\pm0.01$ &$0.44\pm0.01$ &$0.49\pm0.01$ &$0.56\pm0.01$ \\
$10^{-7} \rm ergs/\rm cm^{2}/\rm s$&cutoffpl$_{\rm 2}$&$1.24\pm0.01$ &$1.49\pm0.01$ &$1.61\pm0.01$ &$1.59\pm0.01$ &$1.53\pm0.01$ \\
\hline
\multicolumn{7}{p{0.95\textwidth}}{%
\textbf{Note.} Parameters peg at their limits are denoted by ``$^{*}$".
$N_{\rm disk}$, $N_{\rm c1}$, and $N_{\rm c2}$ denote the normalizations of the {\tt diskbb} and the two {\tt cutoffpl} components, respectively. $N_{\rm Fe}$ is the normalization of the Gaussian Fe line, and $N_{\rm ref}$ corresponds to the reflection component.
In Model 3, the total and the individual component flux in the energy band 2--120~keV are calculated.}\\
\end{tabular}
}
\end{table*}

\begin{table*}
\caption{Best-fit Parameters from Model~4.}
\label{tab:model-reflkerr}
\centering
{\small
\begin{tabular}{ccccccc}
\hline\hline
Component &  Parameter & Epoch 1 &
Epoch 2 &  Epoch 3 & Epoch 4 & Epoch 5\\
\hline
TBabs&$N_{\rm H}(\times10^{22}\rm cm^{-2})$  & &&0.3(fixed)&&\\
\hline
\multicolumn{2}{c}{} & \multicolumn{5}{l}{Model 4: constant*tbabs(diskbb + reflkerr + reflkerr\_bb)} \\
\hline
Diskbb&$kT_{\rm in} (\rm keV)$&$0.43_{-0.01}^{+0.08}$ &$0.34_{-0.02}^{+0.03}$ &$0.34_{-0.00}^{+0.00}$ &$0.32_{-0.02}^{+0.02}$ &$0.38_{-0.05}^{+0.00}$\\
&$N_{\rm disk}(10^{4})$&$2.83_{-1.4}^{+0.8}$ &$8.2_{-3.6}{+5.1}$ &$12.5_{-0.2}^{+0.6}$ &$22.8_{-8.4}^{+9.8}$ &$9.2_{-1.2}^{+5.7}$\\
\cline{2-7}
Thermal &$y_{\rm th}$&$0.58^{+0.02}_{-0.01}$ &$0.56^{+0.01}_{-0.02}$ &$0.53\pm{0.01}$ &$0.52\pm{+0.01}$ &$0.51\pm{0.01}$\\
Comptonization&$kT_{\rm e}$&$6.60_{-0.23}^{+0.17}$ &$6.27_{-0.11}^{+0.19}$ &$5.90_{-0.10}^{+0.03}$ &$5.87_{-0.48}^{+0.10}$ &$5.65_{-0.30}^{+0.10}$\\
&$N_{\rm th}$&$10.05_{-1.39}^{+0.51}$ &$13.12_{-1.15}^{+1.87}$ &$15.08_{-0.86}^{+0.36}$ &$14.57_{-2.54}^{+0.65}$ &$15.18_{-2.84}^{+1.37}$\\
\cline{2-7}
&$y_{h}$&$0.70\pm{0.02}$ &$0.67^{+0.03}_{-0.02}$ &$0.66\pm{0.01}$ &$0.63^{+0.01}_{-0.03}$ &$0.61^{+0.01}_{-0.02}$\\
Hybrid &$kT_{\rm e}$&$20.1_{-1.4}^{+1.7}$ &$18.9_{-2.1}^{+1.4}$ &$16.9_{-0.8}^{+0.2}$ &$17.8_{-1.7}^{+0.8}$&$19.8_{-1.2}^{+1.0}$ \\
Comptonization&$\gamma_{\rm min}$&$1.46_{-0.04}^{+0.05}$ &$1.45_{-0.05}^{+0.02}$  &$1.43_{-0.02}^{+0.01}$ &$1.44_{-0.05}^{+0.02}$ &$1.49_{-0.03}^{+0.04}$\\
&$p$&$2.16_{-0.31}^{+0.40}$ &$2.01_{-0.43}^{+0.09}$ &$1.73_{-0.03}^{+0.02}$ &$1.85_{-0.06}^{+0.30}$ &$2.00_{-0.44}^{+0.09}$\\
and reflection&$R_{\rm in}$(ISCO)&$1.85^{+0.52}_{-0.40}$ &$1.00^{+0.47}_{*}$ &$1.00^{+0.01}_{*}$ &$1.01^{+0.62}_{-0.01}$ &$1.00^{+0.39}_{-0.01}$\\
&$\theta$ (deg)&$50.00_{-4.57}^{+4.10}$ &$50.73_{-5.83}^{+1.98}$ &$50.00_{-2.17}^{+0.53}$ &$40.39_{-4.57}^{+4.68}$ & $50(fixed)$\\
&log($\xi$)&$2.29_{-0.67}^{+0.36}$ &$2.91_{-0.19}^{+0.17}$  &$3.90_{-0.05}^{+0.07}$ &$3.50_{-0.02}^{+0.05}$ &$3.58_{-0.06}^{+0.04}$ \\
&$R_{\rm ref}$&$0.33_{-0.07}^{+0.08}$ &$0.36_{-0.14}^{+0.06}$ &$0.25_{-0.00}^{+0.00}$ &$0.82_{-0.01}^{+0.21}$ &$0.71_{-0.01}^{+0.31}$\\
&$N_{\rm h}$&$6.52_{-1.23}^{+0.74}$ &$8.06_{-1.95}^{+1.30}$ &$9.23_{-0.25}^{+0.63}$ &$9.99_{-0.81}^{+0.87}$ &$10.57_{-0.92}^{+1.80}$\\
\cline{2-7}
Factor&ME&$0.93_{-0.01}^{+0.00}$ &$0.93_{-0.02}^{+0.01}$ &$0.95_{-0.01}^{+0.00}$ &$0.95_{-0.02}^{+0.01}$&$0.94_{-0.00}^{+0.01}$\\
&HE&$0.97_{-0.01}^{+0.00}$ &$0.96_{-0.01}^{+0.01}$ &$0.94_{-0.00}^{+0.01}$ &$0.91_{-0.02}^{+0.01}$ &$0.89_{-0.01}^{+0.01}$\\
\hline
&$\chi^2/d.o.f$ &487/418&467/418&459/418&366/418 &486/419\\
\hline
\multicolumn{7}{p{0.95\textwidth}}{%
\textbf{Note.} The hybrid Comptonization model is fitted over 2--200 keV in order to better constrain the high-energy tail, whereas the other models are restricted to 2--120 keV.
$N_{\rm th,h}$ is the normalization of the Comptonization component, defined as the flux density at 1 keV, and $y_{\rm th,h}$ is the Compton parameter, calculated assuming spherical geometry.
Parameters pegged at their limits are denoted by ``$^{*}$".
}\\
\end{tabular}
}
\end{table*}

\clearpage




\section*{Acknowledgements}

The authors sincerely thank the reviewers for their constructive comments. We also gratefully acknowledge the assistance of Dr.~Michał~Szanecki.
This work made use of the data from the \emph{Insight}-HXMT mission, a project funded by the China National Space Administration (CNSA) and the Chinese Academy of Sciences (CAS), and data and/or software provided by the High Energy Astrophysics Science Archive Research Center (HEASARC), a service of the Astrophysics Science Division at NASA/GSFC. Y.X. acknowledges support by National Science Foundation of China through grants NSFC-12521005 and the Hundred Talents Programme of the Chinese Academy of Sciences. This work is supported by the National Key RD Program of China (2021YFA0718500) and the National Natural Science Foundation of China (NSFC) under grant Nos. U1838202, 12273030, 11733009, 11673023, U1938102, U2038104, U2031205, 12233002, 12133007, 12027803, 12122306, 12333007, 12403048, 12403053, and the Scientific and Technological Innovation Project of IHEP (grant No. Y7515570U1). This work was partially supported by the International Partnership Program of the CAS (grant No. 113111KYSB20190020).

\section*{Data Availability}

This work is primarily based on data from the \textit{Insight}-HXMT mission, a project funded by the China National Space Administration (CNSA) and the Chinese Academy of Sciences (CAS). All data are publicly available at: \url{http://archive.hxmt.cn/proposal}

In addition, this research has made use of data obtained through the HEASARC Online Service, provided by the NASA/Goddard Space Flight Center: \url{https://heasarc.gsfc.nasa.gov/docs/archive.html}

\bibliographystyle{mnras}
\bibliography{example} 

@INPROCEEDINGS{1996Arnaud,
       author = {{Arnaud}, K.~A.},
        title = "{XSPEC: The First Ten Years}",
    booktitle = {Astronomical Data Analysis Software and Systems V},
         year = 1996,
       editor = {{Jacoby}, George H. and {Barnes}, Jeannette},
       series = {Astronomical Society of the Pacific Conference Series},
       volume = {101},
        month = jan,
        pages = {17},
       adsurl = {https://ui.adsabs.harvard.edu/abs/1996ASPC..101...17A}
}

@ARTICLE{2020JHEAp..27...64L,
       author = {{Li}, Xiaobo and {Li}, Xufang and {Tan}, Ying and {Yang}, Yanji and {Ge}, Mingyu and {Zhang}, Juan and {Tuo}, Youli and {Wu}, Baiyang and {Liao}, Jinyuan and {Zhang}, Yifei and {Song}, Liming and {Zhang}, Shu and {Qu}, Jinlu and {Zhang}, Shuang-nan and {Lu}, Fangjun and {Xu}, Yupeng and {Liu}, Congzhan and {Cao}, Xuelei and {Chen}, Yong and {Nie}, Jianyin and {Zhao}, Haisheng and {Li}, Chengkui},
        title = "{In-flight calibration of the Insight-Hard X-ray Modulation Telescope}",
      journal = {Journal of High Energy Astrophysics},
         year = 2020,
        month = aug,
       volume = {27},
        pages = {64-76},
          doi = {10.1016/j.jheap.2020.02.009},
archivePrefix = {arXiv},
       eprint = {2003.06998},
 primaryClass = {astro-ph.IM},
       adsurl = {https://ui.adsabs.harvard.edu/abs/2020JHEAp..27...64L}
}

@ARTICLE{2020JHEAp..27...24L,
       author = {{Liao}, Jin-Yuan and {Zhang}, Shu and {Chen}, Yong and {Zhang}, Juan and {Jin}, Jing and {Chang}, Zhi and {Chen}, Yu-Peng and {Ge}, Ming-Yu and {Guo}, Cheng-Cheng and {Li}, Gang and {Li}, Xiao-Bo and {Lu}, Fang-Jun and {Lu}, Xue-Feng and {Nie}, Jian-Yin and {Song}, Li-Ming and {Yang}, Yan-Ji and {You}, Yuan and {Zhao}, Hai-Sheng and {Zhang}, Shuang-Nan},
        title = "{Background model for the Low-Energy Telescope of Insight-HXMT}",
      journal = {Journal of High Energy Astrophysics},
         year = 2020,
        month = aug,
       volume = {27},
        pages = {24-32},
          doi = {10.1016/j.jheap.2020.02.010},
archivePrefix = {arXiv},
       eprint = {2004.01432},
 primaryClass = {astro-ph.IM},
       adsurl = {https://ui.adsabs.harvard.edu/abs/2020JHEAp..27...24L}
}

@ARTICLE{2020JHEAp..27...44G,
       author = {{Guo}, Cheng-Cheng and {Liao}, Jin-Yuan and {Zhang}, Shu and {Zhang}, Juan and {Tan}, Ying and {Song}, Li-Ming and {Lu}, Fang-Jun and {Cao}, Xue-Lei and {Chang}, Zhi and {Chen}, Yu-Peng and {Du}, Yuan-Yuan and {Ge}, Ming-Yu and {Gu}, Yu-Dong and {Jiang}, Wei-Chun and {Jin}, Jing and {Li}, Gang and {Li}, Xian and {Li}, Xiao-Bo and {Liu}, Shao-Zhen and {Liu}, Xiao-Jing and {Lu}, Xue-Feng and {Luo}, Tao and {Meng}, Bin and {Sun}, Liang and {Yang}, Jia-Wei and {Yang}, Sheng and {You}, Yuan and {Zhang}, Wan-Chang and {Zhao}, Hai-Sheng and {Zhang}, Shuang-Nan},
        title = "{The background model of the medium energy X-ray telescope of Insight-HXMT}",
      journal = {Journal of High Energy Astrophysics},
         year = 2020,
        month = aug,
       volume = {27},
        pages = {44-50},
          doi = {10.1016/j.jheap.2020.02.008},
archivePrefix = {arXiv},
       eprint = {2003.06260},
 primaryClass = {astro-ph.IM},
       adsurl = {https://ui.adsabs.harvard.edu/abs/2020JHEAp..27...44G}
}

@ARTICLE{2020JHEAp..27...14L,
       author = {{Liao}, Jin-Yuan and {Zhang}, Shu and {Lu}, Xue-Feng and {Zhang}, Juan and {Li}, Gang and {Chang}, Zhi and {Chen}, Yu-Peng and {Ge}, Ming-Yu and {Guo}, Cheng-Cheng and {Huang}, Rui and {Jin}, Jing and {Li}, Xiao-Bo and {Li}, Xu-Fang and {Li}, Zheng-Wei and {Liu}, Cong-Zhan and {Lu}, Fang-Jun and {Nie}, Jian-Yin and {Song}, Li-Ming and {Wang}, Si-Fan and {You}, Yuan and {Zhang}, Yi-Fei and {Zhao}, Hai-Sheng and {Zhang}, Shuang-Nan},
        title = "{Background model for the high-energy telescope of Insight-HXMT}",
      journal = {Journal of High Energy Astrophysics},
         year = 2020,
        month = aug,
       volume = {27},
        pages = {14-23},
          doi = {10.1016/j.jheap.2020.04.002},
archivePrefix = {arXiv},
       eprint = {2005.01661},
 primaryClass = {astro-ph.IM},
       adsurl = {https://ui.adsabs.harvard.edu/abs/2020JHEAp..27...14L}
}

@ARTICLE{2020SCPMA..6349505C,
       author = {{Chen}, Yong and {Cui}, WeiWei and {Li}, Wei and {Wang}, Juan and {Xu}, YuPeng and {Lu}, FangJun and {Wang}, YuSa and {Chen}, TianXiang and {Han}, DaWei and {Hu}, Wei and {Zhang}, Yi and {Huo}, Jia and {Yang}, YanJi and {Li}, MaoShun and {Lu}, Bo and {Zhang}, ZiLiang and {Li}, TiPei and {Zhang}, ShuangNan and {Xiong}, ShaoLin and {Zhang}, Shu and {Xue}, RongFeng and {Zhao}, XiaoFan and {Zhu}, Yue and {Zhu}, YuXuan and {Liu}, HongWei and {Yang}, YiJung and {Zhang}, Fan},
        title = "{The Low Energy X-ray telescope (LE) onboard the Insight-HXMT astronomy satellite}",
      journal = {Science China Physics, Mechanics, and Astronomy},
         year = 2020,
        month = apr,
       volume = {63},
       number = {4},
          eid = {249505},
        pages = {249505},
          doi = {10.1007/s11433-019-1469-5},
archivePrefix = {arXiv},
       eprint = {1910.08319},
 primaryClass = {astro-ph.IM},
       adsurl = {https://ui.adsabs.harvard.edu/abs/2020SCPMA..6349505C}
}

@ARTICLE{2020SCPMA..6349504C,
       author = {{Cao}, XueLei and {Jiang}, WeiChun and {Meng}, Bin and {Zhang}, WanChang and {Luo}, Tao and {Yang}, Sheng and {Zhang}, ChunLei and {Gu}, YuDong and {Sun}, Liang and {Liu}, XiaoJing and {Yang}, JiaWei and {Li}, Xian and {Tan}, Ying and {Liu}, ShaoZhen and {Du}, YuanYuan and {Lu}, FangJun and {Xu}, YuPeng and {Guan}, Ju and {Zhang}, ShuangNan and {Wang}, HuanYu and {Li}, TiPei and {Zhang}, ChengMo and {Wen}, XiangYang and {Qu}, JinLu and {Song}, LiMing and {Li}, XiaoBo and {Ge}, MingYu and {Zhou}, YuPeng and {Xiong}, ShaoLin and {Zhang}, Shu and {Zhang}, YongJie and {Cheng}, ZeHao and {Zhang}, Fei and {Li}, MaoShun and {Liang}, XiaoHua and {Gao}, Min and {Yang}, EnBo and {Liu}, XiaoHang and {Liu}, HongWei and {Yang}, YiJung and {Zhang}, Fan},
        title = "{The Medium Energy X-ray telescope (ME) onboard the Insight-HXMT astronomy satellite}",
      journal = {Science China Physics, Mechanics, and Astronomy},
         year = 2020,
        month = apr,
       volume = {63},
       number = {4},
          eid = {249504},
        pages = {249504},
          doi = {10.1007/s11433-019-1506-1},
archivePrefix = {arXiv},
       eprint = {1910.04451},
 primaryClass = {astro-ph.IM},
       adsurl = {https://ui.adsabs.harvard.edu/abs/2020SCPMA..6349504C}
}

@ARTICLE{2020SCPMA..6349503L,
       author = {{Liu}, CongZhan and {Zhang}, YiFei and {Li}, XuFang and {Lu}, XueFeng and {Chang}, Zhi and {Li}, ZhengWei and {Zhang}, AiMei and {Jin}, YongJie and {Yu}, HuiMing and {Zhang}, Zhao and {Fu}, MinXue and {Chen}, YiBao and {Ji}, JianFeng and {Xu}, YuPeng and {Deng}, JingKang and {Shang}, RenCheng and {Liu}, GuoQing and {Lu}, FangJun and {Zhang}, ShuangNan and {Dong}, YongWei and {Li}, TiPei and {Wu}, Mei and {Li}, YanGuo and {Wang}, HuanYu and {Wu}, BoBing and {Zhang}, YongJie and {Zhang}, Zhi and {Xiong}, ShaoLin and {Liu}, Yuan and {Zhang}, Shu and {Liu}, HongWei and {Yang}, YiJung and {Zhang}, Fan},
        title = "{The High Energy X-ray telescope (HE) onboard the Insight-HXMT astronomy satellite}",
      journal = {Science China Physics, Mechanics, and Astronomy},
         year = 2020,
        month = apr,
       volume = {63},
       number = {4},
          eid = {249503},
        pages = {249503},
          doi = {10.1007/s11433-019-1486-x},
archivePrefix = {arXiv},
       eprint = {1910.04955},
 primaryClass = {astro-ph.IM},
       adsurl = {https://ui.adsabs.harvard.edu/abs/2020SCPMA..6349503L}
}

@ARTICLE{2020SCPMA..6349502Z,
       author = {{Zhang}, Shuang-Nan and {Li}, TiPei and {Lu}, FangJun and {Song}, LiMing and {Xu}, YuPeng and {Liu}, CongZhan and {Chen}, Yong and {Cao}, XueLei and {Bu}, QingCui and {Chang}, Zhi and {Chen}, Gang and {Chen}, Li and {Chen}, TianXiang and {Chen}, YiBao and {Chen}, YuPeng and {Cui}, Wei and {Cui}, WeiWei and {Deng}, JingKang and {Dong}, YongWei and {Du}, YuanYuan and {Fu}, MinXue and {Gao}, GuanHua and {Gao}, He and {Gao}, Min and {Ge}, MingYu and {Gu}, YuDong and {Guan}, Ju and {Gungor}, Can and {Guo}, ChengCheng and {Han}, DaWei and {Hu}, Wei and {Huang}, Yue and {Huo}, Jia and {Jia}, ShuMei and {Jiang}, LuHua and {Jiang}, WeiChun and {Jin}, Jing and {Jin}, YongJie and {Li}, Bing and {Li}, ChengKui and {Li}, Gang and {Li}, MaoShun and {Li}, Wei and {Li}, Xian and {Li}, XiaoBo and {Li}, XuFang and {Li}, YanGuo and {Li}, ZiJian and {Li}, ZhengWei and {Liang}, XiaoHua and {Liao}, JinYuan and {Liu}, GuoQing and {Liu}, HongWei and {Liu}, ShaoZhen and {Liu}, XiaoJing and {Liu}, Yuan and {Liu}, YiNong and {Lu}, Bo and {Lu}, XueFeng and {Luo}, Tao and {Ma}, Xiang and {Meng}, Bin and {Nang}, Yi and {Nie}, JianYin and {Ou}, Ge and {Qu}, JinLu and {Sai}, Na and {Shang}, RenCheng and {Shen}, GuoHong and {Sun}, Liang and {Tan}, Ying and {Tao}, Lian and {Tuo}, YouLi and {Wang}, Chen and {Wang}, ChunQin and {Wang}, GuoFeng and {Wang}, HuanYu and {Wang}, Juan and {Wang}, WenShuai and {Wang}, YuSa and {Wen}, XiangYang and {Wu}, BaiYang and {Wu}, BoBing and {Wu}, Mei and {Xiao}, GuangCheng and {Xiong}, ShaoLin and {Yan}, LinLi and {Yang}, JiaWei and {Yang}, Sheng and {Yang}, YanJi and {Yi}, QiBin and {Yuan}, Bin and {Zhang}, AiMei and {Zhang}, ChunLei and {Zhang}, ChengMo and {Zhang}, Fan and {Zhang}, HongMei and {Zhang}, Juan and {Zhang}, Qiang and {Zhang}, ShenYi and {Zhang}, Shu and {Zhang}, Tong and {Zhang}, WanChang and {Zhang}, Wei and {Zhang}, WenZhao and {Zhang}, Yi and {Zhang}, YiFei and {Zhang}, YongJie and {Zhang}, Yue and {Zhang}, Zhao and {Zhang}, Zhi and {Zhang}, ZiLiang and {Zhao}, HaiSheng and {Zhao}, XiaoFan and {Zheng}, ShiJie and {Zhou}, JianFeng and {Zhu}, YuXuan and {Zhu}, Yue and {Zhuang}, RenLin and {Insight-HXMT Team}},
        title = "{Overview to the Hard X-ray Modulation Telescope (Insight-HXMT) Satellite}",
      journal = {Science China Physics, Mechanics, and Astronomy},
         year = 2020,
        month = apr,
       volume = {63},
       number = {4},
          eid = {249502},
        pages = {249502},
          doi = {10.1007/s11433-019-1432-6},
archivePrefix = {arXiv},
       eprint = {1910.09613},
 primaryClass = {astro-ph.IM},
       adsurl = {https://ui.adsabs.harvard.edu/abs/2020SCPMA..6349502Z}
}

@ARTICLE{2019Buisson,
       author = {{Buisson}, D.~J.~K. and {Fabian}, A.~C. and {Barret}, D. and {F{\"u}rst}, F. and {Gandhi}, P. and {Garc{\'\i}a}, J.~A. and {Kara}, E. and {Madsen}, K.~K. and {Miller}, J.~M. and {Parker}, M.~L. and {Shaw}, A.~W. and {Tomsick}, J.~A. and {Walton}, D.~J.},
        title = "{MAXI J1820+070 with NuSTAR I. An increase in variability frequency but a stable reflection spectrum: coronal properties and implications for the inner disc in black hole binaries}",
      journal = {\mnras},
         year = 2019,
        month = nov,
       volume = {490},
       number = {1},
        pages = {1350-1362},
          doi = {10.1093/mnras/stz2681},
archivePrefix = {arXiv},
       eprint = {1909.04688},
 primaryClass = {astro-ph.HE},
       adsurl = {https://ui.adsabs.harvard.edu/abs/2019MNRAS.490.1350B}
}

@ARTICLE{2022RenXQ,
       author = {{Ren}, X.~Q. and {Wang}, Yanan and {Zhang}, S.~N. and {Soria}, R. and {Tao}, L. and {Ji}, L. and {Yang}, Y.~J. and {Qu}, J.~L. and {Zhang}, S. and {Song}, L.~M. and {Ge}, M.~Y. and {Huang}, Y. and {Li}, X.~B. and {Liao}, J.~Y. and {Liu}, H.~X. and {Ma}, R.~C. and {Tuo}, Y.~L. and {Wang}, P.~J. and {Zhang}, W. and {Zhou}, D.~K.},
        title = "{Insight-HXMT Study of the Inner Accretion Disk in the Black Hole Candidate EXO 1846-031}",
      journal = {\apj},
         year = 2022,
        month = jun,
       volume = {932},
       number = {1},
          eid = {66},
        pages = {66},
          doi = {10.3847/1538-4357/ac6dd7},
archivePrefix = {arXiv},
       eprint = {2205.04635},
 primaryClass = {astro-ph.HE},
       adsurl = {https://ui.adsabs.harvard.edu/abs/2022ApJ...932...66R}
}

@ARTICLE{2022ZhangW,
       author = {{Zhang}, W. and {Tao}, L. and {Soria}, R. and {Qu}, J.~L. and {Zhang}, S.~N. and {Weng}, S.~S. and {Zhang}, L. and {Wang}, Y.~N. and {Huang}, Y. and {Ma}, R.~C. and {Zhang}, S. and {Ge}, M.~Y. and {Song}, L.~M. and {Ma}, X. and {Bu}, Q.~C. and {Cai}, C. and {Cao}, X.~L. and {Chang}, Z. and {Chen}, L. and {Chen}, T.~X. and {Chen}, Y.~B. and {Chen}, Y. and {Chen}, Y.~P. and {Cui}, W.~W. and {Du}, Y.~Y. and {Gao}, G.~H. and {Gao}, H. and {Gu}, Y.~D. and {Guan}, J. and {Guo}, C.~C. and {Han}, D.~W. and {Huo}, J. and {Jia}, S.~M. and {Jiang}, W.~C. and {Jin}, J. and {Kong}, L.~D. and {Li}, B. and {Li}, C.~K. and {Li}, G. and {Li}, T.~P. and {Li}, W. and {Li}, X. and {Li}, X.~B. and {Li}, X.~F. and {Li}, Z.~W. and {Liang}, X.~H. and {Liao}, J.~Y. and {Liu}, B.~S. and {Liu}, C.~Z. and {Liu}, H.~X. and {Liu}, H.~W. and {Liu}, X.~J. and {Lu}, F.~J. and {Lu}, X.~F. and {Luo}, Q. and {Luo}, T. and {Meng}, B. and {Nang}, Y. and {Nie}, J.~Y. and {Ou}, G. and {Ren}, X.~Q. and {Sai}, N. and {Song}, X.~Y. and {Sun}, L. and {Tan}, Y. and {Tuo}, Y.~L. and {Wang}, C. and {Wang}, L.~J. and {Wang}, P.~J. and {Wang}, W.~S. and {Wang}, Y.~S. and {Wen}, X.~Y. and {Wu}, B.~Y. and {Wu}, B.~B. and {Wu}, M. and {Xiao}, G.~C. and {Xiao}, S. and {Xiong}, S.~L. and {Chen}, Y.~P. and {Yang}, R.~J. and {Yang}, S. and {Yang}, Y.~J. and {Yang}, Y.~R. and {Yi}, Q.~B. and {Yin}, Q.~Q. and {Yuan}, Y. and {Zhang}, F. and {Zhang}, H.~M. and {Zhang}, P. and {Zhang}, W.~C. and {Zhang}, Y.~F. and {Zhang}, Y.~H. and {Zhao}, H.~S. and {Zhao}, X.~F. and {Zheng}, S.~J. and {Zheng}, Y.~G. and {Zhou}, D.~K.},
        title = "{Peculiar Disk Behaviors of the Black Hole Candidate MAXI J1348-630 in the Hard State Observed by Insight-HXMT and Swift}",
      journal = {\apj},
         year = 2022,
        month = mar,
       volume = {927},
       number = {2},
          eid = {210},
        pages = {210},
          doi = {10.3847/1538-4357/ac4fca},
archivePrefix = {arXiv},
       eprint = {2201.11919},
 primaryClass = {astro-ph.HE},
       adsurl = {https://ui.adsabs.harvard.edu/abs/2022ApJ...927..210Z}
}

@ARTICLE{2020ZdziarskiMN,
       author = {{Zdziarski}, Andrzej A. and {Szanecki}, Micha{\l} and {Poutanen}, Juri and {Gierli{\'n}ski}, Marek and {Biernacki}, Pawe{\l}},
        title = "{Spectral and temporal properties of Compton scattering by mildly relativistic thermal electrons}",
      journal = {\mnras},
         year = 2020,
        month = mar,
       volume = {492},
       number = {4},
        pages = {5234-5246},
          doi = {10.1093/mnras/staa159},
archivePrefix = {arXiv},
       eprint = {1910.04535},
 primaryClass = {astro-ph.HE},
       adsurl = {https://ui.adsabs.harvard.edu/abs/2020MNRAS.492.5234Z}
}

@ARTICLE{2014GarciaApJ,
       author = {{Garc{\'\i}a}, J. and {Dauser}, T. and {Lohfink}, A. and {Kallman}, T.~R. and {Steiner}, J.~F. and {McClintock}, J.~E. and {Brenneman}, L. and {Wilms}, J. and {Eikmann}, W. and {Reynolds}, C.~S. and {Tombesi}, F.},
        title = "{Improved Reflection Models of Black Hole Accretion Disks: Treating the Angular Distribution of X-Rays}",
      journal = {\apj},
         year = 2014,
        month = feb,
       volume = {782},
       number = {2},
          eid = {76},
        pages = {76},
          doi = {10.1088/0004-637X/782/2/76},
archivePrefix = {arXiv},
       eprint = {1312.3231},
 primaryClass = {astro-ph.HE},
       adsurl = {https://ui.adsabs.harvard.edu/abs/2014ApJ...782...76G}
}

@ARTICLE{2024ApJPeng,
       author = {{Peng}, Jing-Qiang and {Zhang}, Shu and {Shui}, Qing-Cang and {Zhang}, Shuang-Nan and {Kong}, Ling-Da and {Chen}, Yu-Peng and {Wang}, Peng-Ju and {Ji}, Long and {Qu}, Jin-Lu and {Tao}, Lian and {Ge}, Ming-Yu and {Chang}, Zhi and {Li}, Jian and {Li}, Zhao-sheng and {Yu}, Zhuo-Li and {Yan}, Zhe},
        title = "{NICER, NuSTAR, and Insight-HXMT Views to the Newly Discovered Black Hole X-Ray Binary Swift J1727.8-1613}",
      journal = {\apjl},
         year = 2024,
        month = jan,
       volume = {960},
       number = {2},
          eid = {L17},
        pages = {L17},
          doi = {10.3847/2041-8213/ad17ca},
       adsurl = {https://ui.adsabs.harvard.edu/abs/2024ApJ...960L..17P}
}

@ARTICLE{1998PASJKubota,
       author = {{Kubota}, Aya and {Tanaka}, Yasuo and {Makishima}, Kazuo and {Ueda}, Yoshihiro and {Dotani}, Tadayasu and {Inoue}, Hajime and {Yamaoka}, Kazutaka},
        title = "{Evidence for a Black Hole in the X-Ray Transient GRS 1009-45}",
      journal = {\pasj},
         year = 1998,
        month = dec,
       volume = {50},
        pages = {667-673},
          doi = {10.1093/pasj/50.6.667},
       adsurl = {https://ui.adsabs.harvard.edu/abs/1998PASJ...50..667K}
}

@ARTICLE{2000ApJWilms,
       author = {{Wilms}, J. and {Allen}, A. and {McCray}, R.},
        title = "{On the Absorption of X-Rays in the Interstellar Medium}",
      journal = {\apj},
         year = 2000,
        month = oct,
       volume = {542},
       number = {2},
        pages = {914-924},
          doi = {10.1086/317016},
archivePrefix = {arXiv},
       eprint = {astro-ph/0008425},
 primaryClass = {astro-ph},
       adsurl = {https://ui.adsabs.harvard.edu/abs/2000ApJ...542..914W}
}

@ARTICLE{1996ApJVerner,
       author = {{Verner}, D.~A. and {Ferland}, G.~J. and {Korista}, K.~T. and {Yakovlev}, D.~G.},
        title = "{Atomic Data for Astrophysics. II. New Analytic FITS for Photoionization Cross Sections of Atoms and Ions}",
      journal = {\apj},
         year = 1996,
        month = jul,
       volume = {465},
        pages = {487},
          doi = {10.1086/177435},
archivePrefix = {arXiv},
       eprint = {astro-ph/9601009},
 primaryClass = {astro-ph},
       adsurl = {https://ui.adsabs.harvard.edu/abs/1996ApJ...465..487V}
}

@ARTICLE{2021NatCoYouBei,
       author = {{You}, Bei and {Tuo}, Yuoli and {Li}, Chengzhe and {Wang}, Wei and {Zhang}, Shuang-Nan and {Zhang}, Shu and {Ge}, Mingyu and {Luo}, Chong and {Liu}, Bifang and {Yuan}, Weimin and {Dai}, Zigao and {Liu}, Jifeng and {Qiao}, Erlin and {Jin}, Chichuan and {Liu}, Zhu and {Czerny}, Bozena and {Wu}, Qingwen and {Bu}, Qingcui and {Cai}, Ce and {Cao}, Xuelei and {Chang}, Zhi and {Chen}, Gang and {Chen}, Li and {Chen}, Tianxiang and {Chen}, Yibao and {Chen}, Yong and {Chen}, Yupeng and {Cui}, Wei and {Cui}, Weiwei and {Deng}, Jingkang and {Dong}, Yongwei and {Du}, Yuanyuan and {Fu}, Minxue and {Gao}, Guanhua and {Gao}, He and {Gao}, Min and {Gu}, Yudong and {Guan}, Ju and {Guo}, Chengcheng and {Han}, Dawei and {Huang}, Yue and {Huo}, Jia and {Jia}, Shumei and {Jiang}, Luhua and {Jiang}, Weichun and {Jin}, Jing and {Jin}, Yongjie and {Kong}, Lingda and {Li}, Bing and {Li}, Chengkui and {Li}, Gang and {Li}, Maoshun and {Li}, Tipei and {Li}, Wei and {Li}, Xian and {Li}, Xiaobo and {Li}, Xufang and {Li}, Yanguo and {Li}, Zhengwei and {Liang}, Xiaohua and {Liao}, Jinyuan and {Liu}, Congzhan and {Liu}, Guoqing and {Liu}, Hongwei and {Liu}, Xiaojing and {Liu}, Yinong and {Lu}, Bo and {Lu}, Fangjun and {Lu}, Xuefeng and {Luo}, Qi and {Luo}, Tao and {Ma}, Xiang and {Meng}, Bin and {Nang}, Yi and {Nie}, Jianyin and {Ou}, Ge and {Qu}, Jinlu and {Sai}, Na and {Shang}, Rencheng and {Song}, Liming and {Song}, Xinying and {Sun}, Liang and {Tan}, Ying and {Tao}, Lian and {Wang}, Chen and {Wang}, Guofeng and {Wang}, Juan and {Wang}, Lingjun and {Wang}, Wenshuai and {Wang}, Yusa and {Wen}, Xiangyang and {Wu}, Baiyang and {Wu}, Bobing and {Wu}, Mei and {Xiao}, Guangcheng and {Xiao}, Shuo and {Xiong}, Shaolin and {Xu}, Yupeng and {Yang}, Jiawei and {Yang}, Sheng and {Yang}, Yanji and {Yi}, Qibin and {Yin}, Qianqing and {You}, Yuan and {Zhang}, Aimei and {Zhang}, Chengmo and {Zhang}, Fan and {Zhang}, Hongmei and {Zhang}, Juan and {Zhang}, Tong and {Zhang}, Wanchang and {Zhang}, Wei and {Zhang}, Wenzhao and {Zhang}, Yi and {Zhang}, Yifei and {Zhang}, Yongjie and {Zhang}, Yue and {Zhang}, Zhao and {Zhang}, Ziliang and {Zhao}, Haisheng and {Zhao}, Xiaofan and {Zheng}, Shijie and {Zhou}, Dengke and {Zhou}, Jianfeng and {Zhu}, Yuxuan and {Zhu}, Yue},
        title = "{Insight-HXMT observations of jet-like corona in a black hole X-ray binary MAXI J1820+070}",
      journal = {Nature Communications},
         year = 2021,
        month = jan,
       volume = {12},
          eid = {1025},
        pages = {1025},
          doi = {10.1038/s41467-021-21169-5},
archivePrefix = {arXiv},
       eprint = {2102.07602},
 primaryClass = {astro-ph.HE},
       adsurl = {https://ui.adsabs.harvard.edu/abs/2021NatCo..12.1025Y}
}

@ARTICLE{2023arXivMereminskiy,
       author = {{Mereminskiy}, I. and {Lutovinov}, A. and {Molkov}, S. and {Krivonos}, R. and {Semena}, A. and {Sazonov}, S. and {Tkachenko}, A. and {Sunyaev}, R.},
        title = "{Hard X-rays and QPO in Swift J1727.8-1613: the rise and plateau of the 2023 outburst}",
      journal = {arXiv e-prints},
         year = 2023,
        month = oct,
          eid = {arXiv:2310.06697},
        pages = {arXiv:2310.06697},
          doi = {10.48550/arXiv.2310.06697},
archivePrefix = {arXiv},
       eprint = {2310.06697},
 primaryClass = {astro-ph.HE},
       adsurl = {https://ui.adsabs.harvard.edu/abs/2023arXiv231006697M}
}

@ARTICLE{2023A&ACangemi,
       author = {{Cangemi}, F. and {Rodriguez}, J. and {Belloni}, T. and {Gouiff{\`e}s}, C. and {Grinberg}, V. and {Laurent}, P. and {Petrucci}, P. -O. and {Wilms}, J.},
        title = "{INTEGRAL study of MAXI J1535{\ensuremath{-}}571, MAXI J1820+070, and MAXI J1348 {\ensuremath{-}} 630 outbursts. I. Detection and polarization properties of the high-energy emission}",
      journal = {\aap},
         year = 2023,
        month = jan,
       volume = {669},
          eid = {A65},
        pages = {A65},
          doi = {10.1051/0004-6361/202243564},
archivePrefix = {arXiv},
       eprint = {2210.08561},
 primaryClass = {astro-ph.HE},
       adsurl = {https://ui.adsabs.harvard.edu/abs/2023A&A...669A..65C}
}

@ARTICLE{2023SciYou,
       author = {{You}, Bei and {Cao}, Xinwu and {Yan}, Zhen and {Hameury}, Jean-Marie and {Czerny}, Bozena and {Wu}, Yue and {Xia}, Tianyu and {Sikora}, Marek and {Zhang}, Shuang-Nan and {Du}, Pu and {Zycki}, Piotr T.},
        title = "{Observations of a black hole x-ray binary indicate formation of a magnetically arrested disk}",
      journal = {Science},
         year = 2023,
        month = sep,
       volume = {381},
       number = {6661},
        pages = {961-964},
          doi = {10.1126/science.abo4504},
archivePrefix = {arXiv},
       eprint = {2309.00200},
 primaryClass = {astro-ph.HE},
       adsurl = {https://ui.adsabs.harvard.edu/abs/2023Sci...381..961Y}
}

@ARTICLE{2023ApJYou,
       author = {{You}, Bei and {Dong}, Yanting and {Yan}, Zhen and {Liu}, Zhu and {Tuo}, Youli and {Yao}, Yuanle and {Cao}, Xinwu},
        title = "{X-Ray Spectral Correlations in a Sample of Low-mass Black Hole X-Ray Binaries in the Hard State}",
      journal = {\apj},
         year = 2023,
        month = mar,
       volume = {945},
       number = {1},
          eid = {65},
        pages = {65},
          doi = {10.3847/1538-4357/acba11},
archivePrefix = {arXiv},
       eprint = {2302.13576},
 primaryClass = {astro-ph.HE},
       adsurl = {https://ui.adsabs.harvard.edu/abs/2023ApJ...945...65Y}
}

@ARTICLE{2020ApJYan,
       author = {{Yan}, Zhen and {Xie}, Fu-Guo and {Zhang}, Wenda},
        title = "{Coronal Properties of Black Hole X-Ray Binaries in the Hard State as Seen by NuSTAR and Swift}",
      journal = {\apjl},
         year = 2020,
        month = jan,
       volume = {889},
       number = {1},
          eid = {L18},
        pages = {L18},
          doi = {10.3847/2041-8213/ab665e},
archivePrefix = {arXiv},
       eprint = {1912.12145},
 primaryClass = {astro-ph.HE},
       adsurl = {https://ui.adsabs.harvard.edu/abs/2020ApJ...889L..18Y}
}

@ARTICLE{2023ApJVeledina,
       author = {{Veledina}, Alexandra and {Muleri}, Fabio and {Dov{\v{c}}iak}, Michal and {Poutanen}, Juri and {Ratheesh}, Ajay and {Capitanio}, Fiamma and {Matt}, Giorgio and {Soffitta}, Paolo and {Tennant}, Allyn F. and {Negro}, Michela and {Kaaret}, Philip and {Costa}, Enrico and {Ingram}, Adam and {Svoboda}, Ji{\v{r}}{\'\i} and {Krawczynski}, Henric and {Bianchi}, Stefano and {Steiner}, James F. and {Garc{\'\i}a}, Javier A. and {Kravtsov}, Vadim and {Nitindala}, Anagha P. and {Ewing}, Melissa and {Mastroserio}, Guglielmo and {Marinucci}, Andrea and {Ursini}, Francesco and {Tombesi}, Francesco and {Tsygankov}, Sergey S. and {Yang}, Yi-Jung and {Weisskopf}, Martin C. and {Trushkin}, Sergei A. and {Egron}, Elise and {Iacolina}, Maria Noemi and {Pilia}, Maura and {Marra}, Lorenzo and {Miku{\v{s}}incov{\'a}}, Romana and {Nathan}, Edward and {Parra}, Maxime and {Petrucci}, Pierre-Olivier and {Podgorn{\'y}}, Jakub and {Tugliani}, Stefano and {Zane}, Silvia and {Zhang}, Wenda and {Agudo}, Iv{\'a}n and {Antonelli}, Lucio A. and {Bachetti}, Matteo and {Baldini}, Luca and {Baumgartner}, Wayne H. and {Bellazzini}, Ronaldo and {Bongiorno}, Stephen D. and {Bonino}, Raffaella and {Brez}, Alessandro and {Bucciantini}, Niccol{\`o} and {Castellano}, Simone and {Cavazzuti}, Elisabetta and {Chen}, Chien-Ting and {Ciprini}, Stefano and {De Rosa}, Alessandra and {Del Monte}, Ettore and {Di Gesu}, Laura and {Di Lalla}, Niccol{\`o} and {Di Marco}, Alessandro and {Donnarumma}, Immacolata and {Doroshenko}, Victor and {Ehlert}, Steven R. and {Enoto}, Teruaki and {Evangelista}, Yuri and {Fabiani}, Sergio and {Ferrazzoli}, Riccardo and {Gunji}, Shuichi and {Hayashida}, Kiyoshi and {Heyl}, Jeremy and {Iwakiri}, Wataru and {Jorstad}, Svetlana G. and {Karas}, Vladimir and {Kislat}, Fabian and {Kitaguchi}, Takao and {Kolodziejczak}, Jeffery J. and {La Monaca}, Fabio and {Latronico}, Luca and {Liodakis}, Ioannis and {Maldera}, Simone and {Manfreda}, Alberto and {Marin}, Fr{\'e}d{\'e}ric and {Marscher}, Alan P. and {Marshall}, Herman L. and {Massaro}, Francesco and {Mitsuishi}, Ikuyuki and {Mizuno}, Tsunefumi and {Ng}, Chi-Yung and {O'Dell}, Stephen L. and {Omodei}, Nicola and {Oppedisano}, Chiara and {Papitto}, Alessandro and {Pavlov}, George G. and {Peirson}, Abel L. and {Perri}, Matteo and {Pesce-Rollins}, Melissa and {Possenti}, Andrea and {Puccetti}, Simonetta and {Ramsey}, Brian D. and {Rankin}, John and {Roberts}, Oliver J. and {Romani}, Roger W. and {Sgr{\`o}}, Carmelo and {Slane}, Patrick and {Spandre}, Gloria and {Swartz}, Douglas A. and {Tamagawa}, Toru and {Tavecchio}, Fabrizio and {Taverna}, Roberto and {Tawara}, Yuzuru and {Thomas}, Nicholas E. and {Trois}, Alessio and {Turolla}, Roberto and {Vink}, Jacco and {Wu}, Kinwah and {Xie}, Fei},
        title = "{Discovery of X-Ray Polarization from the Black Hole Transient Swift J1727.8-1613}",
      journal = {\apjl},
         year = 2023,
        month = nov,
       volume = {958},
       number = {1},
          eid = {L16},
        pages = {L16},
          doi = {10.3847/2041-8213/ad0781},
archivePrefix = {arXiv},
       eprint = {2309.15928},
 primaryClass = {astro-ph.HE},
       adsurl = {https://ui.adsabs.harvard.edu/abs/2023ApJ...958L..16V}
}

@INBOOK{2010LNPBelloni,
       author = {{Belloni}, T.~M.},
        title = "{States and Transitions in Black Hole Binaries}",
    booktitle = {Lecture Notes in Physics, Berlin Springer Verlag},
         year = 2010,
       editor = {{Belloni}, Tomaso},
       volume = {794},
        pages = {53},
          doi = {10.1007/978-3-540-76937-8_3},
       adsurl = {https://ui.adsabs.harvard.edu/abs/2010LNP...794...53B}
}

@ARTICLE{2006ARA&ARemillard,
       author = {{Remillard}, Ronald A. and {McClintock}, Jeffrey E.},
        title = "{X-Ray Properties of Black-Hole Binaries}",
      journal = {\araa},
         year = 2006,
        month = sep,
       volume = {44},
       number = {1},
        pages = {49-92},
          doi = {10.1146/annurev.astro.44.051905.092532},
archivePrefix = {arXiv},
       eprint = {astro-ph/0606352},
 primaryClass = {astro-ph},
       adsurl = {https://ui.adsabs.harvard.edu/abs/2006ARA&A..44...49R}
}

@ARTICLE{2009MNRASMott,
       author = {{Motta}, S. and {Belloni}, T. and {Homan}, J.},
        title = "{The evolution of the high-energy cut-off in the X-ray spectrum of GX 339-4 across a hard-to-soft transition}",
      journal = {\mnras},
         year = 2009,
        month = dec,
       volume = {400},
       number = {3},
        pages = {1603-1612},
          doi = {10.1111/j.1365-2966.2009.15566.x},
archivePrefix = {arXiv},
       eprint = {0908.2451},
 primaryClass = {astro-ph.HE},
       adsurl = {https://ui.adsabs.harvard.edu/abs/2009MNRAS.400.1603M}
}

@ARTICLE{2015ApJGarc,
       author = {{Garc{\'\i}a}, Javier A. and {Steiner}, James F. and {McClintock}, Jeffrey E. and {Remillard}, Ronald A. and {Grinberg}, Victoria and {Dauser}, Thomas},
        title = "{X-Ray Reflection Spectroscopy of the Black Hole GX 339--4: Exploring the Hard State with Unprecedented Sensitivity}",
      journal = {\apj},
         year = 2015,
        month = nov,
       volume = {813},
       number = {2},
          eid = {84},
        pages = {84},
          doi = {10.1088/0004-637X/813/2/84},
archivePrefix = {arXiv},
       eprint = {1505.03607},
 primaryClass = {astro-ph.HE},
       adsurl = {https://ui.adsabs.harvard.edu/abs/2015ApJ...813...84G}
}

@ARTICLE{2018ApJXu,
       author = {{Xu}, Yanjun and {Harrison}, Fiona A. and {Garc{\'\i}a}, Javier A. and {Fabian}, Andrew C. and {F{\"u}rst}, Felix and {Gandhi}, Poshak and {Grefenstette}, Brian W. and {Madsen}, Kristin K. and {Miller}, Jon M. and {Parker}, Michael L. and {Tomsick}, John A. and {Walton}, Dominic J.},
        title = "{Reflection Spectra of the Black Hole Binary Candidate MAXI J1535-571 in the Hard State Observed by NuSTAR}",
      journal = {\apjl},
         year = 2018,
        month = jan,
       volume = {852},
       number = {2},
          eid = {L34},
        pages = {L34},
          doi = {10.3847/2041-8213/aaa4b2},
archivePrefix = {arXiv},
       eprint = {1711.01346},
 primaryClass = {astro-ph.HE},
       adsurl = {https://ui.adsabs.harvard.edu/abs/2018ApJ...852L..34X}
}

@ARTICLE{2023ATel16205Negoro,
       author = {{Negoro}, H. and {Serino}, M. and {Nakajima}, M. and {Kobayashi}, K. and {Tanaka}, M. and {Soejima}, Y. and {Kudo}, Y. and {Mihara}, T. and {Kawamuro}, T. and {Yamada}, S. and {Tamagawa}, T. and {Kawai}, N. and {Matsuoka}, M. and {Sakamoto}, T. and {Sugita}, S. and {Hiramatsu}, H. and {Nishikawa}, H. and {Yoshida}, A. and {Tsuboi}, Y. and {Urabe}, S. and {Nawa}, S. and {Nemoto}, N. and {Shidatsu}, M. and {Takahashi}, I. and {Niwano}, M. and {Sato}, S. and {Higuchi}, N. and {Yatsu}, Y. and {Nakahira}, S. and {Ueno}, S. and {Tomida}, H. and {Ishikawa}, M. and {Ogawa}, S. and {Kurihara}, T. and {Ueda}, Y. and {Setoguchi}, K. and {Yoshitake}, T. and {Nakatani}, Y. and {Yamauchi}, M. and {Hagiwara}, Y. and {Umeki}, Y. and {Otsuki}, Y. and {Yamaoka}, K. and {Kawakubo}, Y. and {Sugizaki}, M. and {Iwakiri}, W.},
        title = "{MAXI/GSC detection of a new hard X-ray transient Swift J1727.8-1613 (GRB 230824A)}",
      journal = {The Astronomer's Telegram},
         year = 2023,
        month = aug,
       volume = {16205},
        pages = {1},
       adsurl = {https://ui.adsabs.harvard.edu/abs/2023ATel16205....1N}
}

@ARTICLE{2023ATel16215Palmer,
       author = {{Palmer}, David M. and {Parsotan}, Tyler M.},
        title = "{Swift J1727.8-1613 reaches 7.6 Crab with strong QPO in Hard X-rays}",
      journal = {The Astronomer's Telegram},
         year = 2023,
        month = aug,
       volume = {16215},
        pages = {1},
       adsurl = {https://ui.adsabs.harvard.edu/abs/2023ATel16215....1P}
}

@ARTICLE{2024ApJIngram,
       author = {{Ingram}, Adam and {Bollemeijer}, Niek and {Veledina}, Alexandra and {Dov{\v{c}}iak}, Michal and {Poutanen}, Juri and {Egron}, Elise and {Russell}, Thomas D. and {Trushkin}, Sergei A. and {Negro}, Michela and {Ratheesh}, Ajay and {Capitanio}, Fiamma and {Connors}, Riley and {Neilsen}, Joseph and {Kraus}, Alexander and {Iacolina}, Maria Noemi and {Pellizzoni}, Alberto and {Pilia}, Maura and {Carotenuto}, Francesco and {Matt}, Giorgio and {Mastroserio}, Guglielmo and {Kaaret}, Philip and {Bianchi}, Stefano and {Garc{\'\i}a}, Javier A. and {Bachetti}, Matteo and {Wu}, Kinwah and {Costa}, Enrico and {Ewing}, Melissa and {Kravtsov}, Vadim and {Krawczynski}, Henric and {Loktev}, Vladislav and {Marinucci}, Andrea and {Marra}, Lorenzo and {Miku{\v{s}}incov{\'a}}, Romana and {Nathan}, Edward and {Parra}, Maxime and {Petrucci}, Pierre-Olivier and {Righini}, Simona and {Soffitta}, Paolo and {Steiner}, James F. and {Svoboda}, Ji{\v{r}}{\'\i} and {Tombesi}, Francesco and {Tugliani}, Stefano and {Ursini}, Francesco and {Yang}, Yi-Jung and {Zane}, Silvia and {Zhang}, Wenda and {Agudo}, Iv{\'a}n and {Antonelli}, Lucio A. and {Baldini}, Luca and {Baumgartner}, Wayne H. and {Bellazzini}, Ronaldo and {Bongiorno}, Stephen D. and {Bonino}, Raffaella and {Brez}, Alessandro and {Bucciantini}, Niccol{\`o} and {Castellano}, Simone and {Cavazzuti}, Elisabetta and {Chen}, Chien-Ting and {Ciprini}, Stefano and {De Rosa}, Alessandra and {Del Monte}, Ettore and {Di Gesu}, Laura and {Di Lalla}, Niccol{\`o} and {Di Marco}, Alessandro and {Donnarumma}, Immacolata and {Doroshenko}, Victor and {Ehlert}, Steven R. and {Enoto}, Teruaki and {Evangelista}, Yuri and {Fabiani}, Sergio and {Ferrazzoli}, Riccardo and {Gunji}, Shuichi and {Hayashida}, Kiyoshi and {Heyl}, Jeremy and {Iwakiri}, Wataru and {Jorstad}, Svetlana G. and {Karas}, Vladimir and {Kislat}, Fabian and {Kitaguchi}, Takao and {Kolodziejczak}, Jeffery J. and {La Monaca}, Fabio and {Latronico}, Luca and {Liodakis}, Ioannis and {Maldera}, Simone and {Manfreda}, Alberto and {Marin}, Fr{\'e}d{\'e}ric and {Marscher}, Alan P. and {Marshall}, Herman L. and {Massaro}, Francesco and {Mitsuishi}, Ikuyuki and {Mizuno}, Tsunefumi and {Muleri}, Fabio and {Ng}, Chi-Yung and {O'Dell}, Stephen L. and {Omodei}, Nicola and {Oppedisano}, Chiara and {Papitto}, Alessandro and {Pavlov}, George G. and {Peirson}, Abel L. and {Perri}, Matteo and {Pesce-Rollins}, Melissa and {Possenti}, Andrea and {Puccetti}, Simonetta and {Ramsey}, Brian D. and {Rankin}, John and {Roberts}, Oliver J. and {Romani}, Roger W. and {Sgr{\`o}}, Carmelo and {Slane}, Patrick and {Spandre}, Gloria and {Swartz}, Douglas A. and {Tamagawa}, Toru and {Tavecchio}, Fabrizio and {Taverna}, Roberto and {Tawara}, Yuzuru and {Tennant}, Allyn F. and {Thomas}, Nicholas E. and {Trois}, Alessio and {Tsygankov}, Sergey S. and {Turolla}, Roberto and {Vink}, Jacco and {Weisskopf}, Martin C. and {Xie}, Fei and {IXPE Collaboration}},
        title = "{Tracking the X-Ray Polarization of the Black Hole Transient Swift J1727.8{\textendash}1613 during a State Transition}",
      journal = {\apj},
         year = 2024,
        month = jun,
       volume = {968},
       number = {2},
          eid = {76},
        pages = {76},
          doi = {10.3847/1538-4357/ad3faf},
archivePrefix = {arXiv},
       eprint = {2311.05497},
 primaryClass = {astro-ph.HE},
       adsurl = {https://ui.adsabs.harvard.edu/abs/2024ApJ...968...76I}
}

@ARTICLE{2024A&AMata,
       author = {{Mata S{\'a}nchez}, D. and {Mu{\~n}oz-Darias}, T. and {Armas Padilla}, M. and {Casares}, J. and {Torres}, M.~A.~P.},
        title = "{Evidence for inflows and outflows in the nearby black hole transient Swift J1727.8{\ensuremath{-}}162}",
      journal = {\aap},
         year = 2024,
        month = feb,
       volume = {682},
          eid = {L1},
        pages = {L1},
          doi = {10.1051/0004-6361/202348754},
archivePrefix = {arXiv},
       eprint = {2401.04107},
 primaryClass = {astro-ph.HE},
       adsurl = {https://ui.adsabs.harvard.edu/abs/2024A&A...682L...1M}
}

@ARTICLE{2024ApJZhao,
       author = {{Zhao}, Qing-Chang and {Tao}, Lian and {Li}, Han-Cheng and {Zhang}, Shuang-Nan and {Feng}, Hua and {Ge}, Ming-Yu and {Ji}, Long and {Wang}, Ya-Nan and {Huang}, Yue and {Ma}, Xiang and {Zhang}, Liang and {Qu}, Jin-Lu and {Xu}, Yan-Jun and {Zhang}, Shu and {Yin}, Qian-Qing and {Shui}, Qing-Cang and {Ma}, Rui-Can and {Zhao}, Shu-Jie and {Li}, Pan-Ping and {Yang}, Zi-Xu and {Liu}, He-Xin and {Yu}, Wei},
        title = "{The First Polarimetric View on Quasiperiodic Oscillations in a Black Hole X-Ray Binary}",
      journal = {\apjl},
         year = 2024,
        month = feb,
       volume = {961},
       number = {2},
          eid = {L42},
        pages = {L42},
          doi = {10.3847/2041-8213/ad1e6c},
archivePrefix = {arXiv},
       eprint = {2401.08970},
 primaryClass = {astro-ph.HE},
       adsurl = {https://ui.adsabs.harvard.edu/abs/2024ApJ...961L..42Z}
}

@ARTICLE{2023ApJLiuhh,
       author = {{Liu}, Honghui and {Bambi}, Cosimo and {Jiang}, Jiachen and {Garc{\'\i}a}, Javier A. and {Ji}, Long and {Kong}, Lingda and {Ren}, Xiaoqin and {Zhang}, Shu and {Zhang}, Shuangnan},
        title = "{The Hard-to-soft Transition of GX 339-4 as Seen by Insight-HXMT}",
      journal = {\apj},
         year = 2023,
        month = jun,
       volume = {950},
       number = {1},
          eid = {5},
        pages = {5},
          doi = {10.3847/1538-4357/acca17},
archivePrefix = {arXiv},
       eprint = {2211.09543},
 primaryClass = {astro-ph.HE},
       adsurl = {https://ui.adsabs.harvard.edu/abs/2023ApJ...950....5L}
}

@ARTICLE{2017ApJWalton,
       author = {{Walton}, D.~J. and {Mooley}, K. and {King}, A.~L. and {Tomsick}, J.~A. and {Miller}, J.~M. and {Dauser}, T. and {Garc{\'\i}a}, J.~A. and {Bachetti}, M. and {Brightman}, M. and {Fabian}, A.~C. and {Forster}, K. and {F{\"u}rst}, F. and {Gandhi}, P. and {Grefenstette}, B.~W. and {Harrison}, F.~A. and {Madsen}, K.~K. and {Meier}, D.~L. and {Middleton}, M.~J. and {Natalucci}, L. and {Rahoui}, F. and {Rana}, V. and {Stern}, D.},
        title = "{Living on a Flare: Relativistic Reflection in V404 Cyg Observed by NuSTAR during Its Summer 2015 Outburst}",
      journal = {\apj},
         year = 2017,
        month = apr,
       volume = {839},
       number = {2},
          eid = {110},
        pages = {110},
          doi = {10.3847/1538-4357/aa67e8},
archivePrefix = {arXiv},
       eprint = {1609.01293},
 primaryClass = {astro-ph.HE},
       adsurl = {https://ui.adsabs.harvard.edu/abs/2017ApJ...839..110W}
}

@ARTICLE{2022SciKrawczynski,
       author = {{Krawczynski}, Henric and {Muleri}, Fabio and {Dov{\v{c}}iak}, Michal and {Veledina}, Alexandra and {Rodriguez Cavero}, Nicole and {Svoboda}, Jiri and {Ingram}, Adam and {Matt}, Giorgio and {Garcia}, Javier A. and {Loktev}, Vladislav and {Negro}, Michela and {Poutanen}, Juri and {Kitaguchi}, Takao and {Podgorn{\'y}}, Jakub and {Rankin}, John and {Zhang}, Wenda and {Berdyugin}, Andrei and {Berdyugina}, Svetlana V. and {Bianchi}, Stefano and {Blinov}, Dmitry and {Capitanio}, Fiamma and {Di Lalla}, Niccol{\`o} and {Draghis}, Paul and {Fabiani}, Sergio and {Kagitani}, Masato and {Kravtsov}, Vadim and {Kiehlmann}, Sebastian and {Latronico}, Luca and {Lutovinov}, Alexander A. and {Mandarakas}, Nikos and {Marin}, Fr{\'e}d{\'e}ric and {Marinucci}, Andrea and {Miller}, Jon M. and {Mizuno}, Tsunefumi and {Molkov}, Sergey V. and {Omodei}, Nicola and {Petrucci}, Pierre-Olivier and {Ratheesh}, Ajay and {Sakanoi}, Takeshi and {Semena}, Andrei N. and {Skalidis}, Raphael and {Soffitta}, Paolo and {Tennant}, Allyn F. and {Thalhammer}, Phillipp and {Tombesi}, Francesco and {Weisskopf}, Martin C. and {Wilms}, Joern and {Zhang}, Sixuan and {Agudo}, Iv{\'a}n and {Antonelli}, Lucio A. and {Bachetti}, Matteo and {Baldini}, Luca and {Baumgartner}, Wayne H. and {Bellazzini}, Ronaldo and {Bongiorno}, Stephen D. and {Bonino}, Raffaella and {Brez}, Alessandro and {Bucciantini}, Niccol{\`o} and {Castellano}, Simone and {Cavazzuti}, Elisabetta and {Ciprini}, Stefano and {Costa}, Enrico and {De Rosa}, Alessandra and {Del Monte}, Ettore and {Di Gesu}, Laura and {Di Marco}, Alessandro and {Donnarumma}, Immacolata and {Doroshenko}, Victor and {Ehlert}, Steven R. and {Enoto}, Teruaki and {Evangelista}, Yuri and {Ferrazzoli}, Riccardo and {Gunji}, Shuichi and {Hayashida}, Kiyoshi and {Heyl}, Jeremy and {Iwakiri}, Wataru and {Jorstad}, Svetlana G. and {Karas}, Vladimir and {Kolodziejczak}, Jeffery J. and {La Monaca}, Fabio and {Liodakis}, Ioannis and {Maldera}, Simone and {Manfreda}, Alberto and {Marscher}, Alan P. and {Marshall}, Herman L. and {Mitsuishi}, Ikuyuki and {Ng}, Chi-Yung and {O{\textquoteright}Dell}, Stephen L. and {Oppedisano}, Chiara and {Papitto}, Alessandro and {Pavlov}, George G. and {Peirson}, Abel L. and {Perri}, Matteo and {Pesce-Rollins}, Melissa and {Pilia}, Maura and {Possenti}, Andrea and {Puccetti}, Simonetta and {Ramsey}, Brian D. and {Romani}, Roger W. and {Sgr{\`o}}, Carmelo and {Slane}, Patrick and {Spandre}, Gloria and {Tamagawa}, Toru and {Tavecchio}, Fabrizio and {Taverna}, Roberto and {Tawara}, Yuzuru and {Thomas}, Nicholas E. and {Trois}, Alessio and {Tsygankov}, Sergey and {Turolla}, Roberto and {Vink}, Jacco and {Wu}, Kinwah and {Xie}, Fei and {Zane}, Silvia},
        title = "{Polarized x-rays constrain the disk-jet geometry in the black hole x-ray binary Cygnus X-1}",
      journal = {Science},
         year = 2022,
        month = nov,
       volume = {378},
       number = {6620},
        pages = {650-654},
          doi = {10.1126/science.add5399},
archivePrefix = {arXiv},
       eprint = {2206.09972},
 primaryClass = {astro-ph.HE},
       adsurl = {https://ui.adsabs.harvard.edu/abs/2022Sci...378..650K}
}

@ARTICLE{2003A&ACorbel,
       author = {{Corbel}, S. and {Nowak}, M.~A. and {Fender}, R.~P. and {Tzioumis}, A.~K. and {Markoff}, S.},
        title = "{Radio/X-ray correlation in the low/hard state of GX 339-4}",
      journal = {\aap},
         year = 2003,
        month = mar,
       volume = {400},
        pages = {1007-1012},
          doi = {10.1051/0004-6361:20030090},
archivePrefix = {arXiv},
       eprint = {astro-ph/0301436},
 primaryClass = {astro-ph},
       adsurl = {https://ui.adsabs.harvard.edu/abs/2003A&A...400.1007C}
}

@ARTICLE{2003MNRASMerloni,
       author = {{Merloni}, Andrea and {Heinz}, Sebastian and {di Matteo}, Tiziana},
        title = "{A Fundamental Plane of black hole activity}",
      journal = {\mnras},
         year = 2003,
        month = nov,
       volume = {345},
       number = {4},
        pages = {1057-1076},
          doi = {10.1046/j.1365-2966.2003.07017.x},
archivePrefix = {arXiv},
       eprint = {astro-ph/0305261},
 primaryClass = {astro-ph},
       adsurl = {https://ui.adsabs.harvard.edu/abs/2003MNRAS.345.1057M}
}

@ARTICLE{2012MNRASGallo,
       author = {{Gallo}, Elena and {Miller}, Brendan P. and {Fender}, Rob},
        title = "{Assessing luminosity correlations via cluster analysis: evidence for dual tracks in the radio/X-ray domain of black hole X-ray binaries}",
      journal = {\mnras},
         year = 2012,
        month = jun,
       volume = {423},
       number = {1},
        pages = {590-599},
          doi = {10.1111/j.1365-2966.2012.20899.x},
archivePrefix = {arXiv},
       eprint = {1203.4263},
 primaryClass = {astro-ph.HE},
       adsurl = {https://ui.adsabs.harvard.edu/abs/2012MNRAS.423..590G}
}

@ARTICLE{1999ApJBeloborodov,
       author = {{Beloborodov}, Andrei M.},
        title = "{Plasma Ejection from Magnetic Flares and the X-Ray Spectrum of Cygnus X-1}",
      journal = {\apjl},
         year = 1999,
        month = jan,
       volume = {510},
       number = {2},
        pages = {L123-L126},
          doi = {10.1086/311810},
archivePrefix = {arXiv},
       eprint = {astro-ph/9809383},
 primaryClass = {astro-ph},
       adsurl = {https://ui.adsabs.harvard.edu/abs/1999ApJ...510L.123B}
}

@ARTICLE{2004A&AFalcke,
       author = {{Falcke}, H. and {K{\"o}rding}, E. and {Markoff}, S.},
        title = "{A scheme to unify low-power accreting black holes. Jet-dominated accretion flows and the radio/X-ray correlation}",
      journal = {\aap},
         year = 2004,
        month = feb,
       volume = {414},
        pages = {895-903},
          doi = {10.1051/0004-6361:20031683},
archivePrefix = {arXiv},
       eprint = {astro-ph/0305335},
 primaryClass = {astro-ph},
       adsurl = {https://ui.adsabs.harvard.edu/abs/2004A&A...414..895F}
}

@ARTICLE{2006MNRASKording,
       author = {{K{\"o}rding}, E.~G. and {Fender}, R.~P. and {Migliari}, S.},
        title = "{Jet-dominated advective systems: radio and X-ray luminosity dependence on the accretion rate}",
      journal = {\mnras},
         year = 2006,
        month = jul,
       volume = {369},
       number = {3},
        pages = {1451-1458},
          doi = {10.1111/j.1365-2966.2006.10383.x},
archivePrefix = {arXiv},
       eprint = {astro-ph/0603731},
 primaryClass = {astro-ph},
       adsurl = {https://ui.adsabs.harvard.edu/abs/2006MNRAS.369.1451K}
}

@ARTICLE{2010ApJTitarchuk,
       author = {{Titarchuk}, Lev and {Shaposhnikov}, Nikolai},
        title = "{Implication of the Observed Spectral Cutoff Energy Evolution in XTE J1550-564}",
      journal = {\apj},
         year = 2010,
        month = dec,
       volume = {724},
       number = {2},
        pages = {1147-1152},
          doi = {10.1088/0004-637X/724/2/1147},
archivePrefix = {arXiv},
       eprint = {1009.4377},
 primaryClass = {astro-ph.HE},
       adsurl = {https://ui.adsabs.harvard.edu/abs/2010ApJ...724.1147T}
}

@ARTICLE{2017ApJXu,
       author = {{Xu}, Yanjun and {Garc{\'\i}a}, Javier A. and {F{\"u}rst}, Felix and {Harrison}, Fiona A. and {Walton}, Dominic J. and {Tomsick}, John A. and {Bachetti}, Matteo and {King}, Ashley L. and {Madsen}, Kristin K. and {Miller}, Jon M. and {Grinberg}, Victoria},
        title = "{Spectral and Timing Properties of IGR J17091-3624 in the Rising Hard State During Its 2016 Outburst}",
      journal = {\apj},
         year = 2017,
        month = dec,
       volume = {851},
       number = {2},
          eid = {103},
        pages = {103},
          doi = {10.3847/1538-4357/aa9ab4},
archivePrefix = {arXiv},
       eprint = {1711.04421},
 primaryClass = {astro-ph.HE},
       adsurl = {https://ui.adsabs.harvard.edu/abs/2017ApJ...851..103X}
}

@ARTICLE{2015A&APepe,
       author = {{Pepe}, Carolina and {Vila}, Gabriela S. and {Romero}, Gustavo E.},
        title = "{Lepto-hadronic model for the broadband emission of Cygnus X-1}",
      journal = {\aap},
         year = 2015,
        month = dec,
       volume = {584},
          eid = {A95},
        pages = {A95},
          doi = {10.1051/0004-6361/201527156},
archivePrefix = {arXiv},
       eprint = {1509.08514},
 primaryClass = {astro-ph.HE},
       adsurl = {https://ui.adsabs.harvard.edu/abs/2015A&A...584A..95P}
}

@ARTICLE{2021NewARMotta,
       author = {{Motta}, S.~E. and {Rodriguez}, J. and {Jourdain}, E. and {Del Santo}, M. and {Belanger}, G. and {Cangemi}, F. and {Grinberg}, V. and {Kajava}, J.~J.~E. and {Kuulkers}, E. and {Malzac}, J. and {Pottschmidt}, K. and {Roques}, J.~P. and {S{\'a}nchez-Fern{\'a}ndez}, C. and {Wilms}, J.},
        title = "{The INTEGRAL view on black hole X-ray binaries}",
      journal = {\nar},
         year = 2021,
        month = dec,
       volume = {93},
          eid = {101618},
        pages = {101618},
          doi = {10.1016/j.newar.2021.101618},
archivePrefix = {arXiv},
       eprint = {2105.05547},
 primaryClass = {astro-ph.HE},
       adsurl = {https://ui.adsabs.harvard.edu/abs/2021NewAR..9301618M}
}

@ARTICLE{2023GCNKennea,
       author = {{Kennea}, Jamie A. and {Swift Team}},
        title = "{GRB 230824A is likely a Galactic Transient: Swift J1727.8-1613}",
      journal = {GRB Coordinates Network},
         year = 2023,
        month = aug,
       volume = {34540},
        pages = {1},
       adsurl = {https://ui.adsabs.harvard.edu/abs/2023GCN.34540....1K}
}

@ARTICLE{2020ApJHare,
       author = {{Hare}, Jeremy and {Tomsick}, John A. and {Buisson}, Douglas J.~K. and {Clavel}, Ma{\"\i}ca and {Gandhi}, Poshak and {Garc{\'\i}a}, Javier A. and {Grefenstette}, Brian W. and {Walton}, Dominic J. and {Xu}, Yanjun},
        title = "{NuSTAR Observations of the Transient Galactic Black Hole Binary Candidate Swift J1858.6-0814: A New Sibling of V404 Cyg and V4641 Sgr?}",
      journal = {\apj},
         year = 2020,
        month = feb,
       volume = {890},
       number = {1},
          eid = {57},
        pages = {57},
          doi = {10.3847/1538-4357/ab6a12},
archivePrefix = {arXiv},
       eprint = {2001.03214},
 primaryClass = {astro-ph.HE},
       adsurl = {https://ui.adsabs.harvard.edu/abs/2020ApJ...890...57H}
}

@ARTICLE{2009ApJCaballero,
       author = {{Caballero-Garc{\'\i}a}, M.~D. and {Miller}, J.~M. and {Trigo}, M. D{\'\i}az and {Kuulkers}, E. and {Fabian}, A.~C. and {Mas-Hesse}, J.~M. and {Steeghs}, D. and {van der Klis}, M.},
        title = "{INTEGRAL and XMM-Newton Spectroscopy of GX 339-4 during Hard/Soft Intermediate and High/Soft States in the 2007 Outburst}",
      journal = {\apj},
         year = 2009,
        month = feb,
       volume = {692},
       number = {2},
        pages = {1339-1353},
          doi = {10.1088/0004-637X/692/2/1339},
archivePrefix = {arXiv},
       eprint = {0810.5470},
 primaryClass = {astro-ph},
       adsurl = {https://ui.adsabs.harvard.edu/abs/2009ApJ...692.1339C}
}

@INPROCEEDINGS{2000HEADCoppi,
       author = {{Coppi}, P.~S.},
        title = "{EQPAIR: A Hybrid Thermal/Non-Thermal Model for the Spectra of X-Ray Binaries}",
    booktitle = {AAS/High Energy Astrophysics Division \#5},
         year = 2000,
       series = {AAS/High Energy Astrophysics Division},
       volume = {5},
        month = oct,
          eid = {23.11},
        pages = {23.11},
       adsurl = {https://ui.adsabs.harvard.edu/abs/2000HEAD....5.2311C}
}

@ARTICLE{2021A&ACangemi,
       author = {{Cangemi}, F. and {Beuchert}, T. and {Siegert}, T. and {Rodriguez}, J. and {Grinberg}, V. and {Belmont}, R. and {Gouiff{\`e}s}, C. and {Kreykenbohm}, I. and {Laurent}, P. and {Pottschmidt}, K. and {Wilms}, J.},
        title = "{Potential origin of the state-dependent high-energy tail in the black hole microquasar Cygnus X-1 as seen with INTEGRAL}",
      journal = {\aap},
         year = 2021,
        month = jun,
       volume = {650},
          eid = {A93},
        pages = {A93},
          doi = {10.1051/0004-6361/202038604},
archivePrefix = {arXiv},
       eprint = {2102.04773},
 primaryClass = {astro-ph.HE},
       adsurl = {https://ui.adsabs.harvard.edu/abs/2021A&A...650A..93C}
}

@ARTICLE{2013MNRASSanto,
       author = {{Del Santo}, M. and {Malzac}, J. and {Belmont}, R. and {Bouchet}, L. and {De Cesare}, G.},
        title = "{The magnetic field in the X-ray corona of Cygnus X-1}",
      journal = {\mnras},
         year = 2013,
        month = mar,
       volume = {430},
       number = {1},
        pages = {209-220},
          doi = {10.1093/mnras/sts574},
archivePrefix = {arXiv},
       eprint = {1212.2040},
 primaryClass = {astro-ph.HE},
       adsurl = {https://ui.adsabs.harvard.edu/abs/2013MNRAS.430..209D}
}

@ARTICLE{2020ApJXu,
       author = {{Xu}, Yanjun and {Harrison}, Fiona A. and {Tomsick}, John A. and {Walton}, Dominic J. and {Barret}, Didier and {Garc{\'\i}a}, Javier A. and {Hare}, Jeremy and {Parker}, Michael L.},
        title = "{Studying the Reflection Spectra of the New Black Hole X-Ray Binary Candidate MAXI J1631-479 Observed by NuSTAR: A Variable Broad Iron Line Profile}",
      journal = {\apj},
         year = 2020,
        month = apr,
       volume = {893},
       number = {1},
          eid = {30},
        pages = {30},
          doi = {10.3847/1538-4357/ab7dc0},
archivePrefix = {arXiv},
       eprint = {2003.03465},
 primaryClass = {astro-ph.HE},
       adsurl = {https://ui.adsabs.harvard.edu/abs/2020ApJ...893...30X}
}

@ARTICLE{1988MNRASGuilbert,
       author = {{Guilbert}, P.~W. and {Rees}, M.~J.},
        title = "{'Cold' material in non-thermal sources.}",
      journal = {\mnras},
         year = 1988,
        month = jul,
       volume = {233},
        pages = {475-484},
          doi = {10.1093/mnras/233.2.475},
       adsurl = {https://ui.adsabs.harvard.edu/abs/1988MNRAS.233..475G}
}

@ARTICLE{1988ApJLightman,
       author = {{Lightman}, Alan P. and {White}, Timothy R.},
        title = "{Effects of Cold Matter in Active Galactic Nuclei: A Broad Hump in the X-Ray Spectra}",
      journal = {\apj},
         year = 1988,
        month = dec,
       volume = {335},
        pages = {57},
          doi = {10.1086/166905},
       adsurl = {https://ui.adsabs.harvard.edu/abs/1988ApJ...335...57L}
}

@ARTICLE{1976ApJShapiro,
       author = {{Shapiro}, S.~L. and {Lightman}, A.~P. and {Eardley}, D.~M.},
        title = "{A two-temperature accretion disk model for Cygnus X-1: structure and spectrum.}",
      journal = {\apj},
         year = 1976,
        month = feb,
       volume = {204},
        pages = {187-199},
          doi = {10.1086/154162},
       adsurl = {https://ui.adsabs.harvard.edu/abs/1976ApJ...204..187S}
}

@ARTICLE{2011SciLaurent,
       author = {{Laurent}, P. and {Rodriguez}, J. and {Wilms}, J. and {Cadolle Bel}, M. and {Pottschmidt}, K. and {Grinberg}, V.},
        title = "{Polarized Gamma-Ray Emission from the Galactic Black Hole Cygnus X-1}",
      journal = {Science},
         year = 2011,
        month = apr,
       volume = {332},
       number = {6028},
        pages = {438},
          doi = {10.1126/science.1200848},
archivePrefix = {arXiv},
       eprint = {1104.4282},
 primaryClass = {astro-ph.HE},
       adsurl = {https://ui.adsabs.harvard.edu/abs/2011Sci...332..438L}
}

@ARTICLE{2006A&AMalzac,
       author = {{Malzac}, J. and {Petrucci}, P.~O. and {Jourdain}, E. and {Cadolle Bel}, M. and {Sizun}, P. and {Pooley}, G. and {Cabanac}, C. and {Chaty}, S. and {Belloni}, T. and {Rodriguez}, J. and {Roques}, J.~P. and {Durouchoux}, P. and {Goldwurm}, A. and {Laurent}, P.},
        title = "{Bimodal spectral variability of <ASTROBJ>Cygnus X-1</ASTROBJ> in an intermediate state}",
      journal = {\aap},
         year = 2006,
        month = mar,
       volume = {448},
       number = {3},
        pages = {1125-1137},
          doi = {10.1051/0004-6361:20053614},
archivePrefix = {arXiv},
       eprint = {astro-ph/0511725},
 primaryClass = {astro-ph},
       adsurl = {https://ui.adsabs.harvard.edu/abs/2006A&A...448.1125M}
}

@ARTICLE{2013MNRASSutton,
       author = {{Sutton}, Andrew D. and {Roberts}, Timothy P. and {Middleton}, Matthew J.},
        title = "{The ultraluminous state revisited: fractional variability and spectral shape as diagnostics of super-Eddington accretion}",
      journal = {\mnras},
         year = 2013,
        month = oct,
       volume = {435},
       number = {2},
        pages = {1758-1775},
          doi = {10.1093/mnras/stt1419},
archivePrefix = {arXiv},
       eprint = {1307.8044},
 primaryClass = {astro-ph.HE},
       adsurl = {https://ui.adsabs.harvard.edu/abs/2013MNRAS.435.1758S}
}

@ARTICLE{2007MNRASPoutanen,
       author = {{Poutanen}, Juri and {Lipunova}, Galina and {Fabrika}, Sergei and {Butkevich}, Alexey G. and {Abolmasov}, Pavel},
        title = "{Supercritically accreting stellar mass black holes as ultraluminous X-ray sources}",
      journal = {\mnras},
         year = 2007,
        month = may,
       volume = {377},
       number = {3},
        pages = {1187-1194},
          doi = {10.1111/j.1365-2966.2007.11668.x},
archivePrefix = {arXiv},
       eprint = {astro-ph/0609274},
 primaryClass = {astro-ph},
       adsurl = {https://ui.adsabs.harvard.edu/abs/2007MNRAS.377.1187P}
}

@ARTICLE{2012ApJKawashima,
       author = {{Kawashima}, T. and {Ohsuga}, K. and {Mineshige}, S. and {Yoshida}, T. and {Heinzeller}, D. and {Matsumoto}, R.},
        title = "{Comptonized Photon Spectra of Supercritical Black Hole Accretion Flows with Application to Ultraluminous X-Ray Sources}",
      journal = {\apj},
         year = 2012,
        month = jun,
       volume = {752},
       number = {1},
          eid = {18},
        pages = {18},
          doi = {10.1088/0004-637X/752/1/18},
       adsurl = {https://ui.adsabs.harvard.edu/abs/2012ApJ...752...18K}
}

@ARTICLE{1971SvABisnovatyi-Kogan,
       author = {{Bisnovatyi-Kogan}, G.~S. and {Zel'dovich}, Ya. B. and {Syunyaev}, R.~A.},
        title = "{Physical Processes in a Low-Density Relativistic Plasma.}",
      journal = {\sovast},
         year = 1971,
        month = aug,
       volume = {15},
        pages = {17},
       adsurl = {https://ui.adsabs.harvard.edu/abs/1971SvA....15...17B}
}

@ARTICLE{2017MNRASBasak,
       author = {{Basak}, Rupal and {Zdziarski}, Andrzej A. and {Parker}, Michael and {Islam}, Nazma},
        title = "{Analysis of NuSTAR and Suzaku observations of Cyg X-1 in the hard state: evidence for a truncated disc geometry}",
      journal = {\mnras},
         year = 2017,
        month = dec,
       volume = {472},
       number = {4},
        pages = {4220-4232},
          doi = {10.1093/mnras/stx2283},
archivePrefix = {arXiv},
       eprint = {1705.06638},
 primaryClass = {astro-ph.HE},
       adsurl = {https://ui.adsabs.harvard.edu/abs/2017MNRAS.472.4220B}
}

@ARTICLE{2021MNRASDzielak,
       author = {{Dzie{\l}ak}, Marta A. and {De Marco}, Barbara and {Zdziarski}, Andrzej A.},
        title = "{A spectrally stratified hot accretion flow in the hard state of MAXI J1820+070}",
      journal = {\mnras},
         year = 2021,
        month = sep,
       volume = {506},
       number = {2},
        pages = {2020-2029},
          doi = {10.1093/mnras/stab1700},
archivePrefix = {arXiv},
       eprint = {2102.11635},
 primaryClass = {astro-ph.HE},
       adsurl = {https://ui.adsabs.harvard.edu/abs/2021MNRAS.506.2020D}
}

@ARTICLE{2001A&AMarkoff,
       author = {{Markoff}, S. and {Falcke}, H. and {Fender}, R.},
        title = "{A jet model for the broadband spectrum of XTE J1118+480. Synchrotron emission from radio to X-rays in the Low/Hard spectral state}",
      journal = {\aap},
         year = 2001,
        month = jun,
       volume = {372},
        pages = {L25-L28},
          doi = {10.1051/0004-6361:20010420},
archivePrefix = {arXiv},
       eprint = {astro-ph/0010560},
 primaryClass = {astro-ph},
       adsurl = {https://ui.adsabs.harvard.edu/abs/2001A&A...372L..25M}
}

@ARTICLE{2003A&AMarkoff,
       author = {{Markoff}, S. and {Nowak}, M. and {Corbel}, S. and {Fender}, R. and {Falcke}, H.},
        title = "{Exploring the role of jets in the radio/X-ray correlations of GX 339-4}",
      journal = {\aap},
         year = 2003,
        month = jan,
       volume = {397},
        pages = {645-658},
          doi = {10.1051/0004-6361:20021497},
archivePrefix = {arXiv},
       eprint = {astro-ph/0210439},
 primaryClass = {astro-ph},
       adsurl = {https://ui.adsabs.harvard.edu/abs/2003A&A...397..645M}
}

@ARTICLE{2018MNRASZdziarski,
       author = {{Zdziarski}, Andrzej A. and {Malyshev}, Denys and {Dubus}, Guillaume and {Pooley}, Guy G. and {Johnson}, Tyrel and {Frankowski}, Adam and {De Marco}, Barbara and {Chernyakova}, Maria and {Rao}, A.~R.},
        title = "{A comprehensive study of high-energy gamma-ray and radio emission from Cyg X-3}",
      journal = {\mnras},
         year = 2018,
        month = oct,
       volume = {479},
       number = {4},
        pages = {4399-4415},
          doi = {10.1093/mnras/sty1618},
archivePrefix = {arXiv},
       eprint = {1804.07460},
 primaryClass = {astro-ph.HE},
       adsurl = {https://ui.adsabs.harvard.edu/abs/2018MNRAS.479.4399Z}
}

@ARTICLE{2019ARA&ABlandford,
       author = {{Blandford}, Roger and {Meier}, David and {Readhead}, Anthony},
        title = "{Relativistic Jets from Active Galactic Nuclei}",
      journal = {\araa},
         year = 2019,
        month = aug,
       volume = {57},
        pages = {467-509},
          doi = {10.1146/annurev-astro-081817-051948},
archivePrefix = {arXiv},
       eprint = {1812.06025},
 primaryClass = {astro-ph.HE},
       adsurl = {https://ui.adsabs.harvard.edu/abs/2019ARA&A..57..467B}
}

@ARTICLE{2002ApJMcConnell,
       author = {{McConnell}, M.~L. and {Zdziarski}, A.~A. and {Bennett}, K. and {Bloemen}, H. and {Collmar}, W. and {Hermsen}, W. and {Kuiper}, L. and {Paciesas}, W. and {Phlips}, B.~F. and {Poutanen}, J. and {Ryan}, J.~M. and {Sch{\"o}nfelder}, V. and {Steinle}, H. and {Strong}, A.~W.},
        title = "{The Soft Gamma-Ray Spectral Variability of Cygnus X-1}",
      journal = {\apj},
         year = 2002,
        month = jun,
       volume = {572},
       number = {2},
        pages = {984-995},
          doi = {10.1086/340436},
archivePrefix = {arXiv},
       eprint = {astro-ph/0112326},
 primaryClass = {astro-ph},
       adsurl = {https://ui.adsabs.harvard.edu/abs/2002ApJ...572..984M}
}

@ARTICLE{2009MNRASMotta,
       author = {{Motta}, S. and {Belloni}, T. and {Homan}, J.},
        title = "{The evolution of the high-energy cut-off in the X-ray spectrum of GX 339-4 across a hard-to-soft transition}",
      journal = {\mnras},
         year = 2009,
        month = dec,
       volume = {400},
       number = {3},
        pages = {1603-1612},
          doi = {10.1111/j.1365-2966.2009.15566.x},
archivePrefix = {arXiv},
       eprint = {0908.2451},
 primaryClass = {astro-ph.HE},
       adsurl = {https://ui.adsabs.harvard.edu/abs/2009MNRAS.400.1603M}
}

@ARTICLE{2016ApJWalton,
       author = {{Walton}, D.~J. and {Tomsick}, J.~A. and {Madsen}, K.~K. and {Grinberg}, V. and {Barret}, D. and {Boggs}, S.~E. and {Christensen}, F.~E. and {Clavel}, M. and {Craig}, W.~W. and {Fabian}, A.~C. and {Fuerst}, F. and {Hailey}, C.~J. and {Harrison}, F.~A. and {Miller}, J.~M. and {Parker}, M.~L. and {Rahoui}, F. and {Stern}, D. and {Tao}, L. and {Wilms}, J. and {Zhang}, W.},
        title = "{The Soft State of Cygnus X-1 Observed with NuSTAR: A Variable Corona and a Stable Inner Disk}",
      journal = {\apj},
         year = 2016,
        month = jul,
       volume = {826},
       number = {1},
          eid = {87},
        pages = {87},
          doi = {10.3847/0004-637X/826/1/87},
archivePrefix = {arXiv},
       eprint = {1605.03966},
 primaryClass = {astro-ph.HE},
       adsurl = {https://ui.adsabs.harvard.edu/abs/2016ApJ...826...87W}
}

@INCOLLECTION{2022hxga.bookYuan,
       author = {{Yuan}, Weimin and {Zhang}, Chen and {Chen}, Yong and {Ling}, Zhixing},
        title = "{The Einstein Probe Mission}",
    booktitle = {Handbook of X-ray and Gamma-ray Astrophysics},
         year = 2022,
          eid = {86},
        pages = {86},
          doi = {10.1007/978-981-16-4544-0_151-1},
       adsurl = {https://ui.adsabs.harvard.edu/abs/2022hxga.book...86Y}
}

@ARTICLE{2014ARA&AYuan,
       author = {{Yuan}, Feng and {Narayan}, Ramesh},
        title = "{Hot Accretion Flows Around Black Holes}",
      journal = {\araa},
         year = 2014,
        month = aug,
       volume = {52},
        pages = {529-588},
          doi = {10.1146/annurev-astro-082812-141003},
archivePrefix = {arXiv},
       eprint = {1401.0586},
 primaryClass = {astro-ph.HE},
       adsurl = {https://ui.adsabs.harvard.edu/abs/2014ARA&A..52..529Y}
}

@ARTICLE{2023MNRASMaRuican,
       author = {{Ma}, Ruican and {M{\'e}ndez}, Mariano and {Garc{\'\i}a}, Federico and {Sai}, Na and {Zhang}, Liang and {Zhang}, Yuexin},
        title = "{A variable corona during the transition from type-C to type-B quasi-periodic oscillations in the black hole X-ray binary MAXI J1820+070}",
      journal = {\mnras},
         year = 2023,
        month = oct,
       volume = {525},
       number = {1},
        pages = {854-875},
          doi = {10.1093/mnras/stad2284},
archivePrefix = {arXiv},
       eprint = {2307.12728},
 primaryClass = {astro-ph.HE},
       adsurl = {https://ui.adsabs.harvard.edu/abs/2023MNRAS.525..854M}
}

@ARTICLE{2022NatAsMendez,
       author = {{M{\'e}ndez}, Mariano and {Karpouzas}, Konstantinos and {Garc{\'\i}a}, Federico and {Zhang}, Liang and {Zhang}, Yuexin and {Belloni}, Tomaso M. and {Altamirano}, Diego},
        title = "{Coupling between the accreting corona and the relativistic jet in the microquasar GRS 1915+105}",
      journal = {Nature Astronomy},
         year = 2022,
        month = mar,
       volume = {6},
        pages = {577-583},
          doi = {10.1038/s41550-022-01617-y},
archivePrefix = {arXiv},
       eprint = {2203.02963},
 primaryClass = {astro-ph.HE},
       adsurl = {https://ui.adsabs.harvard.edu/abs/2022NatAs...6..577M}
}

@ARTICLE{2022ApJLiuHX,
       author = {{Liu}, H.~X. and {Huang}, Y. and {Bu}, Q.~C. and {Yu}, W. and {Yang}, Z.~X. and {Zhang}, L. and {Kong}, L.~D. and {Xiao}, G.~C. and {Qu}, J.~L. and {Zhang}, S.~N. and {Zhang}, S. and {Song}, L.~M. and {Jia}, S.~M. and {Ma}, X. and {Tao}, L. and {Ge}, M.~Y. and {Liu}, Q.~Z. and {Yan}, J.~Z. and {Ma}, R.~C. and {Ren}, X.~Q. and {Zhou}, D.~K. and {Li}, T.~M. and {Wu}, B.~Y. and {Xu}, Y.~C. and {Du}, Y.~F. and {Fu}, Y.~C. and {Xiao}, Y.~X. and {Ding}, G.~Q. and {Yu}, X.~X.},
        title = "{Transitions and Origin of the Type-B Quasi-periodic Oscillations in the Black Hole X-Ray Binary MAXI J1348-630}",
      journal = {\apj},
         year = 2022,
        month = oct,
       volume = {938},
       number = {2},
          eid = {108},
        pages = {108},
          doi = {10.3847/1538-4357/ac88c6},
archivePrefix = {arXiv},
       eprint = {2208.07066},
 primaryClass = {astro-ph.HE},
       adsurl = {https://ui.adsabs.harvard.edu/abs/2022ApJ...938..108L}
}

@ARTICLE{2023MNRASYangZX,
       author = {{Yang}, Zi-Xu and {Zhang}, Liang and {Zhang}, S.~N. and {M{\'e}ndez}, M. and {Garc{\'\i}a}, Federico and {Huang}, Yue and {Bu}, Qingcui and {Liu}, He-Xin and {Yu}, Wei and {Wang}, P.~J. and {Tao}, L. and {Altamirano}, D. and {Qu}, Jin-Lu and {Zhang}, S. and {Ma}, X. and {Song}, L.~M. and {Jia}, S.~M. and {Ge}, M.~Y. and {Liu}, Q.~Z. and {Yan}, J.~Z. and {Li}, T.~M. and {Ren}, X.~Q. and {Ma}, R.~C. and {Zhang}, Yuexin and {Xu}, Y.~C. and {Ma}, B.~Y. and {Du}, Y.~F. and {Fu}, Y.~C. and {Xiao}, Y.~X. and {Li}, P.~P. and {Jin}, P. and {Zhao}, S.~J. and {Zhao}, Q.~C.},
        title = "{Fast transitions of X-ray variability in the black hole transient GX 339-4: comparison with MAXI J1820+070 and MAXI J1348-630}",
      journal = {\mnras},
         year = 2023,
        month = may,
       volume = {521},
       number = {3},
        pages = {3570-3584},
          doi = {10.1093/mnras/stad795},
archivePrefix = {arXiv},
       eprint = {2303.07588},
 primaryClass = {astro-ph.HE},
       adsurl = {https://ui.adsabs.harvard.edu/abs/2023MNRAS.521.3570Y}
}

@ARTICLE{2024ApJYang,
       author = {{Yang}, Zi-Xu and {Zhang}, Liang and {Zhang}, Shuang-Nan and {Tao}, Lian and {Zhang}, Shu and {Ma}, Ruican and {Bu}, Qing-Cui and {Huang}, Yue and {Liu}, He-Xin and {Yu}, Wei and {Xiao}, Guangcheng and {Wang}, Peng-Ju and {Feng}, Hua and {Song}, Li-Ming and {Ma}, Xiang and {Ge}, Mingyu and {Zhao}, Qing-Chang and {Qu}, Jin-Lu},
        title = "{A Timing View of the Additional High-energy Spectral Component Discovered in the Black Hole Candidate Swift J1727.8-1613}",
      journal = {\apjl},
         year = 2024,
        month = aug,
       volume = {970},
       number = {2},
          eid = {L33},
        pages = {L33},
          doi = {10.3847/2041-8213/ad60bd},
archivePrefix = {arXiv},
       eprint = {2407.05236},
 primaryClass = {astro-ph.HE},
       adsurl = {https://ui.adsabs.harvard.edu/abs/2024ApJ...970L..33Y}
}

@ARTICLE{2024A&ABouchet,
       author = {{Bouchet}, T. and {Rodriguez}, J. and {Cangemi}, F. and {Thalhammer}, P. and {Laurent}, P. and {Grinberg}, V. and {Wilms}, J. and {Pottschmidt}, K.},
        title = "{INTEGRAL/IBIS polarization detection in the hard and soft intermediate states of Swift J1727.8{\ensuremath{-}}1613}",
      journal = {\aap},
         year = 2024,
        month = aug,
       volume = {688},
          eid = {L5},
        pages = {L5},
          doi = {10.1051/0004-6361/202450826},
archivePrefix = {arXiv},
       eprint = {2407.05871},
 primaryClass = {astro-ph.HE},
       adsurl = {https://ui.adsabs.harvard.edu/abs/2024A&A...688L...5B}
}

@ARTICLE{2024MNRASYW,
       author = {{Yu}, Wei and {Bu}, Qing-Cui and {Zhang}, Shuang-Nan and {Liu}, He-Xin and {Zhang}, Liang and {Ducci}, Lorenzo and {Tao}, Lian and {Santangelo}, Andrea and {Doroshenko}, Victor and {Huang}, Yue and {Yang}, Zi-Xu and {Qu}, Jin-Lu},
        title = "{Timing analysis of the newly discovered black hole candidate Swift J1727.8-1613 with Insight-HXMT}",
      journal = {\mnras},
         year = 2024,
        month = apr,
       volume = {529},
       number = {4},
        pages = {4624-4632},
          doi = {10.1093/mnras/stae835},
archivePrefix = {arXiv},
       eprint = {2403.13127},
 primaryClass = {astro-ph.HE},
       adsurl = {https://ui.adsabs.harvard.edu/abs/2024MNRAS.529.4624Y}
}

@ARTICLE{2025ApJCao,
       author = {{Cao}, Jia-Ying and {Liao}, Jin-Yuan and {Zhang}, Shuang-Nan and {Feng}, Hua and {Qu}, Jin-Lu and {Zhang}, Liang and {Liu}, He-Xin and {Yu}, Wei and {Zhao}, Qing-Chang and {Peng}, Jing-Qiang and {Ge}, Ming-Yu and {Tao}, Lian and {Xu}, Yan-Jun and {Zhang}, Shu and {Yang}, Zi-Xu},
        title = "{Spectral Analysis of the X-Ray Flares in the 2023 Outburst of the New Black Binary Transient Swift J1727.8{\textendash}1613 Observed with Insight-HXMT}",
      journal = {\apj},
         year = 2025,
        month = apr,
       volume = {983},
       number = {1},
          eid = {23},
        pages = {23},
          doi = {10.3847/1538-4357/adbd0f},
archivePrefix = {arXiv},
       eprint = {2503.05411},
 primaryClass = {astro-ph.HE},
       adsurl = {https://ui.adsabs.harvard.edu/abs/2025ApJ...983...23C}
}

@ARTICLE{2021ApJZdziarski,
       author = {{Zdziarski}, Andrzej A. and {Jourdain}, Elisabeth and {Lubi{\'n}ski}, Piotr and {Szanecki}, Micha{\l} and {Nied{\'z}wiecki}, Andrzej and {Veledina}, Alexandra and {Poutanen}, Juri and {Dzie{\l}ak}, Marta A. and {Roques}, Jean-Pierre},
        title = "{Hybrid Comptonization and Electron-Positron Pair Production in the Black-hole X-Ray Binary MAXI J1820+070}",
      journal = {\apjl},
         year = 2021,
        month = jun,
       volume = {914},
       number = {1},
          eid = {L5},
        pages = {L5},
          doi = {10.3847/2041-8213/ac0147},
archivePrefix = {arXiv},
       eprint = {2104.04316},
 primaryClass = {astro-ph.HE},
       adsurl = {https://ui.adsabs.harvard.edu/abs/2021ApJ...914L...5Z}
}

@ARTICLE{1996ApJPoutanen,
       author = {{Poutanen}, Juri and {Svensson}, Roland},
        title = "{The Two-Phase Pair Corona Model for Active Galactic Nuclei and X-Ray Binaries: How to Obtain Exact Solutions}",
      journal = {\apj},
         year = 1996,
        month = oct,
       volume = {470},
        pages = {249},
          doi = {10.1086/177865},
archivePrefix = {arXiv},
       eprint = {astro-ph/9605073},
 primaryClass = {astro-ph},
       adsurl = {https://ui.adsabs.harvard.edu/abs/1996ApJ...470..249P}
}

@ARTICLE{2010ApJGarca,
       author = {{Garc{\'\i}a}, J. and {Kallman}, T.~R.},
        title = "{X-ray Reflected Spectra from Accretion Disk Models. I. Constant Density Atmospheres}",
      journal = {\apj},
         year = 2010,
        month = aug,
       volume = {718},
       number = {2},
        pages = {695-706},
          doi = {10.1088/0004-637X/718/2/695},
archivePrefix = {arXiv},
       eprint = {1006.0485},
 primaryClass = {astro-ph.HE},
       adsurl = {https://ui.adsabs.harvard.edu/abs/2010ApJ...718..695G}
}

@ARTICLE{2018ApJGarca,
       author = {{Garc{\'\i}a}, Javier A. and {Steiner}, James F. and {Grinberg}, Victoria and {Dauser}, Thomas and {Connors}, Riley M.~T. and {McClintock}, Jeffrey E. and {Remillard}, Ronald A. and {Wilms}, J{\"o}rn and {Harrison}, Fiona A. and {Tomsick}, John A.},
        title = "{Reflection Spectroscopy of the Black Hole Binary XTE J1752-223 in Its Long-stable Hard State}",
      journal = {\apj},
         year = 2018,
        month = sep,
       volume = {864},
       number = {1},
          eid = {25},
        pages = {25},
          doi = {10.3847/1538-4357/aad231},
archivePrefix = {arXiv},
       eprint = {1807.01949},
 primaryClass = {astro-ph.HE},
       adsurl = {https://ui.adsabs.harvard.edu/abs/2018ApJ...864...25G}
}

@ARTICLE{1995MNRASMagdziarz,
       author = {{Magdziarz}, Pawel and {Zdziarski}, Andrzej A.},
        title = "{Angle-dependent Compton reflection of X-rays and gamma-rays}",
      journal = {\mnras},
         year = 1995,
        month = apr,
       volume = {273},
       number = {3},
        pages = {837-848},
          doi = {10.1093/mnras/273.3.837},
       adsurl = {https://ui.adsabs.harvard.edu/abs/1995MNRAS.273..837M}
}

@ARTICLE{2021ApJLZdziarskiA,
       author = {{Zdziarski}, Andrzej A. and {Dzie{\l}ak}, Marta A. and {De Marco}, Barbara and {Szanecki}, Micha{\l} and {Nied{\'z}wiecki}, Andrzej},
        title = "{Accretion Geometry in the Hard State of the Black Hole X-Ray Binary MAXI J1820+070}",
      journal = {\apjl},
         year = 2021,
        month = mar,
       volume = {909},
       number = {1},
          eid = {L9},
        pages = {L9},
          doi = {10.3847/2041-8213/abe7ef},
archivePrefix = {arXiv},
       eprint = {2101.04482},
 primaryClass = {astro-ph.HE},
       adsurl = {https://ui.adsabs.harvard.edu/abs/2021ApJ...909L...9Z}
}

@ARTICLE{2025ApJHughes,
       author = {{Hughes}, Andrew K. and {Carotenuto}, Francesco and {Russell}, Thomas D. and {Tetarenko}, Alexandra J. and {Miller-Jones}, James C.~A. and {Bahramian}, Arash and {Bright}, Joe S. and {Cowie}, Fraser J. and {Fender}, Rob and {Gurwell}, Mark A. and {Khaulsay}, Jasvinderjit K. and {Kirby}, Anastasia and {Jones}, Serena and {Lescure}, Elodie and {McCollough}, Michael and {Plotkin}, Richard M. and {Rao}, Ramprasad and {Vrtilek}, Saeqa D. and {Williams-Baldwin}, David R.~A. and {Wood}, Callan M. and {Sivakoff}, Gregory R. and {Altamirano}, Diego and {Casella}, Piergiorgio and {Corbel}, St{\'e}phane and {DeBoer}, David R. and {Del Santo}, Melania and {Echibur{\'u}-Trujillo}, Constanza and {Farah}, Wael and {Gandhi}, Poshak and {Koljonen}, Karri I.~I. and {Maccarone}, Thomas and {Matthews}, James H. and {Markoff}, Sera B. and {Pollak}, Alexander W. and {Russell}, David M. and {Saikia}, Payaswini and {Castro Segura}, Noel and {Shaw}, Aarran W. and {Siemion}, Andrew and {Soria}, Roberto and {Tomsick}, John A. and {van den Eijnden}, Jakob},
        title = "{Comprehensive Radio Monitoring of the Black Hole X-Ray Binary Swift J1727.8‑1613 during Its 2023{\textendash}2024 Outburst}",
      journal = {\apj},
         year = 2025,
        month = jul,
       volume = {988},
       number = {1},
          eid = {109},
        pages = {109},
          doi = {10.3847/1538-4357/ade2e6},
archivePrefix = {arXiv},
       eprint = {2506.07798},
 primaryClass = {astro-ph.HE},
       adsurl = {https://ui.adsabs.harvard.edu/abs/2025ApJ...988..109H}
}

@ARTICLE{1978MNRASBell,
       author = {{Bell}, A.~R.},
        title = "{The acceleration of cosmic rays in shock fronts - I.}",
      journal = {\mnras},
         year = 1978,
        month = jan,
       volume = {182},
        pages = {147-156},
          doi = {10.1093/mnras/182.2.147},
       adsurl = {https://ui.adsabs.harvard.edu/abs/1978MNRAS.182..147B}
}

@ARTICLE{1978ApJBlandford,
       author = {{Blandford}, R.~D. and {Ostriker}, J.~P.},
        title = "{Particle acceleration by astrophysical shocks.}",
      journal = {\apjl},
         year = 1978,
        month = apr,
       volume = {221},
        pages = {L29-L32},
          doi = {10.1086/182658},
       adsurl = {https://ui.adsabs.harvard.edu/abs/1978ApJ...221L..29B}
}

@ARTICLE{2014ApJSironi,
       author = {{Sironi}, Lorenzo and {Spitkovsky}, Anatoly},
        title = "{Relativistic Reconnection: An Efficient Source of Non-thermal Particles}",
      journal = {\apjl},
         year = 2014,
        month = mar,
       volume = {783},
       number = {1},
          eid = {L21},
        pages = {L21},
          doi = {10.1088/2041-8205/783/1/L21},
archivePrefix = {arXiv},
       eprint = {1401.5471},
 primaryClass = {astro-ph.HE},
       adsurl = {https://ui.adsabs.harvard.edu/abs/2014ApJ...783L..21S}
}

@ARTICLE{2026MNRASSuriano,
       author = {{Suriano}, Alessio and {Nurisso}, Matteo and {Celotti}, Annalisa and {Mignone}, Andrea and {Bodo}, Gianluigi},
        title = "{Particles acceleration with magnetic reconnection in large scale RMHD simulations─II. Particle spectra}",
      journal = {\mnras},
         year = 2026,
        month = mar,
       volume = {546},
       number = {4},
          eid = {stag322},
        pages = {stag322},
          doi = {10.1093/mnras/stag322},
       adsurl = {https://ui.adsabs.harvard.edu/abs/2026MNRAS.546ag322S}
}

@ARTICLE{1970RvMPBlumenthal,
       author = {{Blumenthal}, George R. and {Gould}, Robert J.},
        title = "{Bremsstrahlung, Synchrotron Radiation, and Compton Scattering of High-Energy Electrons Traversing Dilute Gases}",
      journal = {Reviews of Modern Physics},
         year = 1970,
        month = jan,
       volume = {42},
       number = {2},
        pages = {237-271},
          doi = {10.1103/RevModPhys.42.237},
       adsurl = {https://ui.adsabs.harvard.edu/abs/1970RvMP...42..237B}
}

@ARTICLE{1962SvAKardashev,
       author = {{Kardashev}, N.~S.},
        title = "{Nonstationarity of Spectra of Young Sources of Nonthermal Radio Emission}",
      journal = {\sovast},
         year = 1962,
        month = dec,
       volume = {6},
        pages = {317},
       adsurl = {https://ui.adsabs.harvard.edu/abs/1962SvA.....6..317K}
}

@ARTICLE{2015MNRASDrappeau,
       author = {{Drappeau}, S. and {Malzac}, J. and {Belmont}, R. and {Gandhi}, P. and {Corbel}, S.},
        title = "{Internal shocks driven by accretion flow variability in the compact jet of the black hole binary GX 339-4}",
      journal = {\mnras},
         year = 2015,
        month = mar,
       volume = {447},
       number = {4},
        pages = {3832-3839},
          doi = {10.1093/mnras/stu2711},
archivePrefix = {arXiv},
       eprint = {1412.5819},
 primaryClass = {astro-ph.HE},
       adsurl = {https://ui.adsabs.harvard.edu/abs/2015MNRAS.447.3832D}
}

@ARTICLE{2002MNRASWardziski,
       author = {{Wardzi{\'n}ski}, Grzegorz and {Zdziarski}, Andrzej A. and {Gierli{\'n}ski}, Marek and {Grove}, J. Eric and {Jahoda}, Keith and {Johnson}, W. Neil},
        title = "{X-ray and {\ensuremath{\gamma}}-ray spectra and variability of the black hole candidate GX 339-4}",
      journal = {\mnras},
         year = 2002,
        month = dec,
       volume = {337},
       number = {3},
        pages = {829-839},
          doi = {10.1046/j.1365-8711.2002.05914.x},
archivePrefix = {arXiv},
       eprint = {astro-ph/0207598},
 primaryClass = {astro-ph},
       adsurl = {https://ui.adsabs.harvard.edu/abs/2002MNRAS.337..829W}
}

\bsp	
\label{lastpage}
\end{document}